\documentclass[aps,prd,twocolumn,showpacs,groupedaddress]{revtex4}  
\usepackage{graphicx}  
\usepackage{dcolumn}   
\usepackage{bm}        
\usepackage{amssymb}   

\newcommand{\MET}   {\mbox{$\not\!\!E_T$}}
\newcommand{\met}   {\mbox{$\not\!\!E_T$}}

\newcommand{\ppbar} {\mbox{$p\bar{p}$}}
\newcommand{\ttbar} {\mbox{$t\bar{t}$}}
\newcommand{\bbar}  {\mbox{$\bar{b}$}}
\newcommand{\bbbar} {\mbox{$b\bar{b}$}}

\newcommand{\rar}   {\mbox{$\rightarrow$}}

\newcommand{\dzero}     {D\O\ }

\newcommand{\comphep}   {C\rm{omp}\sc{hep}}
\newcommand{\herwig}    {\sc{herwig}}
\newcommand{\pythia}    {\sc{pythia}}
\newcommand{\alpgen}    {\sc{alpgen}}

\begin{document}

\begin{widetext}


\title{Multivariate searches for single top quark production with the \dzero detector}

%
\author{                                                                      
V.M.~Abazov,$^{36}$                                                           
B.~Abbott,$^{76}$                                                             
M.~Abolins,$^{66}$                                                            
B.S.~Acharya,$^{29}$                                                          
M.~Adams,$^{52}$                                                              
T.~Adams,$^{50}$                                                              
M.~Agelou,$^{18}$                                                             
J.-L.~Agram,$^{19}$                                                           
S.H.~Ahn,$^{31}$                                                              
M.~Ahsan,$^{60}$
% ...
\\                                                                            
\vskip 0.30cm                                                                 
\centerline{(D\O\ Collaboration)}                                             
\vskip 0.30cm                                                                 
}                                                                             
\affiliation{                                                                 
\centerline{$^{1}$Universidad de Buenos Aires, Buenos Aires, Argentina}       
\centerline{$^{2}$LAFEX, Centro Brasileiro de Pesquisas F{\'\i}sicas,         
                  Rio de Janeiro, Brazil}                                     
\centerline{$^{3}$Universidade do Estado do Rio de Janeiro,                   
                  Rio de Janeiro, Brazil}                                     
\centerline{$^{4}$Instituto de F\'{\i}sica Te\'orica, Universidade            
                  Estadual Paulista, S\~ao Paulo, Brazil}                     
\centerline{$^{5}$University of Alberta, Edmonton, Alberta, Canada,           
                  Simon Fraser University, Burnaby, British Columbia, Canada,}
\centerline{York University, Toronto, Ontario, Canada, and                    
                  McGill University, Montreal, Quebec, Canada}                
\centerline{$^{6}$Institute of High Energy Physics, Beijing,                  
                  People's Republic of China}                                 
\centerline{$^{7}$University of Science and Technology of China, Hefei,       
                  People's Republic of China}                                 
\centerline{$^{8}$Universidad de los Andes, Bogot\'{a}, Colombia}             
\centerline{$^{9}$Center for Particle Physics, Charles University,            
                  Prague, Czech Republic}                                     
\centerline{$^{10}$Czech Technical University, Prague, Czech Republic}        
\centerline{$^{11}$Center for Particle Physics, Institute of Physics,         
                   Academy of Sciences of the Czech Republic,                 
                   Prague, Czech Republic}                                    
\centerline{$^{12}$Universidad San Francisco de Quito, Quito, Ecuador}        
\centerline{$^{13}$Laboratoire de Physique Corpusculaire, IN2P3-CNRS,         
                   Universit\'e Blaise Pascal, Clermont-Ferrand, France}      
\centerline{$^{14}$Laboratoire de Physique Subatomique et de Cosmologie,      
                   IN2P3-CNRS, Universite de Grenoble 1, Grenoble, France}    
\centerline{$^{15}$CPPM, IN2P3-CNRS, Universit\'e de la M\'editerran\'ee,     
                   Marseille, France}                                         
\centerline{$^{16}$IN2P3-CNRS, Laboratoire de l'Acc\'el\'erateur              
                   Lin\'eaire, Orsay, France}                                 
\centerline{$^{17}$LPNHE, IN2P3-CNRS, Universit\'es Paris VI and VII,         
                   Paris, France}                                             
\centerline{$^{18}$DAPNIA/Service de Physique des Particules, CEA, Saclay,    
                   France}                                                    
\centerline{$^{19}$IReS, IN2P3-CNRS, Universit\'e Louis Pasteur, Strasbourg,  
                    France, and Universit\'e de Haute Alsace,                 
                    Mulhouse, France}                                         
\centerline{$^{20}$Institut de Physique Nucl\'eaire de Lyon, IN2P3-CNRS,      
                   Universit\'e Claude Bernard, Villeurbanne, France}         
\centerline{$^{21}$III. Physikalisches Institut A, RWTH Aachen,               
                   Aachen, Germany}                                           
\centerline{$^{22}$Physikalisches Institut, Universit{\"a}t Bonn,             
                   Bonn, Germany}                                             
\centerline{$^{23}$Physikalisches Institut, Universit{\"a}t Freiburg,         
                   Freiburg, Germany}                                         
\centerline{$^{24}$Institut f{\"u}r Physik, Universit{\"a}t Mainz,            
                   Mainz, Germany}                                            
\centerline{$^{25}$Ludwig-Maximilians-Universit{\"a}t M{\"u}nchen,            
                   M{\"u}nchen, Germany}                                      
\centerline{$^{26}$Fachbereich Physik, University of Wuppertal,               
                   Wuppertal, Germany}                                        
\centerline{$^{27}$Panjab University, Chandigarh, India}                      
\centerline{$^{28}$Delhi University, Delhi, India}                            
\centerline{$^{29}$Tata Institute of Fundamental Research, Mumbai, India}     
\centerline{$^{30}$University College Dublin, Dublin, Ireland}                
\centerline{$^{31}$Korea Detector Laboratory, Korea University,               
                   Seoul, Korea}                                              
\centerline{$^{32}$SungKyunKwan University, Suwon, Korea}                     
\centerline{$^{33}$CINVESTAV, Mexico City, Mexico}                            
\centerline{$^{34}$FOM-Institute NIKHEF and University of                     
                   Amsterdam/NIKHEF, Amsterdam, The Netherlands}              
\centerline{$^{35}$Radboud University Nijmegen/NIKHEF, Nijmegen, The          
                  Netherlands}                                                
\centerline{$^{36}$Joint Institute for Nuclear Research, Dubna, Russia}       
\centerline{$^{37}$Institute for Theoretical and Experimental Physics,        
                   Moscow, Russia}                                            
\centerline{$^{38}$Moscow State University, Moscow, Russia}                   
\centerline{$^{39}$Institute for High Energy Physics, Protvino, Russia}       
\centerline{$^{40}$Petersburg Nuclear Physics Institute,                      
                   St. Petersburg, Russia}                                    
\centerline{$^{41}$Lund University, Lund, Sweden, Royal Institute of          
                   Technology and Stockholm University, Stockholm,            
                   Sweden, and}                                               
\centerline{Uppsala University, Uppsala, Sweden}                              
\centerline{$^{42}$Physik Institut der Universit{\"a}t Z{\"u}rich,            
                   Z{\"u}rich, Switzerland}                                   
\centerline{$^{43}$Lancaster University, Lancaster, United Kingdom}           
\centerline{$^{44}$Imperial College, London, United Kingdom}                  
\centerline{$^{45}$University of Manchester, Manchester, United Kingdom}      
\centerline{$^{46}$University of Arizona, Tucson, Arizona 85721, USA}         
\centerline{$^{47}$Lawrence Berkeley National Laboratory and University of    
                   California, Berkeley, California 94720, USA}               
\centerline{$^{48}$California State University, Fresno, California 93740, USA}
\centerline{$^{49}$University of California, Riverside, California 92521, USA}
\centerline{$^{50}$Florida State University, Tallahassee, Florida 32306, USA} 
\centerline{$^{51}$Fermi National Accelerator Laboratory,                     
            Batavia, Illinois 60510, USA}                                     
\centerline{$^{52}$University of Illinois at Chicago,                         
            Chicago, Illinois 60607, USA}                                     
\centerline{$^{53}$Northern Illinois University, DeKalb, Illinois 60115, USA} 
\centerline{$^{54}$Northwestern University, Evanston, Illinois 60208, USA}    
\centerline{$^{55}$Indiana University, Bloomington, Indiana 47405, USA}       
\centerline{$^{56}$University of Notre Dame, Notre Dame, Indiana 46556, USA}  
\centerline{$^{57}$Purdue University Calumet, Hammond, Indiana 46323, USA}    
\centerline{$^{58}$Iowa State University, Ames, Iowa 50011, USA}              
\centerline{$^{59}$University of Kansas, Lawrence, Kansas 66045, USA}         
\centerline{$^{60}$Kansas State University, Manhattan, Kansas 66506, USA}     
\centerline{$^{61}$Louisiana Tech University, Ruston, Louisiana 71272, USA}   
\centerline{$^{62}$University of Maryland, College Park, Maryland 20742, USA} 
\centerline{$^{63}$Boston University, Boston, Massachusetts 02215, USA}       
\centerline{$^{64}$Northeastern University, Boston, Massachusetts 02115, USA} 
\centerline{$^{65}$University of Michigan, Ann Arbor, Michigan 48109, USA}    
\centerline{$^{66}$Michigan State University,                                 
            East Lansing, Michigan 48824, USA}                                
\centerline{$^{67}$University of Mississippi,                                 
            University, Mississippi 38677, USA}                               
\centerline{$^{68}$University of Nebraska, Lincoln, Nebraska 68588, USA}      
\centerline{$^{69}$Princeton University, Princeton, New Jersey 08544, USA}    
\centerline{$^{70}$State University of New York, Buffalo, New York 14260, USA}
\centerline{$^{71}$Columbia University, New York, New York 10027, USA}        
\centerline{$^{72}$University of Rochester, Rochester, New York 14627, USA}   
\centerline{$^{73}$State University of New York,                              
            Stony Brook, New York 11794, USA}                                 
\centerline{$^{74}$Brookhaven National Laboratory, Upton, New York 11973, USA}
\centerline{$^{75}$Langston University, Langston, Oklahoma 73050, USA}        
\centerline{$^{76}$University of Oklahoma, Norman, Oklahoma 73019, USA}       
\centerline{$^{77}$Oklahoma State University, Stillwater, Oklahoma 74078, USA}
\centerline{$^{78}$Brown University, Providence, Rhode Island 02912, USA}     
\centerline{$^{79}$University of Texas, Arlington, Texas 76019, USA}          
\centerline{$^{80}$Southern Methodist University, Dallas, Texas 75275, USA}   
\centerline{$^{81}$Rice University, Houston, Texas 77005, USA}                
\centerline{$^{82}$University of Virginia, Charlottesville,                   
            Virginia 22901, USA}                                              
\centerline{$^{83}$University of Washington, Seattle, Washington 98195, USA}  
}                                                                             
\date{April 10, 2006}

\begin{abstract}


We present a search for electroweak production of single top quarks in
the $s$-channel (${\ppbar}{\rar}t \bar{b}$+$X$) and $t$-channel 
(${\ppbar}{\rar}tq \bar{b}$+$X$) modes. We have analyzed 230 pb$^{-1}$
of data collected with the \dzero detector at the Fermilab Tevatron
collider at a center-of-mass energy of $\sqrt{s} = 1.96$ TeV. Two
separate analysis methods are used: neural networks and a cut-based
analysis. No evidence for a single top quark signal is found. We set 95\%
confidence level upper limits on the production cross
sections using Bayesian statistics, based on event counts and
binned likelihoods formed from the neural network output.
The limits from the neural network (cut-based) analysis are
6.4~pb (10.6~pb) in the $s$-channel and 5.0~pb (11.3~pb) in the
$t$-channel.
\end{abstract}

\maketitle 
\end{widetext}


\section{Introduction}
The top quark, discovered in 1995 at the Fermilab Tevatron Collider
by the CDF and \dzero
collaborations~\cite{topdiscovery}, is by far the heaviest
elementary particle found to date. Its large mass and corresponding
coupling strength to the Higgs boson of order unity suggest that the
physics of electroweak symmetry breaking might be visible in the top quark sector. 

Top quarks are produced at the Tevatron mainly in top-antitop pairs
through the strong interaction. This mode led to the
discovery of the top quark and has been the only top quark production
mode observed to date.  The top quark decays predominantly to a $W$~boson
and a $b$~quark, but little else is known experimentally about its
electroweak interactions.

All previous studies of the top quark
electroweak interaction and the $Wtb$ vertex have been done either in 
the low-energy regime using virtual top quarks (in studies of
$b$ quark decays), or in the decay of real top quarks. Both of these
types of studies presuppose the unitarity of the CKM matrix and
are thus constrained to studying the standard model with three
generations of quarks.
This restriction can be overcome by exploring the production of
single top quarks through electroweak interactions. This production
mode is becoming accessible at the Tevatron and promises
the first direct measurement of the electroweak coupling strength
of the top quark as well as a first glimpse at possible top quark
interactions beyond the standard model (SM).

\subsection{Physics with Single Top Quarks}
The study of single top quark production provides the possibility of
investigating top quark related properties that cannot be measured in top
quark pair production. The most relevant of these is a direct measurement of the 
CKM matrix element $|V_{tb}|$ from the single top quark production cross sections. 
This provides the only measurement of $|V_{tb}|$ without having to assume 
three quark generations or CKM matrix unitarity. Together with the other
CKM matrix measurements~\cite{Hagiwara:fs}, we will be able to 
test the unitarity of the CKM matrix.

Single top quarks are produced through a left-handed interaction.
Therefore, they are expected to be highly polarized. Since the top quark
decays before hadronization can occur, the spin correlations are retained
in the final decay
products. Hence, single top quark production offers an opportunity to
observe the polarization and to test the corresponding SM predictions. 

Measurements of the charged-current couplings of the top
quark probe any nonstandard structure of the couplings and can
therefore provide hints of new physics.  Any deviation in
the $(V$--$A)$ structure of the $Wtb$ coupling would lead to a
violation of the spin correlation properties~\cite{Heinson:1996zm}.  
Furthermore, combining single top quark measurements with $W$ helicity measurements 
in top quark decays provides the most stringent information on the 
$Wtb$ coupling~\cite{Chen:2005vr}. 

Finally, rather than manifesting itself in a modified $Wtb$ coupling, 
new physics could produce a single top quark final state through
other processes. There are several
models of new physics that would increase the single top quark production
cross sections~\cite{Tait:2000sh}. Thus, constraints on physics
beyond the standard model are possible even
before an actual observation of single top quark production.

\subsection{Single Top Quark Production}
\begin{figure}[!h!tbp]
\includegraphics[width=0.4\textwidth]
{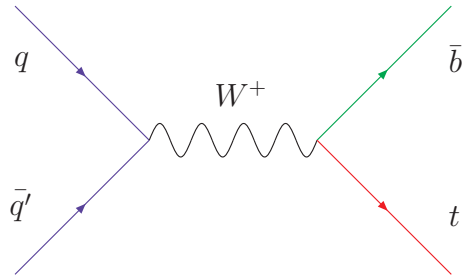}
\caption{Feynman diagram for leading order
$s$-channel single top quark production.}
\label{fig:stop_diag_tb}
\end{figure}

\begin{figure}[!h!tbp]
\includegraphics[width=0.30\textwidth]
{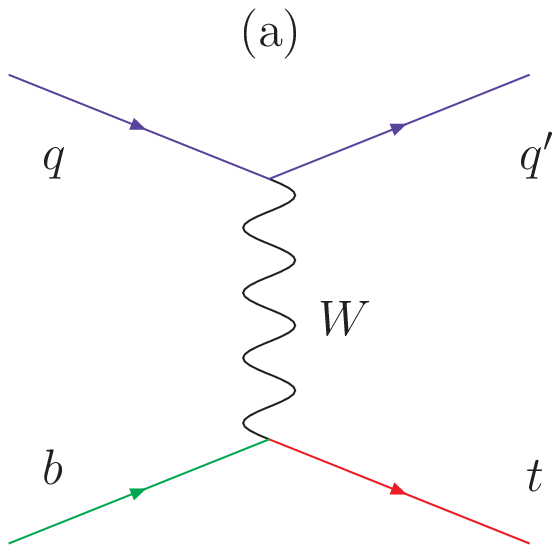}
\includegraphics[width=0.36\textwidth]
{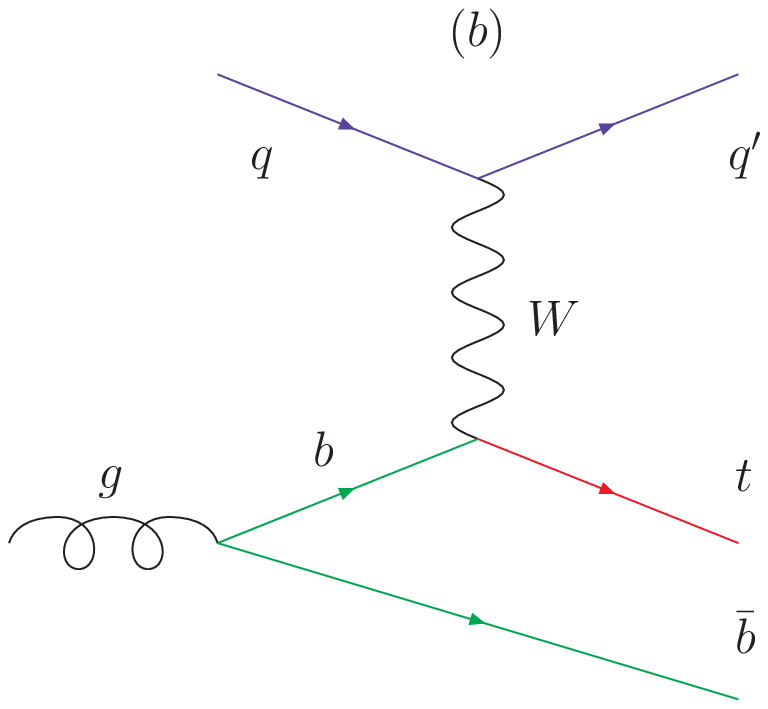}
\caption{Representative Feynman diagrams for $t$-channel single top
quark production. Shown is the (a) leading order and 
(b) the $O(\alpha_s)$ $W$-gluon fusion diagram.}
\label{fig:stop_diag_tqb}
\end{figure}

There are three standard model modes of single top quark production at hadron
colliders. Each of these modes may be characterized by the four-momentum squared
$Q^2_W$, the virtuality, of the participating $W$~boson:
\begin{itemize}
\item
$s$-channel $W$~boson exchange ($Q^2_W>0$): This process,
${\ppbar}{\rar}t\bar{b}$+$X$, 
is referred to as ``$tb$,'' which includes both $t\bar{b}$ and $\bar{t}b$ (see Fig.~\ref{fig:stop_diag_tb}).
\item
$t$-channel and $u$-channel $W$~boson exchange ($Q^2_W<0$): This process,
${\ppbar}{\rar}tq\bar{b}$+$X$, has the largest cross section of the
three. It includes the leading order diagram (Fig.~\ref{fig:stop_diag_tqb}a)
with a $b$~quark from
the proton sea in the initial state, and a second diagram (Fig.~\ref{fig:stop_diag_tqb}b)
where an extra ${\bbar}$~quark appears in the final state explicitly.
This latter mode is of order $O(\alpha_s)$ in the strong coupling $\alpha_s$,
but nevertheless provides the largest contribution to the total cross section.
Historically, $t$-channel production has also been referred to as
$W$-gluon fusion, since the
${\bbar}$~quark in the final state arises from a gluon splitting to a
${\bbbar}$ pair. We refer to the $t$-channel process as ``$tqb$,'' which
includes $tq\bar{b}$, $\bar{t}\bar{q}b$, $tq$, and $\bar{t}\bar{q}$.
\item
Real $W$~boson production ($Q^2_W=m_W^2$): In this process, $p\bar{p}\to tW$+$X$, a single top quark
appears in association with a real $W$~boson in the final state.  This process has a
negligible cross section at the Tevatron~\cite{Heinson:1996zm} and will not be
addressed in this paper. 
\end{itemize}

The next-to-leading order (NLO) production rates at the Tevatron ($\sqrt{s}=1.96$~TeV) for the
$s$- and $t$-channel single top quark modes have been 
calculated~\cite{Smith:1996ij, Stelzer:1997ns, Harris:2002md, sintop-xsec2,
Campbell:2004ch, sintop-nlo-sch, sintop-nlo-tch} and the results for
cross sections are shown in Table~\ref{tab:sigma}. The uncertainties include
components from the choice of scale and the
parton distribution functions, but not for the top quark mass.

\begin{table}[!h!tbp]
\begin{center}
\caption[table-xsecs]{Theoretically calculated total cross sections for single top quark
production at a $p\bar{p}$ collider with 
$\sqrt{s} = 1.96$~TeV, using $m_t=175$~GeV.}
\begin{ruledtabular}
\begin{tabular}{lc}
Process & Cross Section [pb] \\
\hline
 \\
$s$-channel ($tb$)  & $0.88^{+0.07}_{-0.06}$ \vspace{0.05in} \\
\vspace{0.05in}
$t$-channel ($tqb$) & $1.98^{+0.23}_{-0.18}$ \vspace{0.05in} \\
$tW$ production     & $0.093 \pm 0.024$
\end{tabular}
\end{ruledtabular}
\label{tab:sigma}
\end{center}
\end{table}

For comparison, the calculated top quark pair production cross section at the
Tevatron at 1.96~TeV is $6.77 \pm 0.42$~pb~\cite{Kidonakis:2003qe}. This already
makes it clear that it is more difficult to isolate the single top quark signal
than the top quark pair signal.

Under the assumption that all top quarks decay to a $W$~boson and a $b$~quark,
and only using $W$~boson decays to electron and muon final states,
the final state signature of a single top quark event detected in this analysis
is characterized by a high transverse momentum ($p_T$), centrally produced,
isolated lepton ($e^\pm$ or $\mu^\pm)$ and missing transverse energy ({\met}),
together with two or three jets. One of the jets comes from a high-$p_T$ central
$b$~quark from the top quark decay.

\begin{figure}[!h!tbp]
\includegraphics[width=0.40\textwidth]{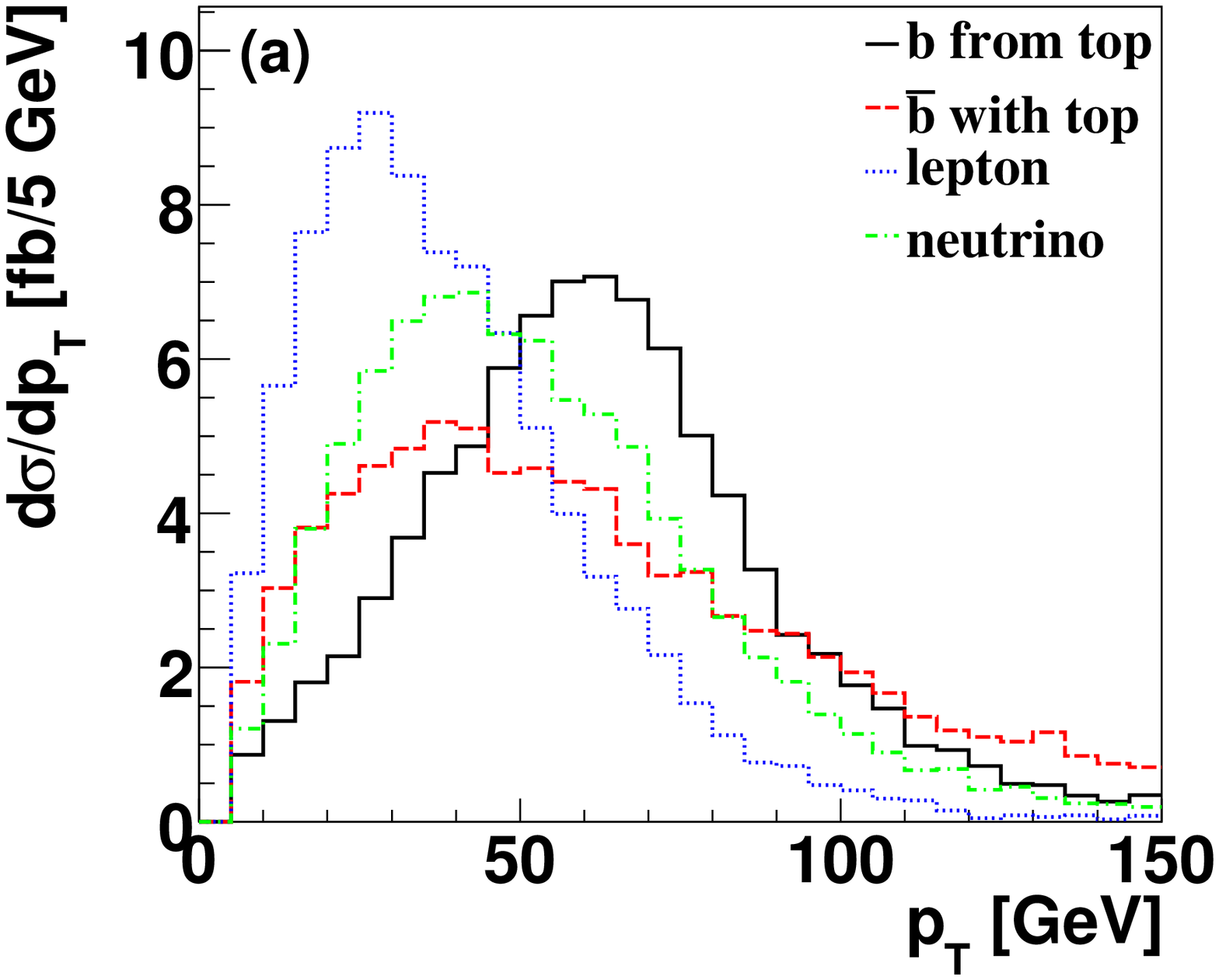}
\includegraphics[width=0.40\textwidth]{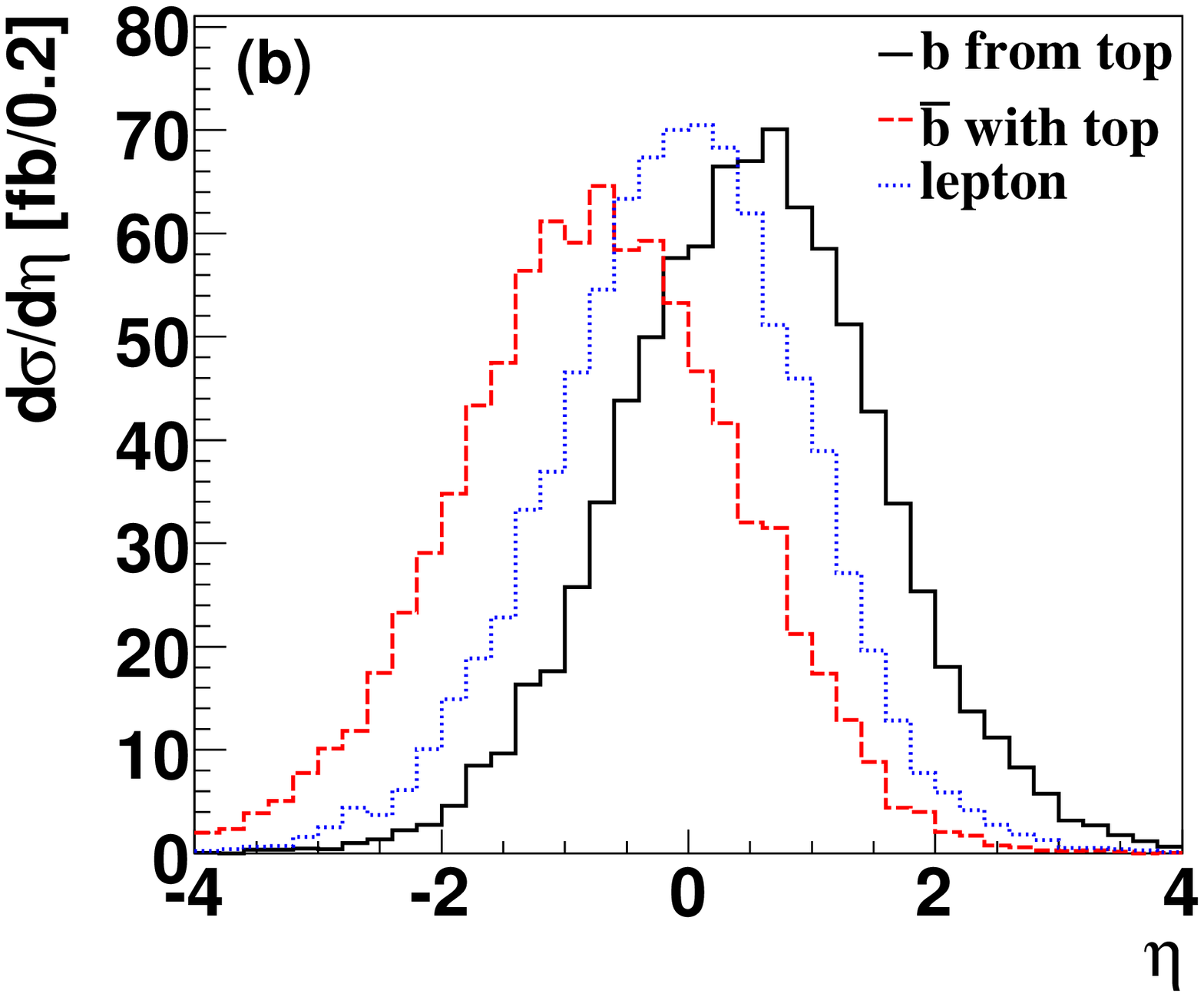}
\caption{Distributions of transverse momenta (a) 
and pseudorapidity (b) for the final state partons in 
$s$-channel single top quark events. 
The histograms only include the final state of $t$, not $\bar{t}$.}
\label{fig:parton-pt-eta-tb}
\end{figure}

\begin{figure}[!h!tbp]
\includegraphics[width=0.40\textwidth]{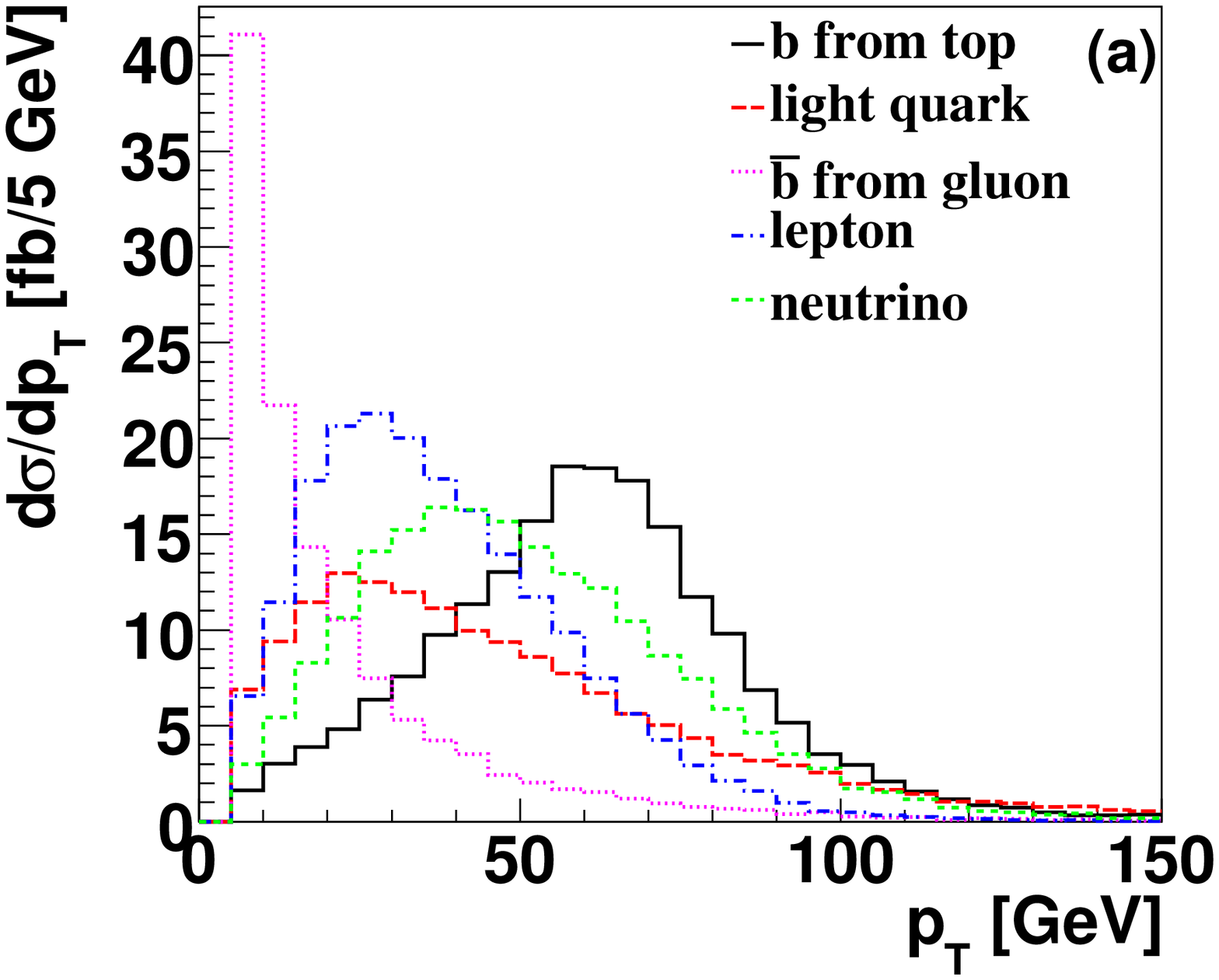}
\includegraphics[width=0.40\textwidth]{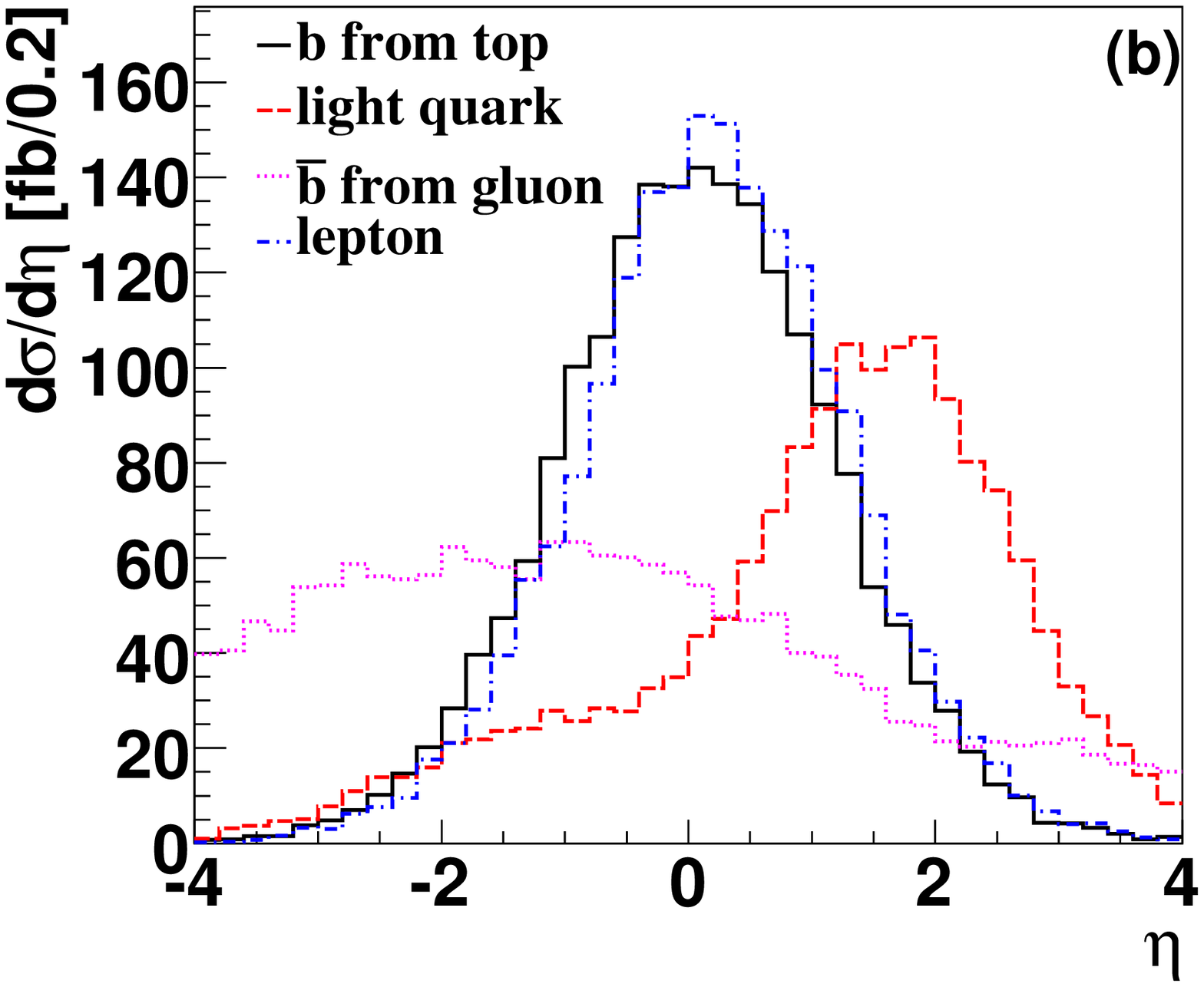}
\caption{Distributions of transverse momenta (a)
and pseudorapidity (b) for the final state partons in 
$t$-channel single top quark events. 
The histograms only include the final state of $t$, not $\bar{t}$.}
\label{fig:parton-pt-eta-tqb}
\end{figure}

Figures~\ref{fig:parton-pt-eta-tb} and~\ref{fig:parton-pt-eta-tqb} shows the 
transverse momenta and pseudorapidities $\eta$~\cite{fiducial_endnote} for 
the partons in our modeling of the
$s$-channel and $t$-channel single top quark processes, after decay of the
top quark and $W$~boson.

The final state fermions from the top quark decay have relatively high
transverse momenta and central rapidities. Since the $s$-channel process
involves the decay of a heavy virtual object, the $\bbar$ quark
produced with the top quark is also at high transverse momentum and central
pseudorapidity. By contrast, the light quark in the $t$-channel appears
at lower transverse momentum and at more forward pseudorapidities because
it is produced when an initial state parton emits a virtual $W$~boson.
The $\bbar$ quark from $t$-channel initial state radiation appears typically
at very low $p_T$ and with large pseudorapidities and is thus often not reconstructed
experimentally.

Due to its electroweak nature, single top quark production results in
a polarized final state top quark. It has been shown~\cite{Mahlon:1995zn} that
the top quark spin follows the direction of the
down-type quark momentum in the top quark rest frame.  This is the
direction of the initial $\bar{d}$~quark for the $s$-channel and
close to the direction of the final state $d$ quark for the $t$-channel. 
The above
result follows directly from the properties of the polarized top quark
decays when single top quark production is considered as top quark decay
going ``backwards in time''~\cite{Boos:2002xw}.

\subsection{Overview of the Backgrounds}
\label{intro:backgrounds}
Searches for single top quark production are challenging because of the very
large backgrounds. The situation is significantly different from top pair
production not just because of the smaller production rate, but more
importantly because of the smaller multiplicity of final state particles
(leptons or jets). Single top quark events are typically less energetic (because there
is only one heavy object), less spherical (because of the production mechanism),
and typically have two or three jets, not four as do
{\ttbar} events. 

Processes that can have the same single top quark experimental signature
include in order of importance $W$+jets, {\ttbar}, multijet
production, and some smaller contributions from $Z$+jets and diboson events.
\begin{itemize}
\item
$W$+jets events form the dominant part of the background. The cross
section for $W$+2 jets production is over
1000~pb~\cite{Mangano:2002ea,Mrenna:2003if} with $W{\bbbar}$
contributing about 1\%.
\item
The second largest background is due to {\ttbar} production. 
This process has a
larger multiplicity of final state particles than single top quark events. 
However,
when some of the jets or a lepton are not identified, the kinematics of the
remaining particles are very similar to those of the signal.
\item
Multijet events form a background in the electron channel when a jet
is misidentified as an electron. The probability of such
misidentification is rather small, but the $\ge$3~jet
cross section is so large that the overall contribution is
significant.

Additionally, {\bbbar} production contributes
to the background when one of the $b$'s decays semileptonically.  This
background in the electron channel is very small. In the muon channel,
{\bbbar} events form a background when the muon is away from the jet axis
or when the jet is not reconstructed. 
\item
$Z$/Drell-Yan+jets production can mimic the single top quark signals if one of
the leptons is misidentifed. 
\item
$WW$, $WZ$, and $ZZ$ processes are the electroweak part of the
$W$+jets and $Z$+jets backgrounds, but with different kinematics. 
\end{itemize}

Single top quark events are kinematically and topologically similar
to $W$+jets and {\ttbar} events. 
Therefore, extracting the signal from the backgrounds is challenging in
a search for single top quark production.

\subsection{Status of Searches}
Both the CDF and \dzero collaborations have previously performed
searches for single top quark production~\cite{d0runI,Acosta:2001un}. 
Recently, CDF performed a search using 160~pb$^{-1}$ of data and obtained upper
limits of 13.6~pb ($s$-channel),
10.1~pb ($t$-channel), and 17.8~pb ($s$+$t$ combined) at the 95\% 
confidence level~\cite{RunII:cdf_result}. 
\dzero has published a neural network search for single top quark
production using 230~pb$^{-1}$ of data~\cite{RunII:d0_result}, 
which is described in more detail in this article.



\subsection{Outline of the Analysis}
We have performed a search for the electroweak production of single top quarks
in the $s$-channel and $t$-channel production modes with the
\dzero detector at the Fermilab Tevatron collider. We consider lepton+jets in the final state, 
where the lepton is either an electron or a muon.

To take advantage of the differences between $s$-~and $t$-channel final state
topologies, we differentiate the $s$-channel search from the $t$-channel search
by requiring at least one untagged jet in the $t$-channel search.  For both
$s$-channel and $t$-channel searches, we separate the data into independent
analysis sets based on the lepton flavor ($e$ or $\mu$) and the multiplicity of
identified $b$~quarks (one tagged jet or more than one).

We use two different multivariate methods to extract the signal from the large
backgrounds: a cut-based analysis, first presented here,
and an analysis based on neural networks that was first presented in brief
form in Ref.~\cite{RunII:d0_result}. In
the absence of any significant evidence for signal, we set upper limits at the
95$\%$ C.L. on the single top quark production cross sections.
 
Finally, we present limit contours in a two-dimensional plane of the $s$-channel 
signal cross section versus the $t$-channel signal cross section. 

\subsection{Outline of the Paper}
This paper is organized as follows. Section~\ref{sec:detector} describes the
\dzero detector and the reconstruction of the final state
objects. Section~\ref{sec:triggers} summarizes the triggers for the data samples
used in the search and Section~\ref{selection-section} describes the selection
requirements. Section~\ref{sec:SigBkgModeling} explains the modeling of signals
and backgrounds, and Section~\ref{sec:EventYields} presents the numbers of
events passing all selections. Section~\ref{sec:EventAnalysis} discusses the
most important variables that offer discrimination between the signals and
backgrounds, and provides details of the cut-based and the neural network
analyses. Section~\ref{systematics} lists the systematic uncertainties in this
measurement. Section~\ref{section_limits} discusses the procedure for setting
limits on the signal cross section using Bayesian statistics. The limits are
presented in Section~\ref{results}, and we summarize the results in
Section~\ref{sec:conclusions}.

\section{The \dzero Detector and Object Reconstruction}
\label{sec:detector}
\subsection{The \dzero Detector}
\begin{figure*}[h!tbp]

\includegraphics[width=0.95\textwidth]{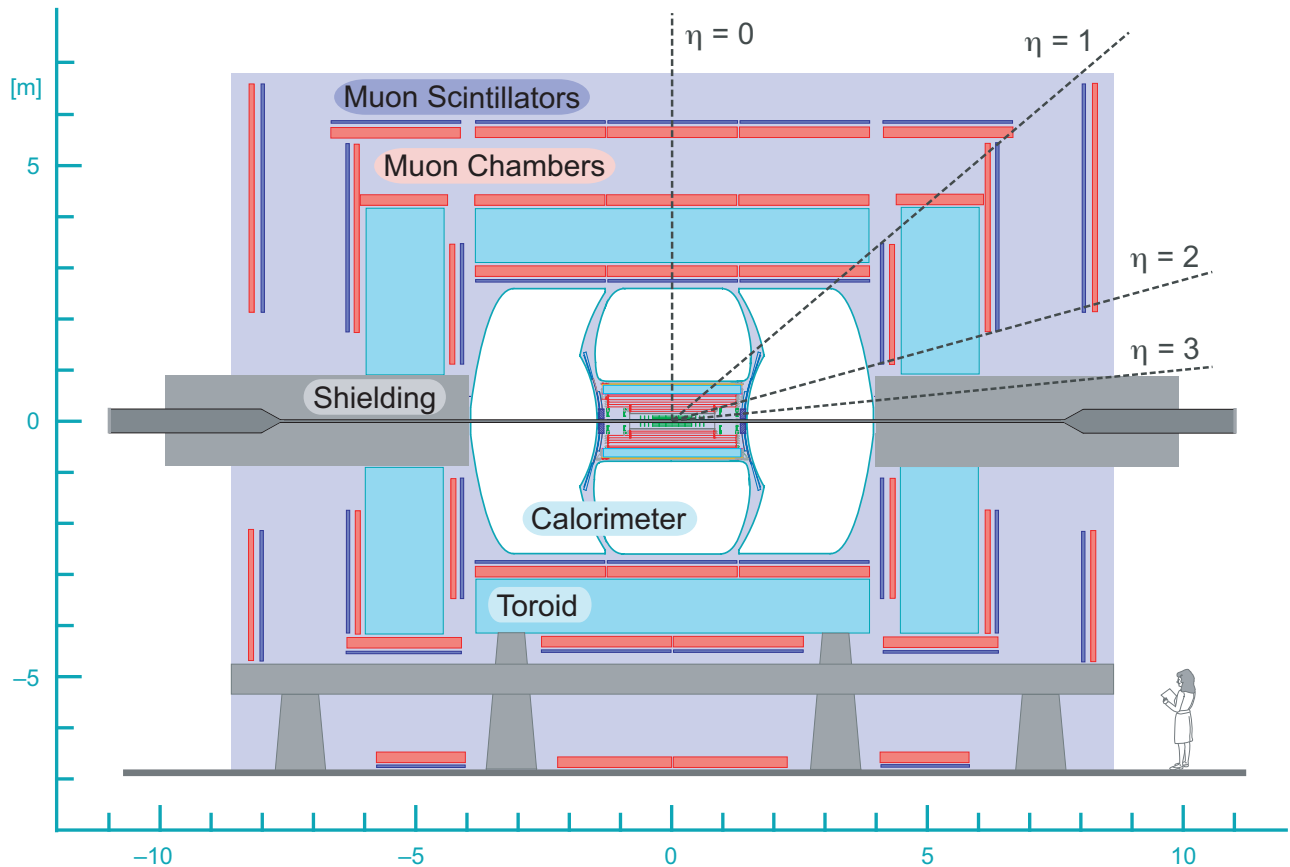}
\caption{General view of the \dzero detector. The proton beam travels from left to   
right and the antiproton beam from right to left in this figure.}
\label{fig:detector}
\end{figure*}

\begin{figure*}[h!tbp]

\includegraphics[width=0.95\textwidth]{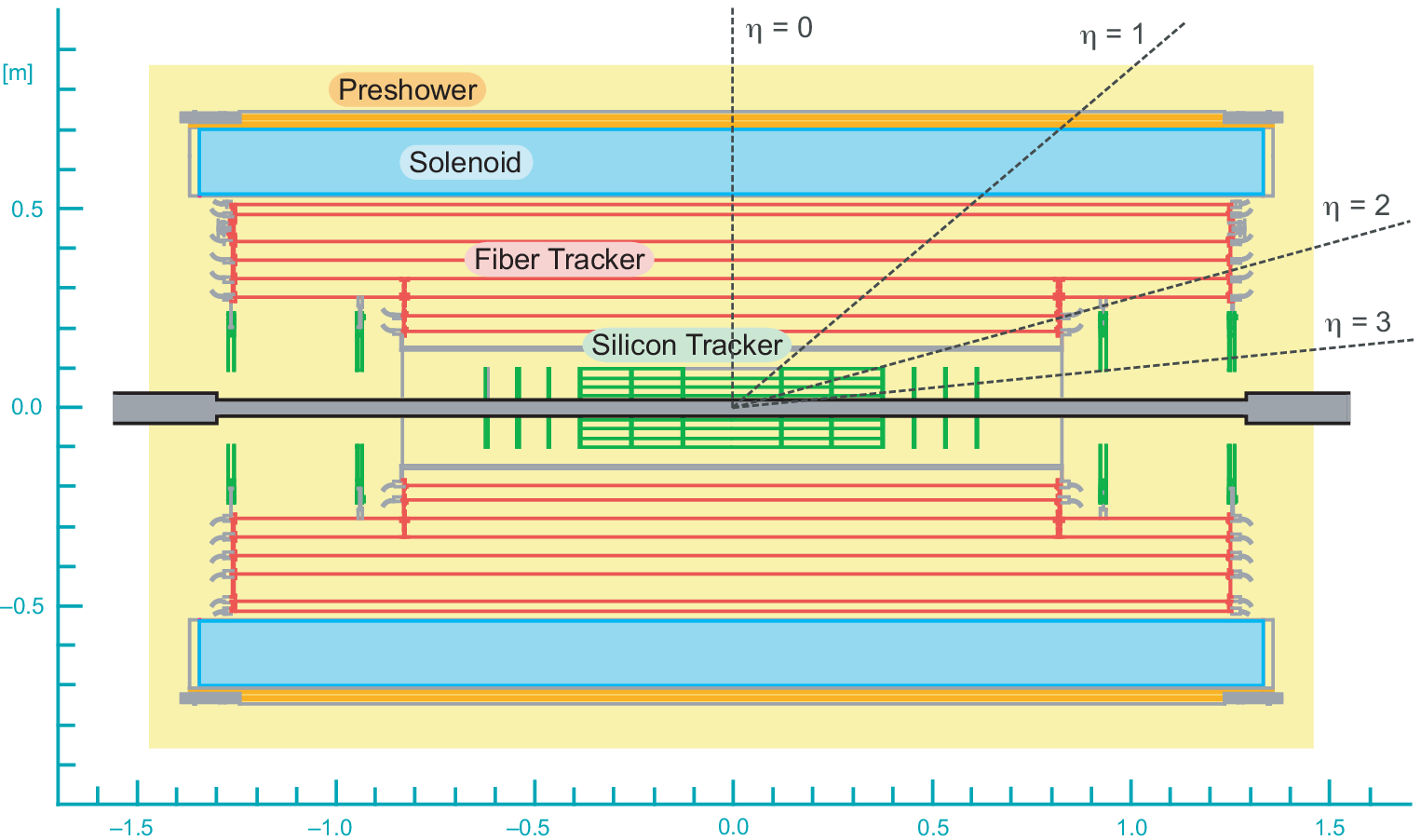}
\caption{Close view of the tracking systems.}
\label{fig:tracking}
\end{figure*}

The \dzero detector~\cite{Abazov:2005pn} is shown in
Figs.~\ref{fig:detector} and~\ref{fig:tracking} and consists of several layered
elements. The first is a magnetic central-tracking system, which includes a
silicon microstrip tracker (SMT) and a central fiber tracker (CFT), both located
within a 2~T superconducting solenoidal magnet. The SMT has $\approx 800,000$
individual strips, with a typical pitch of $50-80$~$\mu$m, and a design
optimized for tracking and vertexing capability at pseudorapidities of
$|\eta|<3.0$. The system has a six-barrel longitudinal structure, each with a
set of four layers arranged axially around the beam pipe, and interspersed with
16 radial disks. The CFT has eight thin coaxial barrels, each supporting two
doublets of overlapping scintillating fibers of 0.835~mm diameter, one doublet
being parallel to the collision axis, and the other alternating by $\pm
3^{\circ}$ relative to the axis. Light signals are transferred via clear light
fibers to solid-state photon counters (visible light photon counters, VLPCs)
that have $\approx 80\%$ quantum efficiency.

Central and forward preshower detectors are located just outside of the
superconducting coil (in front of the calorimetry). These are constructed of several
layers of extruded triangular scintillator strips that are read out using
wavelength-shifting fibers and VLPCs. The next layer of detection involves three
liquid-argon/uranium calorimeters: a central section (CC) covering $|\eta|$ up
to $\approx 1$, and two end calorimeters (EC) extending coverage to
$|\eta|\approx 4$, all housed in separate cryostats~\cite{run1det}. In addition
to the preshower detectors, scintillators between the CC and EC cryostats
provide sampling of developing showers for $1.1<|\eta|<1.4$.

A muon system resides beyond the calorimetry, and consists of a layer of
tracking detectors and scintillation trigger counters before 1.8~T iron toroids,
followed by two more similar layers after the toroids. Tracking for $|\eta|<1$
relies on 10~cm wide drift tubes~\cite{run1det}, while 1~cm mini drift tubes are 
used for $1<|\eta|<2$.

The luminosity is obtained from the rate of inelastic
collisions measured using plastic scintillator arrays located in front of the
EC cryostats, covering $2.7 < |\eta| < 4.4$.

\subsection{Object Reconstruction}
\label{sec:objectreco}
Physics objects are reconstructed from the digital signals
recorded in each part of the detector. Particles can be identified by certain
patterns and, when correlated with other objects in the same event, they provide the
basis for understanding the physics that produced such signatures in the detector. 

\subsubsection{Primary Vertex}
The position of the hard scatter interaction is determined at \dzero by
clustering tracks into seed vertices using a Kalman filter
algorithm~\cite{kalman}. The primary vertex is then selected using a probability
function based on the $p_T$ values of the tracks assigned to each vertex. The hard
scatter vertex is distinguished from other soft interaction vertices by the
higher average $p_T$ of its tracks. In multijet data events, the position
resolution of the primary vertex in the transverse plane (perpendicular to the
beam pipe) is around 40~$\mu$m, convoluted with a typical beam spot size of
around 30~$\mu$m. For the longitudinal direction (along the beam pipe), the
typical resolution is about 1~cm.

\subsubsection{Electrons} 
Electron candidates are initially identified as energy clusters in the
central region of the electromagnetic calorimeter, $|\eta| \leq 1.1$. 
We define two classes of electron candidates: {\it loose} and {\it tight}. 
{\it Loose} electrons are required to have the fraction
of their total energy deposited in the electromagnetic (EM) calorimeter $f_{\rm EM} > 0.9$
and a shower-shape chi-squared, based on seven variables that compare the values of the energy
deposited in each layer of the electromagnetic calorimeter with average
distributions from simulated electrons, to be $\chi^2_{\rm cal} <
75$. Finally, loose electron candidates are also required to be isolated by
measuring the total deposited energy and the energy from the EM calorimeter only
around the electron track: 
$E_{\rm Total}(R<0.4) < 1.15 \times E_{\rm EM}(R<0.2)$, where
$R=\sqrt{(\Delta\phi)^2+(\Delta\eta)^2}$ is the radius of a cone defined by the
azimuthal angle $\phi$ and the pseudorapidty $\eta$.

For an electron candidate to be included in the {\it tight} class, a track must
be matched to the {\it loose} cluster within $|\Delta\eta|<0.05$ and
$|\Delta\phi|<0.05$, and additionally pass a cut on a
seven-variable likelihood built to separate real electrons from
backgrounds. 
The following variables are used in the likelihood: 
(i)~$f_{\rm EM}$; 
(ii)~$\chi^2_{\rm cal}$;
(iii)~$E_T^{\rm cal}/p_T^{\rm track}$, transverse energy of the cluster divided by the
transverse momentum of the matched track; (iv)~$\chi^2$ probability of the track
match; (v)~distance of closest approach between
the track and the primary vertex in the transverse plane; (vi)~ $N_{\rm
tracks}$, the number of tracks inside a cone of $R < 0.05$ around the
matched track; and (vii)~$\sum{p_T}$ of tracks in an $R < 0.4$ cone around
the matched track. {\it Tight} electrons are obtained by applying a cut on the
likelihood of $\mathcal{L}>0.85$.
The overall tight electron identification efficiency in data is around 75\%. 

A comparison between the dielectron invariant mass distributions for
$Z\to ee$ simulated events and data shows that the position of the simulated $Z$~boson
peak is shifted from that in data, and that the electron energy
resolution is better than in data. 
We apply small corrections to the identification efficiency and
electromagnetic energy of simulated electrons and smear
their energies to agree with data. 

\subsubsection{Muons} 
Muons are reconstructed in \dzero up to $|\eta| = 2$ by first finding
hits in all three layers of the muon spectrometers and requiring that the timing of
these hits is consistent with the hard scatter, thus rejecting cosmic rays. 
Secondly, all muon candidates must be matched  to a
track in the central tracker. That central track must pass the following
criteria: (i) chi-squared per degree of freedom less than 4; (ii) 
the distance of closest approach to the primary vertex in the
transverse plane must be less than three standard deviations; and (iii) the distance
in $z$ between the track and the primary vertex must be less than 1~cm. 

As for electrons, we similarly define two classes: {\it loose} and {\it tight},
but this time based solely on the muon's isolation from other objects. A {\it
loose} isolated muon must comply with $R$(muon, jet) $> 0.5$, which is the
distance between the muon and the jet axis.  A {\it
tight} isolated muon must be {\it loose} and additionally satisfy track-based
and calorimeter-based criteria: $|\sum^{\rm tracks} p_T / p_T(\mu)| < 0.06$
where the sum is over tracks within a cone of $R$({\rm track, muon})$<0.5$; and
$|\sum^{\rm cells} E_T / p_T(\mu)| < 0.08$ where the sum is over calorimeter
cells within an anulus of $0.1 < R$({\rm calorimeter cell, muon})$ < 0.4$.  The
overall tight muon identification efficiency in data is around 65\%.

Similarly to electrons in the simulation, we correct the energy scale for
simulated muons and smear their energies to reproduce the data in $Z\to\mu\mu$.

\subsubsection{Jets} 
We reconstruct jets based on calorimeter cell energies, using the improved legacy cone
algorithm~\cite{jet_def} with radius $R=0.5$. Noisy calorimeter cells are
ignored in the reconstruction algorithm by imposing the requirement that neighboring cells
have signals above the noise level. 

Jet identification is based on a set of cuts to reject poor quality jets or
noisy jets: (i) $0.05 < f_{\rm EM}< 0.95$; (ii) fraction of jet $E_T$ in the
coarse hadronic calorimeter layers $< 0.4$; (iii) ratio of $E_T$'s of the most
energetic cell to the second most energetic cell in the jet $< 10$; and (iv)
smallest number of towers that make up $90\%$ of the jet $E_T$, $n_{90} > 1$.

Jet energy scale corrections are applied 
to convert jet energies from the reconstructed level into particle-level 
energies.  The reconstructed fully-corrected energy of jets from the simulation of the
detector performance does not exactly match that seen in data. Similar to
electrons and muons, we smear jet energies by a small amount in the simulation
to reproduce the resolution measured in data.

\subsubsection{Missing Energy} 
We infer the transverse energy of the
neutrino in the event as the opposite of the vector sum of all the energy
deposited in the calorimeter. 
This calorimeter-only missing transverse energy
is then corrected with the jet energy scale, the electromagnetic scale, and the
energy loss from isolated muons in the calorimeter and their momenta. 

\subsection{Identification of $b$-Quark Jets}
\label{sec:bID}
The presence of $b$~quarks can be inferred from the long lifetime
of $B$ hadrons, which typically travel a few millimeters before
hadronization. 
Thus $b$-quark jets contain a displaced vertex inside a
jet whereas light-quark jets do not. 
The Secondary Vertex Tagger (SVT), described below, makes use of this fact to
identify, or tag, $b$-quark jets by fitting tracks in the jet
into a secondary vertex.

\subsubsection{Taggability} 
Before the $b$-quark tagging algorithm is applied to identify displaced vertices
in the jet, a set of cuts is applied to ensure a good quality jet and factor out
detector geometry effects. Thus the final probability to identify a $b$-quark
jet is factored into two parts: a {\it taggability} part, or
jet-quality-sensitive component, and a {\it tagger} part, or
heavy-flavor-sensitive component. A taggable jet requires at least two tracks
within a cone of $R=0.5$. At least one of these tracks must have $p_T > 1.0$~GeV,
and additional tracks must have $p_T > 0.5$~GeV. All tracks must have at least
one SMT hit, an $xy$ distance-of-closest-approach (DCA) of $<0.2$~cm, and a $z$
DCA of $< 0.4$~cm with respect to the primary vertex. The {\it taggability} 
is the number of taggable jets divided
by the number of good jets. Only jets satisfying jet identification
requirements, with $p_T > 15$~GeV (after jet energy corrections) and $|\eta|\leq
2.5$ are considered to be good for the definition of taggability.

In simulated events, the taggability is higher than in data mainly due to a
non-comprehensive description of the tracking detectors (dead detector elements,
other inefficiencies, noise, etc.) resulting in a higher tracking efficiency (in
particular within jets). Therefore, the Monte Carlo taggability must be
calibrated to that observed in the data. A {\em taggability-rate function} is
utilized to do this by parametrizing the taggability as a function of jet $p_T$
and $\eta$. Thus, the taggability per jet is determined in data and applied to
the Monte Carlo as:
\begin{eqnarray}
P^{\rm taggable}(p_T, \eta) =
\frac{{\rm \#\:taggable\:jets \:in\:}(p_T,\eta){\rm \: bin}}
{{\rm \#\: jets\: in\:}(p_T,\eta){\rm \: bin}}.
\end{eqnarray}
Central jets with momenta above 40~GeV have taggabilities of around 85\%. For
simulated jets the taggability is $\approx 90$\%.

\subsubsection {Secondary Vertex Tagger}
The SVT algorithm is designed to reconstruct a displaced vertex inside a jet by
fitting tracks that have a large impact parameter from the hard scatter vertex.
A simple algorithm is applied to the tracks to remove most $K_S^0$'s,
$\Lambda$'s, and photon conversions.  Tracks are then required to have at least
two SMT hits, $p_T>1.0$~GeV, transverse impact parameter significance
($d_{ca}/\sigma_{d_{ca}}$) greater than 3.5, and a track $\chi^2>10$. A simple
cone jet-algorithm is used to cluster the tracks into {\em track-jets}, and then
a Kalman filter algorithm is used to find vertices with the tracks in each
track-jet. The distance between the primary vertex and the found secondary
vertex, the decay length $L_{xy}$, and its error $\sigma_{L_{xy}}$ are
calculated taking into account the uncertainty on the primary vertex position.
The decay length is a signed parameter, defined by the sign of the cosine of the
angle between the vector from the primary vertex to the decay point and the
total momentum of the tracks attached to the secondary vertex.
If the decay length significance
$L_{xy}/\sigma_{L_{xy}}$ is more than 7, then the found vertex is
considered a tag. A calorimeter jet is considered tagged if the distance between
the jet axis and the line joining the primary vertex and the secondary vertex is
$R<0.5$ in $\eta , \phi$ space. 
This set of cuts has been tuned to obtain a probability for a light quark
mistag of 0.25\%. Note that gluon jets are included in the light quark
category. 

\begin{figure}[htbp]
\vspace{-0.15in}
\includegraphics[width=0.45\textwidth]{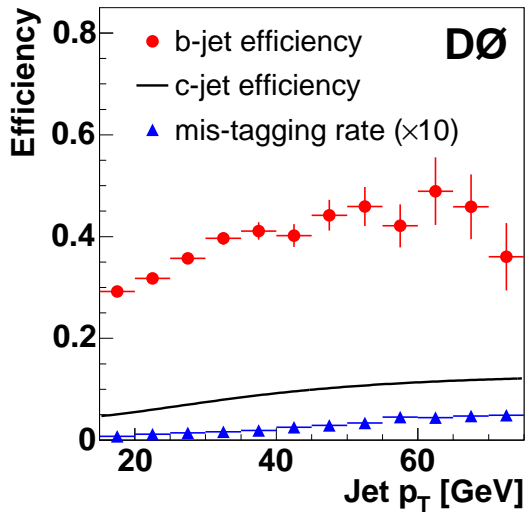} 
\vspace{-0.1in}
\caption{Measured $b$-tagging efficiency (circles) and mis-tagging rate
(triangles), and estimated $c$-tagging efficiency (solid line) as a function of
jet $p_T$.}
\label{svt_data_eff}
\end{figure}

We estimate the $b$ tagging efficiency in a dijet data
sample. The heavy flavor content of the sample is enhanced by requiring one of
the jets to have a high-$p_T$ muon relative to the jet
axis. The SVT efficiency to tag the other jet can then be inferred. We estimate
the $c$ quark tagging efficiency from a Monte Carlo simulation. The mis-tagging
rate, or how often a light-flavor jet (from $u, d, s$ quarks or gluons) is
identified as a $b$ jet, is also measured in a dijet data sample. We count
the number of found secondary vertices with $L_{xy}/\sigma_{L_{xy}}<-7$ and
correct for the contribution of heavy-flavor jets in the sample and the presence
of long-lived particles in light-flavor jets. The sign in the decay length
measurement comes from the scalar product of the decay length vector and the
unit vector defined by the  
Figure~\ref{svt_data_eff} shows the
tagging efficiency as a function of jet $p_T$ for the different types of jets. 

To calculate the probability for a
simulated jet to be tagged, a {\em tag-rate function} (TRF) derived from data is used
similarly to the taggability parametrized in $p_T$ and $\eta$: 
\begin{eqnarray}
P^{\rm tagging}(p_T, \eta) =
\frac{{\rm \#\:SVT\:tagged\:jets \:in\:} (p_T,\eta){\rm \: bin}}
{{\rm \#\:taggable\:jets\: in\:}(p_T,\eta){\rm \: bin}}.
\end{eqnarray}
Separate functions are determined for $b$-quark jets, $c$-quark jets, and
light-quark jets, as in Fig.~\ref{svt_data_eff}.  

The TRFs are applied to the Monte Carlo samples in the following way. First,
for each jet in the event (with $p_T>15$~GeV and $|\eta|<3.4$) a
taggability-rate function is applied. 
Next, each jet's lineage is determined. If the jet contains a
$B$~meson within $R < 0.5$ of the jet axis it is labeled a
$b$-quark jet. If a $D$~meson is within $R < 0.5$ of the jet
axis, it is labeled a $c$-quark jet. If no $B$ or $D$ meson is found in the 
jet, the jet is labeled a light-quark jet. The probability determined
from the appropriate TRF is then applied. The taggability and tagging
probability are multiplied together to determine the probability of
the simulated jet to be tagged. 

In data we apply the secondary vertex algorithm directly and can identify which
jet is tagged and which is not. 
The situation in simulated events is different; the TRFs return a probability
(or weight) rather than a tagged/not-tagged answer per
jet. Since many of the discriminant variables used later on in the analysis (see
Sec.~\ref{discrim_vars}) need to know which jet was tagged, each
possible combination of tagged and untagged jets is considered for every simulated
event. Thus each event is used repeatedly in the analysis, considering each time
a different jet as tagged. The probability of
each combination is calculated using the tag rate functions, and
combined with the overall event weight. 
The sum of the 
weights for all the possible combinations of each event is equal to
the original probability for an event to have at least one tagged jet.

The use of all permissible tagged jet combinations in each simulated event
is a very powerful tool. It ensures that the
kinematic distributions in histograms of tagged events have the
correct shape, and it allows tagged jet information to be used in
variables for signal/background separation, since the final classifiers
are trained with weighted events.

\section{Triggers and Data Set}
\label{sec:triggers}
The \dzero trigger system is composed of three levels.  The first level consists
of hardware and firmware components, the second level uses information from the
first level to construct simple physics objects, and the third level is software
based and performs full event reconstruction.  

The \dzero calorimeter is used to trigger events based on the energy deposited
in towers of size $\Delta \eta \times \Delta \phi = 0.2 \times 0.2$ that are
segmented longitudinally into electromagnetic and hadronic sections.  The
level~1 electron trigger requires electrons to be above a certain threshold:
$E_{T} \equiv E \sin \theta > T$ where $E$ is the energy deposited in the tower,
$\theta$ is the angle between the beam and the trigger tower from the center of
the detector, and $T$ is the programmed threshold.  The level~2 electron trigger
uses a seed-based clustering algorithm that sums the energy deposited in two
neighboring towers and has the ability to make a decision based on the threshold
of the cluster, the electromagnetic fraction, and isolation of the electron.
The level~3 electron trigger uses a simple cone algorithm with $R < 0.25$ and
requirements on the $E_{T}$, the electromagnetic fraction, and the quality of
the transverse shower shape.

The level~1 jet trigger is similar to the electron trigger tower algorithm, but
includes the energy deposited in the hadronic portion of the calorimeter.  The
level~2 jet trigger uses a seed-based clustering algorithm summing the energy
deposition in a $5 \times 5$ tower array.  The level~3 jet algorithm is similar
to the level~3 electron algorithm, but does not include a requirement on the
electromagnetic fraction or shower shape.

The level~1 muon trigger examines hits from the muon wire chambers, muon
scintillation counters, and tracks from the level~1 track trigger for patterns
consistent with those coming from a muon.  The level~2
muon trigger reconstructs muon tracks from both wire and scintillator elements
in the muon system.  It can impose requirements on the number of muons, the
$p_{T}$ and $\eta$ of the muons, and the overall quality of the muons.  The
level~3 muon trigger uses wire and scintillator hits to reconstruct tracks using
segments inside and outside the toroid.

The output of the first level of the trigger is used to limit the rate for
accepted events to $\approx$~1.5~kHz. At the next trigger stage, with more
refined information, the rate is reduced further to $\approx$~800~Hz. The third
level of the trigger, with access to all the event information, reduces the
output rate to $\approx$~50~Hz, which is written to tape.

The data were acquired in the period between August 2002 and March 2004.
Tables~\ref{trigger_description_ejets_prd} and
\ref{trigger_description_mujets_prd} show the triggers used to collect the data
for the electron plus jets ($e$+jets) and muon plus jets ($\mu$+jets) triggers
and give the integrated luminosity for each trigger.

\begin{table*}[!h!t!bp]
\caption[trigger_description_ejets_prd]{Trigger conditions at levels~1, 2, and 3
for the electron plus jets trigger.}
\begin{center}
\begin{scriptsize}
\begin{ruledtabular}
\begin{tabular}{cccc} 
Level 1 & Level 2 & Level 3 & Luminosity \\ Condition & Condition & Condition &
\\ \hline 1 EM tower, $E_{T} > 10$ GeV & 1 $e$, $E_{T} > 10$ GeV, EM fraction $>
0.85$ & 1 tight $e$, $E_{T} > 15$ GeV & 19.4 pb$^{-1}$ \\ 2 jet towers, $E_{T}
> 5$ GeV & 2 jets, $E_{T} > 10$ GeV & 2 jets, $E_{T} > 15$ GeV \\
\hline
1 EM tower, $E_{T} > 10$ GeV & 1 $e$, $E_{T} > 10$ GeV, EM fraction $> 0.85$ & 1
loose $e$, $E_{T} > 15$ GeV & 91.2 pb$^{-1}$ \\ 2 jet towers, $E_{T} > 5$ GeV &
2 jets, $E_{T} > 10$ GeV & 2 jets, $E_{T} > 15$ GeV \\
\hline
1 EM tower, $E_{T} > 11$ GeV & & 1 tight $e$, $E_{T} > 15$ GeV & 115.4
pb$^{-1}$ \\ & & 2 jets, $E_{T} > 20$ GeV \\
\end{tabular}
\end{ruledtabular}
\end{scriptsize}
\label{trigger_description_ejets_prd}
\end{center}
\end{table*}

\begin{table*}[!h!t!bp]
\caption[trigger_description_mujets_prd]{Trigger conditions at levels~1, 2, and 3
for the muon plus jets trigger.}
\begin{center}
\begin{scriptsize}
\begin{ruledtabular}
\begin{tabular}{cccc} 
Level 1 & Level 2 & Level 3 & Luminosity \\ Condition & Condition & Condition &
\\ \hline 1 $\mu$, $\mid \eta \mid < 2.0$ & 1 $\mu$, $\mid \eta \mid < 2.0$ & 1
jet, $E_{T} > 20$ GeV & 113.7 pb$^{-1}$ \\ 1 jet tower, $E_{T} > 5$ GeV & & \\
\hline
1 $\mu$, $\mid \eta \mid < 2.0$ & 1 $\mu$, $\mid \eta \mid < 2.0$ & 1 jet,
$E_{T} > 25$ GeV & 113.7 pb$^{-1}$ \\ 1 jet tower, $E_{T} > 3$ GeV & 1 jet,
$E_{T} > 10$ GeV & \\
\end{tabular}
\end{ruledtabular}
\end{scriptsize}
\label{trigger_description_mujets_prd}
\end{center}
\end{table*}

\section{Event Selection}
\label{selection-section}
Event selection begins after all corrections have been applied
to the data.  These corrections include the jet energy
and the EM energy calibrations.   The primary vertex, $z_{\rm vertex}$, 
for the event must be within the tracking fiducial region, 
$|z_{\rm vertex}| < 60$~cm, which allows
for a sufficient number of tracks, $N_{\rm tracks} \ge 3$,
associated with it to be properly reconstructed.

As discussed in Sec.~\ref{intro:backgrounds}, the single top
quark signature is characterized by one isolated high-$p_{T}$ charged
lepton, {\met}, and two to four jets.  We accept events with
three or four jets in order to include contributions from extra gluons
and quarks.  The $b$ jet from the single
top quark decay tends to be more energetic than the other jets associated
with the event, so we require a higher $E_{T}$ for the leading jet.
Table~\ref{event_selection} lists the 
requirements of the initial selection.

\begin{table}[!h!tbp]
\caption[event_selection]{Initial event selection requirements.}
\begin{ruledtabular}
\begin{tabular}{lccc} 
Selection Cut                                 & $e$+jets  &                         & $\mu$+jets \\ \hline
tight $e$, $E_{T} \ge$ 15 GeV                 & =1         &                        & =0           \\            
tight $\mu$, $E_{T} \ge$ 15 GeV               & =0          &                       & =1           \\            
{\met}                                        &           & $\ge$ 15 GeV            & \\
$N_{\rm jets}$                                &           & $2 \le N_{\rm jets}\le 4$   & \\
$E_{T}({\rm jet})$                            &           & $\ge 15$ GeV            & \\
$|\eta({\rm jet})|$                           &           & $\le 3.2$               & \\
$E_{T}({\rm jet1})$                           &           & $\ge 25$ GeV            & \\
$|\eta({\rm jet1})|$                          &           & $\le 2.4$               & \\
\end{tabular}
\end{ruledtabular}
\label{event_selection}
\end{table}

\begin{figure*}[!h!tbp]
\includegraphics[width=0.40\textwidth]
{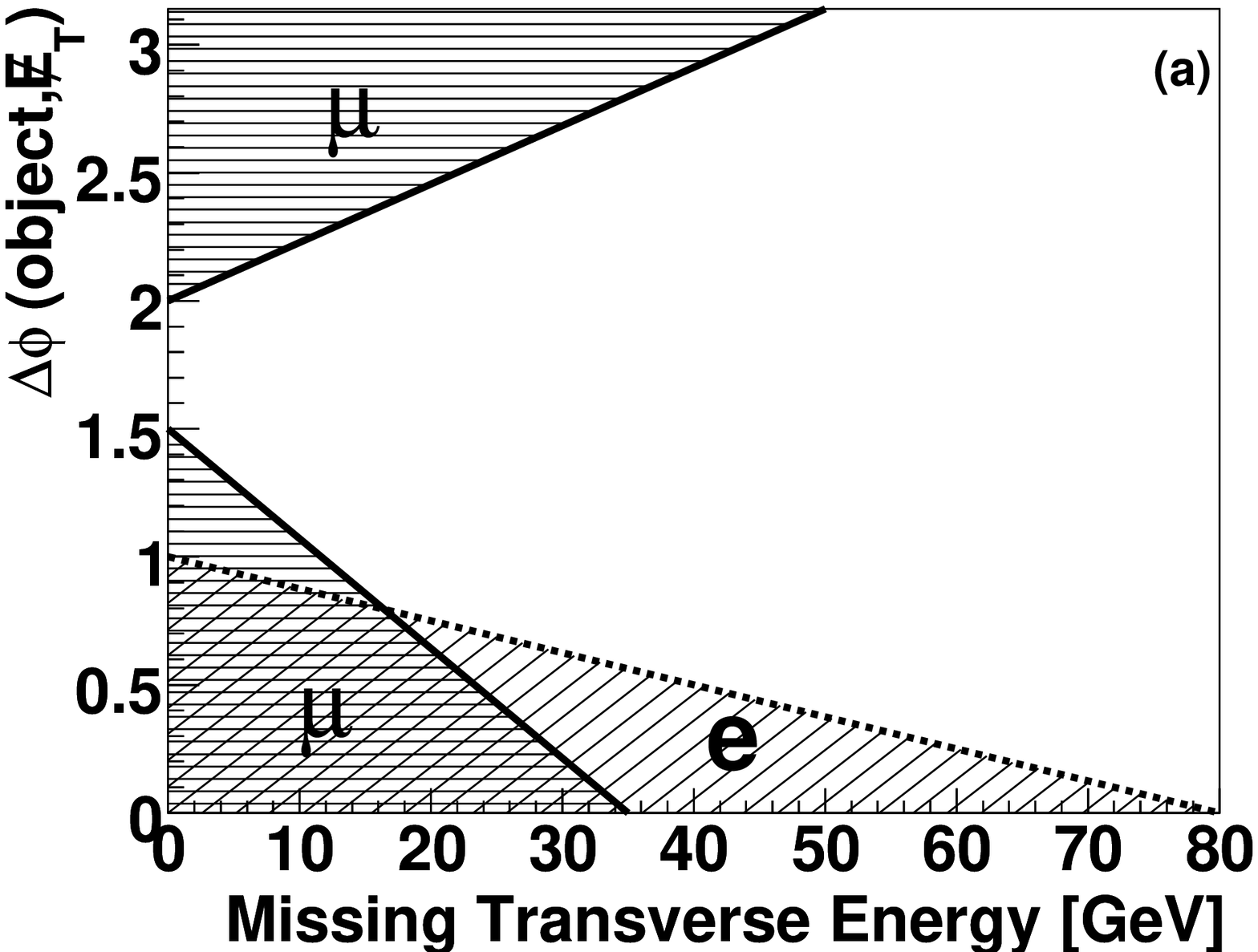}
\includegraphics[width=0.40\textwidth]
{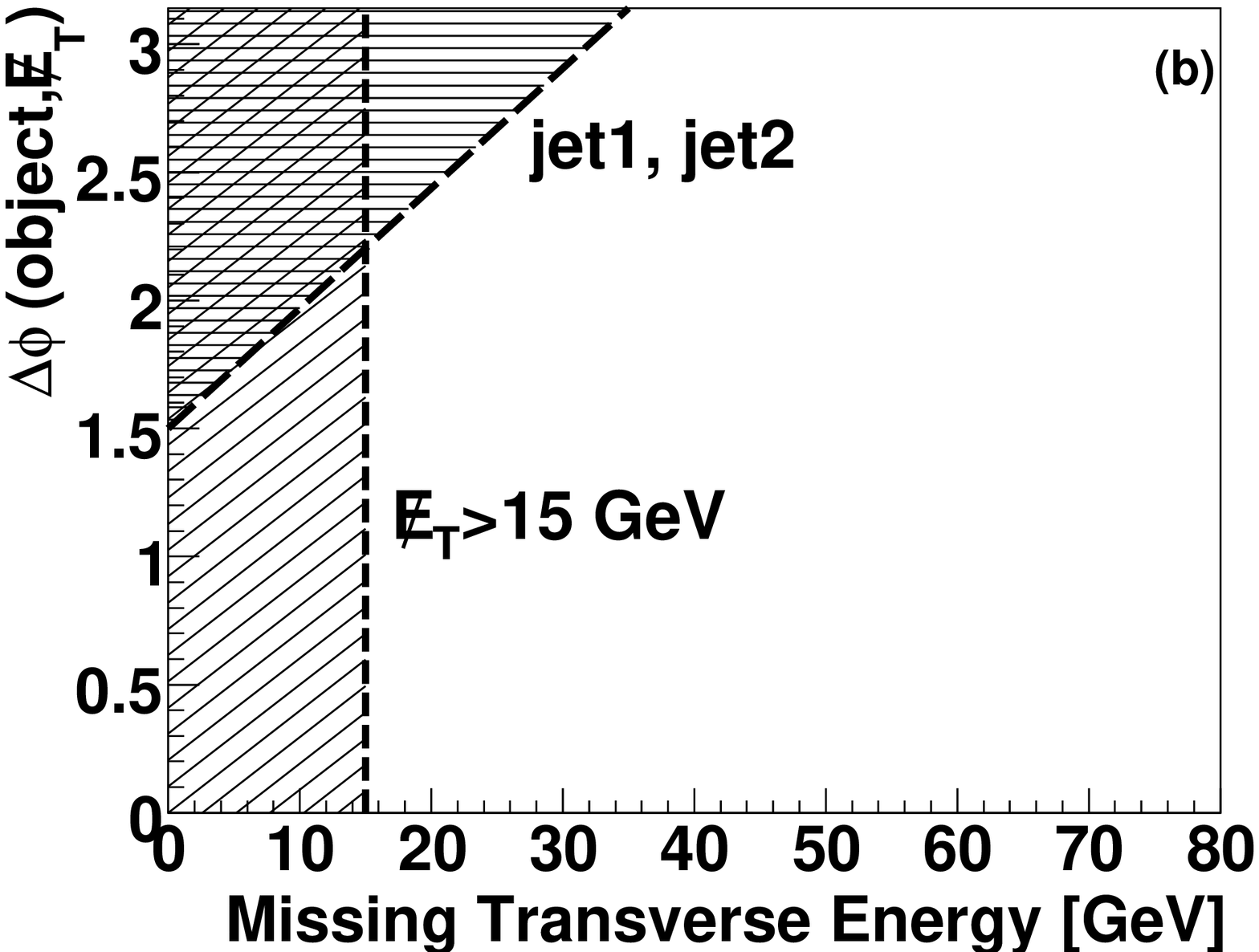}
\caption[triangle-cuts]{Kinematic regions excluded in the $e$+jets and
$\mu$+jets analyses by the triangle cuts applied in the $(\Delta\phi({\rm
object},\MET),\MET)$ plane, where each object can be: the tight isolated
electron or muon (a), and the leading and second leading jets (b). The shaded
areas are excluded.}
\label{triangle-cuts}
\end{figure*}

In addition, we make a set of cuts that remove misreconstructed events, also
known as ``triangle cuts.'' If the transverse energy of an object
is mismeasured, this tends to create false missing energy in a parallel or
antiparallel direction. The triangle cuts remove these mismeasured events,
which are difficult to model, but do not affect the signal appreciably because
there is very small signal acceptance in these kinematic regions.  In
Fig.~\ref{triangle-cuts}, we show the kinematic regions that are removed by the
triangle cuts.

From the selected jets in the event, at least one $b$-tagged jet
must be found. For the $t$-channel analysis, at least one jet
must be untagged.  This requirement comes from the fact that 
one of the main features of the 
$t$-channel signal is that a light-quark jet exists in the final
state.  The events are then divided into subsets consisting of the
number of tagged jets found in the event: single-tagged or double-tagged.
Since the $t$-channel requires at least one untagged jet, there are
no two-jet events in the double-tagged sample in the double-tagged
$t$-channel search.

\section{Signal and Background Modeling}
\label{sec:SigBkgModeling}
In order to compare the observed event yield in data with our expectation, and
to set limits on the single top quark production cross sections, we determine
acceptances and event yields for the single top quark signals and the various SM
background contributions. This estimation is based primarily on simulated
samples for shapes of distributions, except for the multijet background where we
use data samples. The yield normalization is based on theoretical cross
sections, except for the $W$+jets and multijet backgrounds which are normalized
to data.

\subsection{Acceptance and Yield for Simulated Samples}
The acceptance $\alpha$ for a particular simulated signal or background sample
is calculated as:
\begin{equation}
 {\alpha} = \frac{1}{N^{\rm MC}} \sum_i w_i  \label{eq:acceptance}
\end{equation}
where the sum is over simulated events that pass the selection cuts
and is normalized to the total number of simulated events in the sample $N^{\rm MC}$. The 
event weight $w_i$ is given by:
\begin{equation}
 w_i = \epsilon^{\rm lepton\; ID}_i \times \epsilon^{\rm jet\; ID}_i \times
 \epsilon^{\rm trigger}_i \times \epsilon^{b\;{\rm tagging}}_i
 \label{eq:event_weight}
\end{equation}
and includes correction factors $\epsilon$ to account for effects not modeled
and for cuts not applied to the simulated samples.
Trigger requirements are not made in the simulation (see
Sec.~\ref{sec:triggers}) and the correction factors $\epsilon^{\rm trigger}_i$
are about 90\%. Furthermore, we do not require $b$~tagging in simulated events,
and the correction factor $\epsilon^{b\;{\rm tagging}}_i$ averages about 55\%
for $s$-channel events and about 40\% for $t$-channel events.

The yield estimate ${\cal Y}$ is given by the product of acceptance,
integrated luminosity $\cal L$, theory cross section $\sigma^{\rm
theory}$, and branching fraction {\it B}:
\begin{equation}
 {\cal Y} = \alpha \times {\cal L} \times \sigma^{\rm theory} \times {\it B} . \label{eq:yield} 
\end{equation}
The branching fraction factor gives the fraction of events that result in the
final state lepton of interest ($e$ or $\mu$). The yield includes a small
contribution from $W \rightarrow \tau$ decays where the $\tau$ decays to $e$~or
$\mu$.

\subsection{Single Top Quark Signals}
The {\comphep} matrix element generator~\cite{comphepref} has been used to model
single top quark $s$-channel and $t$-channel signal events. We include not only
the leading order Feynman diagrams in the event generation, but also the
$O(\alpha_s)$ diagrams with real gluon radiation in order to reproduce NLO
distributions. For the $t$-channel sample, we include both the leading order
diagram (Fig.~\ref{fig:stop_diag_tqb} (a)) and the $W$-gluon fusion diagram
(Fig.~\ref{fig:stop_diag_tqb} (b)) explicitly, generating $W$-gluon fusion
events for the region of phase space where the $\bbar$ quark from gluon
splitting has $p_T(\bbar)>17$~GeV and leading order events otherwise.

\subsection{$\ttbar$ Background}
Top quark pair production contributes as a background both in the lepton+jets
and in the dilepton decay channels. This background is modeled using 
{\alpgen}~\cite{Mangano:2002ea}, and the yields are normalized
to the theory cross section (see Sec.~\ref{intro:backgrounds}).

\subsection{$WW$ and $WZ$ Backgrounds}
The backgrounds from diboson production are modeled using 
{\alpgen}, and the yields are normalized
to the theory cross sections~\cite{Campbell:1999ah}.

\subsection{Multijet and $W$+jets Backgrounds}
The backgrounds from multijet (fake lepton) and $W$+jets production are normalized to
the data sample before $b$~tagging~\cite{top-cs-topo}. We start from 
a data sample passing all selection cuts including the {\it loose} lepton requirements
(see Sec.~\ref{sec:objectreco}). From that sample, we select a subset of events
that also pass the {\it tight} lepton requirements. In addition, we determine
the probabilities for real and fake leptons to pass the tight lepton requirement.
These two probabilities together with
the numbers of events in the two samples then allow us to calculate
the number of real and fake lepton events in the $W$+jets and multijet
background samples~\cite{Abbott:1999tt}. 

The shapes of the distributions for
the multijet background are modeled using a data sample that passes all selection
cuts but fails the tight lepton identification requirements. 
The shapes of distributions for the $W$+jets background are modeled using 
{\alpgen} $W$+2jets events.

\subsubsection{Multijet Background}
A part of the background comes from events in which jets are misidentified as isolated
leptons. In the electron channel, this background is typically produced by jets that
contain a $\pi^0$, which, together with a randomly associated track, 
is misreconstructed as an isolated electron since it decays to two photons.
In the muon channel, this background
is typically produced by heavy-flavor jets in which a muon from a semileptonic
decay is misreconstructed as an isolated high-$p_T$ muon. 

The multijet background is estimated purely from data.  We use multijet data
samples that pass all event selection requirements, but fail the requirement on
tight muon isolation or tight electron quality (see Sec.~\ref{sec:objectreco})
to determine the kinematic shape of distributions. These samples are normalized
to the multijet background estimate in the data sample after event selection,
but before requiring a $b$~tag.

\subsubsection{$W$+Jets Background}
\begin{figure}[!h!tbp]
\includegraphics[width=0.40\textwidth]
{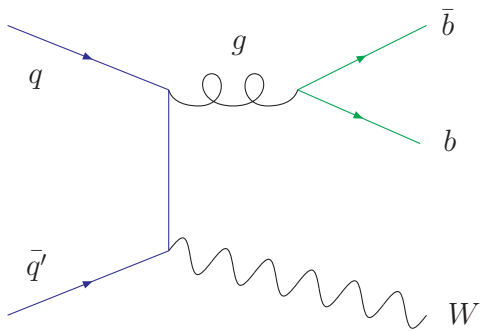}
\caption{Representative Feynman diagram for $Wb\bar{b}$ production.}
\label{fig:w2jet_diags}
\end{figure}

An example Feynman diagram for $W$+2~jet production is shown in
Fig.~\ref{fig:w2jet_diags}. This background is modeled from a simulated $Wjj$
sample ($j = u, d, s, c, g$), which includes not just light-quark flavors but
also $c$ quarks (considered massless in this model).  
We use a separate sample for $Wb\bar{b}$ and explicitly exclude events with $b$
quarks from the $Wjj$ sample. The parton level samples were generated with
{\alpgen}.

Since the $W$+jets background is normalized to data (after subtraction of the
small $\ttbar$ and diboson content), it includes all sources
of $W$+jets events with a similar flavor composition, in particular $Z$+jets events where one
of the leptons from the $Z$~boson decay is not identified.

\subsection{Detector Simulation}
\label{sec:detectorsim}
The parton-level samples for the single top quark signals,
$\ttbar$, $W$+jets, $WW$, and $WZ$ backgrounds
are processed with {\sc
pythia}~\cite{pythiaref} for hadronization and
modeling of the underlying event, 
using the {\sc cteq5l}~\cite{cteqref} parton
distribution functions. 
{\sc tauola}~\cite{Jadach:1990mz} is used for tau lepton decays and {\sc
evtgen}~\cite{Lange:2001uf} for $B$~hadron decays.
The generated events are processed through a 
{\sc geant}-based~\cite{geantref} simulation of the \dzero
detector.

\section{Event Yields}
\label{sec:EventYields}
The expected event yields for the various 
background contributions are calculated from both simulated samples and
data. The expected event yield for the single top quark signal is 
calculated from simulated samples and normalized to the theoretical cross
sections.

The total background event yield $\cal{Y}$ is given by the sum over all backgrounds:
\begin{equation}
  {\cal{Y}}    = \sum_{i} {\cal{Y}}_i
\end{equation}
where each individual yield ${\cal{Y}}_i$ is given by Eq.~\ref{eq:yield}
for the various MC samples.

Table~\ref{tab:yields} shows the numbers of events for each of the signals,
combinations of signals, backgrounds, and data, after event selection and
$b$~tagging.
The background sum reproduces the data within uncertainties for
all samples after $b$~tagging. 
\begin{table*}[!h!tbp]
\begin{center}
\caption{Event yields after selection in the electron and muon channels.}
\begin{ruledtabular}
\begin{tabular}{lcccccccc}
 & \multicolumn{4}{c}{Electron Channel}
 & \multicolumn{4}{c}{Muon Channel}\\
 & before tag & =1 tag & $\ge$2 tags & $\ge$2 tags
 & before tag & =1 tag & $\ge$2 tags & $\ge$2 tags \\
 &           &        & $s$-channel & $t$-channel
 &           &        & $s$-channel & $t$-channel \\
\hline
{\bf Signals}           &       &       &       &       
                         &       &       &       &       \\
~~$tb$                   &     5 &   2.3 &  0.53 &   --- 
                         &     5 &   2.2 &  0.50 &   --- \\
~~$tqb$                  &    12 &   4.1 &   --- &  0.25 
                         &    11 &   3.9 &   --- &  0.23 \\
{\bf Backgrounds}             &       &       &       &       
                         &       &       &       &       \\
~~$tb$                   &   --- &   --- &   --- &  0.14 
                         &   --- &   --- &   --- &  0.13 \\
~~$tqb$                  &   --- &   --- &  0.32 &   --- 
                         &   --- &   --- &  0.29 &   --- \\
~~${\ttbar}{\rar}\ell$+jets &    59 &  25.0 &  6.12 &  5.74 
                         &    58 &  24.2 &  5.78 &  5.48 \\
~~${\ttbar}{\rar}\ell\ell$     &    16 &   6.8 &  1.58 &  0.74 
                         &    17 &   7.2 &  1.62 &  0.75 \\
~~$Wb\bar{b}$            &    40 &  15.1 &  2.49 &  0.59 
                         &    33 &  12.7 &  2.15 &  0.55 \\
~~$Wjj$                  & 3,211 &  68.2 &  1.53 &  0.66 
                         & 2,898 &  62.9 &  1.29 &  0.68 \\
~~$WW$                   &    13 &   0.7 &  0.00 &  0.00 
                         &    14 &   0.7 &  0.00 &  0.00 \\
~~$WZ$                   &     4 &   0.6 &  0.11 &  0.02 
                         &     5 &   0.5 &  0.09 &  0.01 \\
~~Multijet               &   478 &  13.7 &  0.31 &  0.14 
                         &   256 &  17.2 &  0.20 &  0.20 \\
{\bf Summed signals}           &    17 &   6.4 &  0.53 &  0.25 
                         &    16 &   6.0 &  0.50 &  0.23 \\
{\bf Summed backgrounds}       & 3,821 & 130.1 &   --- &   --- 
                          & 3,280 & 125.4 &   --- &   --- \\
{\bf Summed backgrounds+$tqb$} & 3,833 & 134.2 & 12.47 &   --- 
                         & 3,291 & 129.3 & 11.43 &   --- \\
{\bf Summed backgrounds+$tb$}  & 3,826 & 132.4 &   --- &  8.03 
                         & 3,285 & 127.6 &   --- &  7.80 \\
{\bf Data}              & 3,821 & 134   & 15    & 11 
                         & 3,280 & 118   & 16    &  8
\end{tabular}
\end{ruledtabular}
\label{tab:yields}
\end{center}
\end{table*}

A summary of the yield estimates for the signal and backgrounds and the numbers
of observed events in data after selection, including the systematic
uncertainties as described in Sec.~\ref{systematics}, is shown in
Table~\ref{tab:yield_sys}.

\begin{table}[!h!tbp]
\begin{center}
\caption{Estimates for signal and background yields and the numbers
of observed events in data after event selection for the electron and muon,
single-tagged and double-tagged analysis sets combined. The $W$+jets yields
include the diboson backgrounds.  The total background for the $s$-channel
($t$-channel) search includes the $tqb$ ($tb$) yield. The quoted yield
uncertainties include systematic uncertainties taking into account correlations
between the different analysis channels and samples.}
\begin{ruledtabular}
\begin{tabular}{lr@{$\,\pm\!\!\!\!$}lr@{$\,\pm\!\!\!\!$}l} 
Source           & \multicolumn{2}{c}{$s$-channel search} &
\multicolumn{2}{c}{$t$-channel search} \\ \hline
$tb$             &   5.5 & 1.2  &   4.8 & 1.0  \\ 
$tqb$            &   8.6 & 1.9  &   8.5 & 1.9  \\ 
$W$+jets         & 169.1 & 19.2 & 163.9 & 17.8 \\
$\ttbar$         &  78.3 & 17.6 &  75.9 & 17.0 \\
Multijet         &  31.4 & 3.3  &  31.3 & 3.2  \\ \hline
Total background & ~~~287.4 & 31.4 & ~~~275.8 & 31.5 \\
Observed events  & \multicolumn{2}{c}{283}  &  \multicolumn{2}{c}{271}     \\
\end{tabular}
\end{ruledtabular}
\label{tab:yield_sys}
\end{center}
\end{table}

After $b$~tagging, the $W$+jets background makes up around 60\% of the total
background model (48\% $Wjj$, 12\% $Wb\bar{b}$), the {\ttbar} background is
around 27\% (21\% lepton+jets, 6\% dilepton), 10\% is mainly multijet
background, and $s$-channel single top quark production provides
$3\%$ in the $t$-channel search and vice versa.

\section{Event Analysis}
\label{sec:EventAnalysis}
Table~\ref{tab:yields} shows that even after event selection and $b$~tagging,
the expected single top quark signal yield is small compared to the overwhelming
backgrounds.  Additional steps are necessary in order to separate the signal and
background.  In this section, we first present kinematic variables that allow us
to separate the $s$-channel or $t$-channel single top quark signal from the
backgrounds. We then describe a cut-based analysis and a neural networks
analysis that use these variables. 

\subsection{Discriminating Variables}
\label{discrim_vars}
\vspace{-0.2cm}
In this section we
introduce the variables that we found to be most effective in separating
the single top quark signals from the backgrounds.
The list of discriminating variables has been chosen based on an 
analysis of Feynman diagrams of signals and backgrounds~\cite{boos-dudko} 
and on a study of single top quark production at NLO~\cite{sintop-nlo-sch,sintop-nlo-tch}. 

The variables fall into three categories: individual object
kinematics, global event kinematics, and variables based on angular
correlations.
The list of variables is shown in Table~\ref{tab:variable-sets}.

In order to get optimum separation between signal and background, the single top
quark final state is reconstructed according to whether a variable is primarily
used in the $s$-channel or the $t$-channel search. The $W$~boson from the top
quark decay is reconstructed from the isolated lepton and the missing transverse
energy. The $z$-component of the neutrino momentum is calculated using a
$W$~boson mass constraint, choosing the solution with smaller $|p^\nu_z|$ from
the two possible solutions. The candidate top quark is reconstructed from this
$W$~boson and a jet. This jet is chosen to be either the leading $b$-tagged jet
or the $best$ jet.  In the $t$-channel analysis, there is typically only one
high-$p_T$ $b$~quark jet in the final state, thus the leading $b$-tagged jet is
chosen to reconstruct the top quark. By contrast, in the $s$-channel there are
two high-$p_T$ $b$~quark jets in the final state, and a choice needs to be made
between them. Furthermore, typically only one of the two is identified as a
$b$-tagged jet.  We use the best-jet algorithm~\cite{d0runI} to identify this
jet without using $b$ tagging information.  The best jet is defined as
the jet in each event which gives, together with the reconstructed $W$~boson, an
invariant mass closest to 175~GeV.
Jets that have not been identified by the $b$~tagging algorithm are called
``untagged'' jets.

\begin{table*}[!h!tbp]
\begin{center}
\caption[tab:variable-sets]{List of discriminating variables.
A check mark in the final four columns indicates in which signal-background pair
of the neural net analysis the variable is used.
}
\begin{footnotesize}
\begin{ruledtabular}
\begin{tabular}{lp{0.60\textwidth}cccc} 
\multicolumn{6}{r}{Signal-Background Pairs} \\
& & \multicolumn{2}{c}{$tb$} &  \multicolumn{2}{c}{$tqb$} \\
\multicolumn{1}{c}{Variable}&\multicolumn{1}{c}{Description} &      $W\bbbar$  & ${\ttbar}$ &  $W\bbbar$ & ${\ttbar}$ \\
\hline
\multicolumn{6}{c}{\bf{Individual object kinematics}} \\
$p_T({\rm jet1}_{\rm tagged})$     & 
Transverse momentum of the leading tagged jet     & $\surd$ & $\surd$ & $\surd$ & --- \\            
$p_{T}({\rm jet1}_{\rm untagged})$ & 
Transverse momentum of the leading untagged jet   & --- & --- & $\surd$ & $\surd$ \\            
$p_{T}({\rm jet2}_{\rm untagged})$ & 
Transverse momentum of the second untagged jet    & --- & --- & --- & $\surd$ \\            
$p_{T}({\rm jet1}_{\rm non-best})$ & 
Transverse momentum of the leading non-best jet   & $\surd$ & $\surd$ & --- & --- \\            
$p_{T}({\rm jet2}_{\rm non-best})$ & 
Transverse momentum of the second non-best jet    & $\surd$ & $\surd$ & --- & --- \\            
\multicolumn{6}{c}{\bf{Global event kinematics}} \\
$\sqrt{\hat{s}}$ &
Invariant mass of all final state objects 
                                       & $\surd$ & --- & $\surd$ & $\surd$ \\    
$p_T({\rm jet1},{\rm jet2})$        & 
Transverse momentum of the two leading jets& $\surd$ & --- & $\surd$ & --- \\            
$M_T({\rm jet1},{\rm jet2})$        & 
Transverse mass of the two leading jets     & $\surd$ & --- & --- & --- \\
$M({\rm alljets})$           & 
Invariant mass of all jets           & $\surd$ & $\surd$ & $\surd$ & $\surd$ \\ 
$H_T({\rm alljets})$         & 
Sum of the transverse energies of all jets      & --- & --- & $\surd$ & --- \\
$p_T({\rm alljets}-{\rm jet1}_{\rm tagged})$ & 
Transverse momentum of all jets excluding the leading tagged jet    & --- & $\surd$ & --- & $\surd$ \\
$M({\rm alljets}-{\rm jet1}_{\rm tagged})$ & 
Invariant mass of all jets excluding the leading tagged jet   & --- & --- & --- & $\surd$ \\  
$H({\rm alljets}-{\rm jet1}_{\rm tagged})$ & 
Sum of the energies of all jets excluding the leading tagged jet & --- & $\surd$ & --- & $\surd$ \\ 
$H_T({\rm alljets}-{\rm jet1}_{\rm tagged})$ & 
Sum of the transverse energies of all jets excluding the leading tagged jet         & --- & --- & --- & $\surd$ \\ 
$M(W,{\rm jet1}_{\rm tagged})$ & 
Invariant mass of the reconstructed top quark using the leading tagged jet & $\surd$ & $\surd$ & $\surd$ & $\surd$ \\
$M({\rm alljets - jet_{best}})$ & 
Invariant mass of all jets excluding the best jet          & --- & $\surd$ & --- & --- \\            
$H({\rm alljets}-{\rm jet}_{\rm best})$ & 
Sum of the energies of all jets excluding the best jet    & --- & $\surd$ & --- & --- \\      
$H_T({\rm alljets}-{\rm jet}_{\rm best})$ & 
Sum of the transverse energies of all jets excluding the best jet     & --- & $\surd$ & --- & --- \\
$M(W,{\rm jet}_{\rm best})$ & 
Invariant mass of the reconstructed top quark using the best jet  & $\surd$ & --- & --- & --- \\            
\multicolumn{6}{c}{\bf{Angular variables}} \\
$\eta({\rm jet1}_{\rm untagged}) \times Q_{\ell}$ & 
Pseudorapidity of the leading untagged jet  $\times$ lepton charge     & --- & --- & $\surd$ & $\surd$ \\
$\Delta \cal{R}({\rm jet1},{\rm jet2})$ &
Angular separation between the leading two jets     & $\surd$ & --- & $\surd$ & --- \\ 
$\cos({\rm \ell},{\rm jet1}_{\rm untagged})_{\rm top_{tagged}}$      &
Top quark spin correlation in the optimal basis for the $t$-channel~\cite{Mahlon:1995zn}, reconstructing the top quark with the leading tagged jet & --- & --- & $\surd$ & --- \\            
$\cos({\rm \ell},Q_{\ell}$$\times$$z)_{\rm top_{best}}$ &
Top quark spin correlation in the optimal basis for the $s$-channel~\cite{Mahlon:1995zn}, reconstructing the top quark with the best jet  & $\surd$ & --- & --- & --- \\ 
$\cos({\rm alljets},{\rm jet1}_{\rm tagged})_{\rm alljets}$          &            
Cosine of the angle between the leading tagged jet and the alljets system in the alljets rest frame     & --- & --- & $\surd$ & $\surd$ \\ 
$\cos({\rm alljets},{\rm jet}_{\rm non-best})_{\rm alljets}$  &
Cosine of the angle between the leading non-best jet and the alljets system in the alljets rest frame  & --- & $\surd$ & --- & --- \\ 
\end{tabular}
\end{ruledtabular}
\end{footnotesize}
\label{tab:variable-sets}
\end{center}
\end{table*}

\begin{figure*}[!h!tbp]  
\begin{center}
\includegraphics[width=0.32\textwidth]
{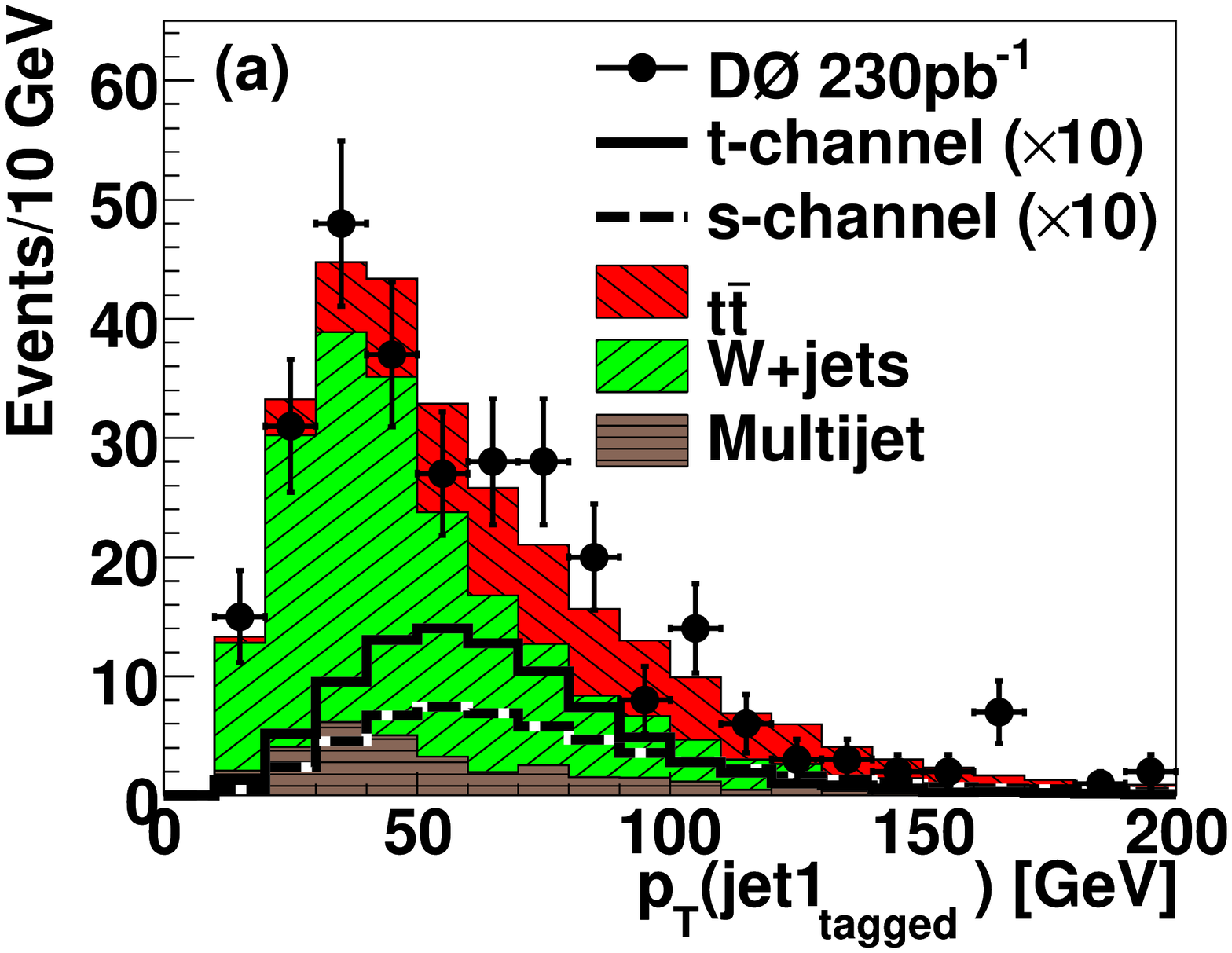}
\includegraphics[width=0.32\textwidth]
{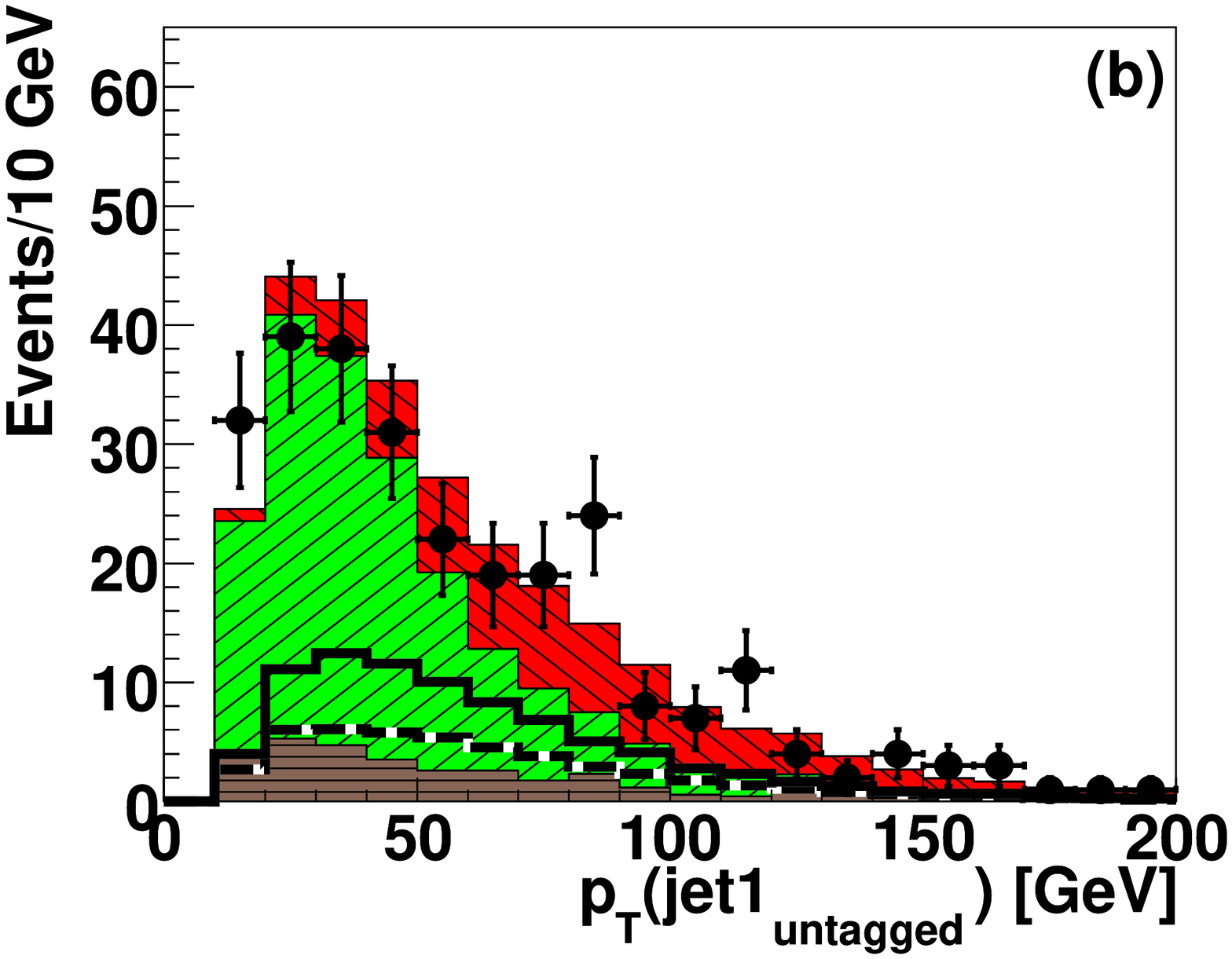}
\includegraphics[width=0.32\textwidth]
{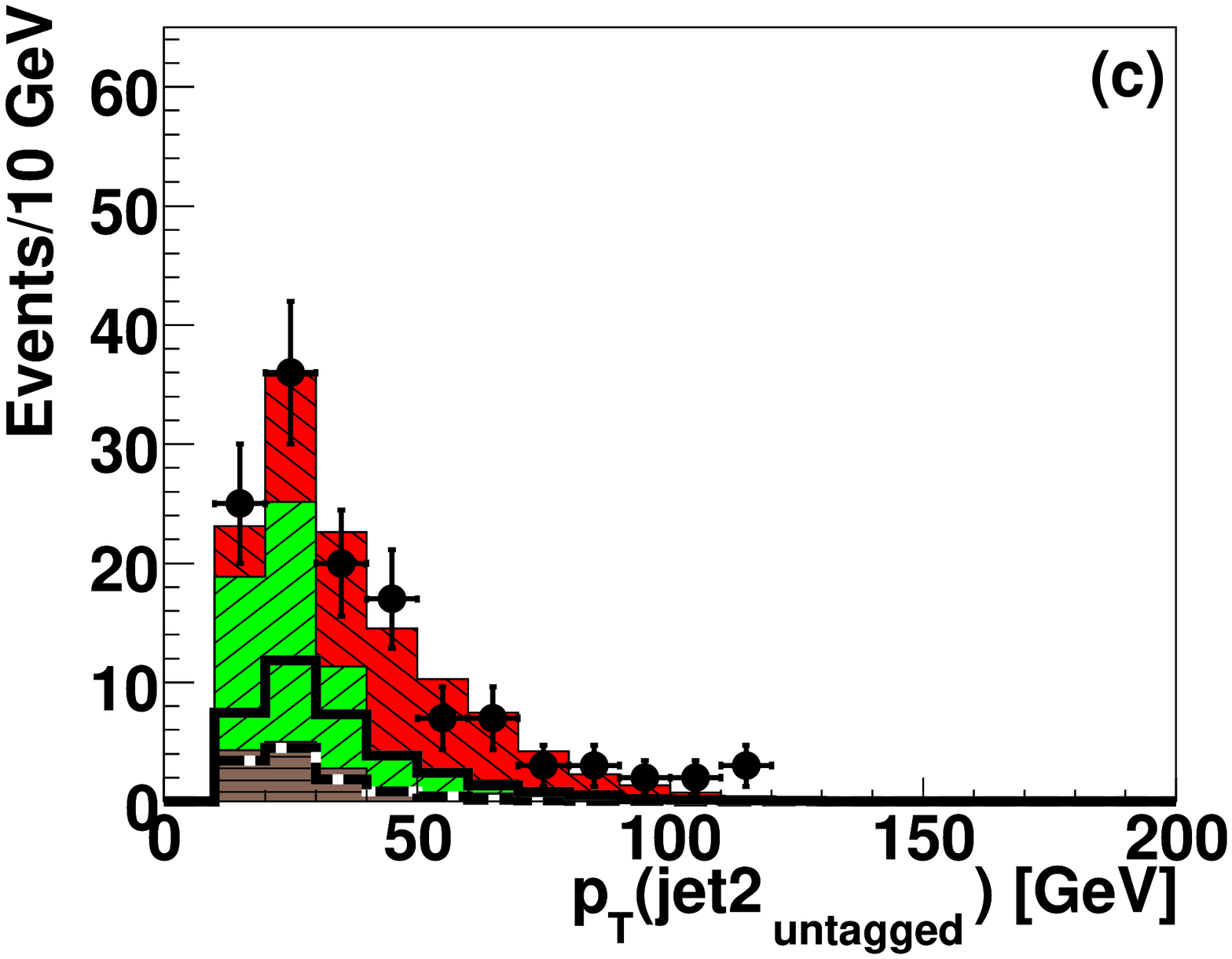}
\includegraphics[width=0.32\textwidth]
{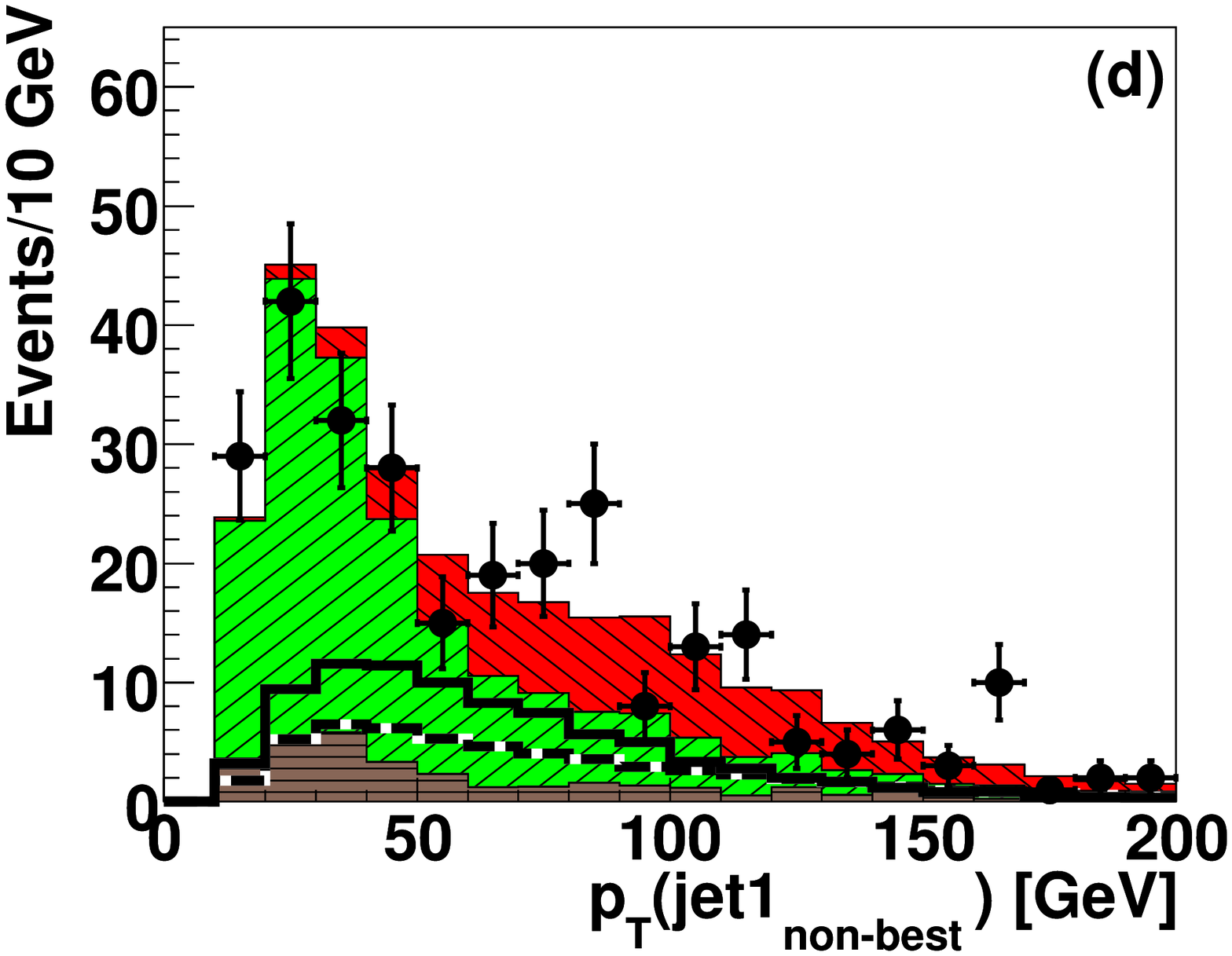}
\includegraphics[width=0.32\textwidth]
{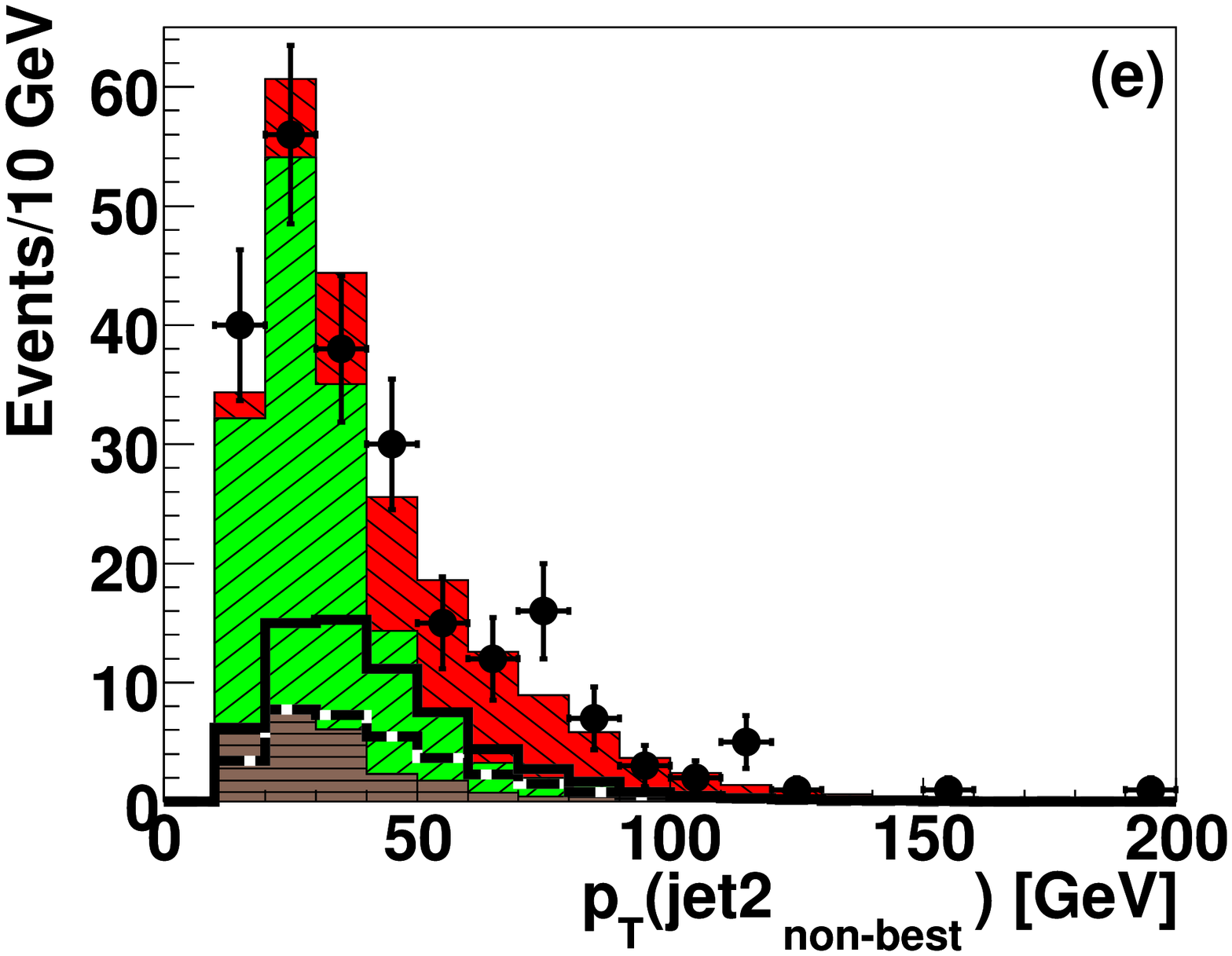}
\end{center}
\caption{Comparison of signal, backgrounds, and 
data after selection and requiring at least one $b$-tagged jet for five individual
object variables. Electron and muon channels are combined. The transverse
momentum is shown for (a) the leading tagged jet; (b) the leading untagged jet;
(c) the second untagged jet, for those events that contain at least two untagged
jets; (d) the leading non-best jet; and (e) the second non-best jet.  Signals
are multiplied by ten.}
\label{fig:variables-object}
\end{figure*}

\begin{figure*}[!h!tbp]  
\begin{center}
\includegraphics[width=0.32\textwidth]
{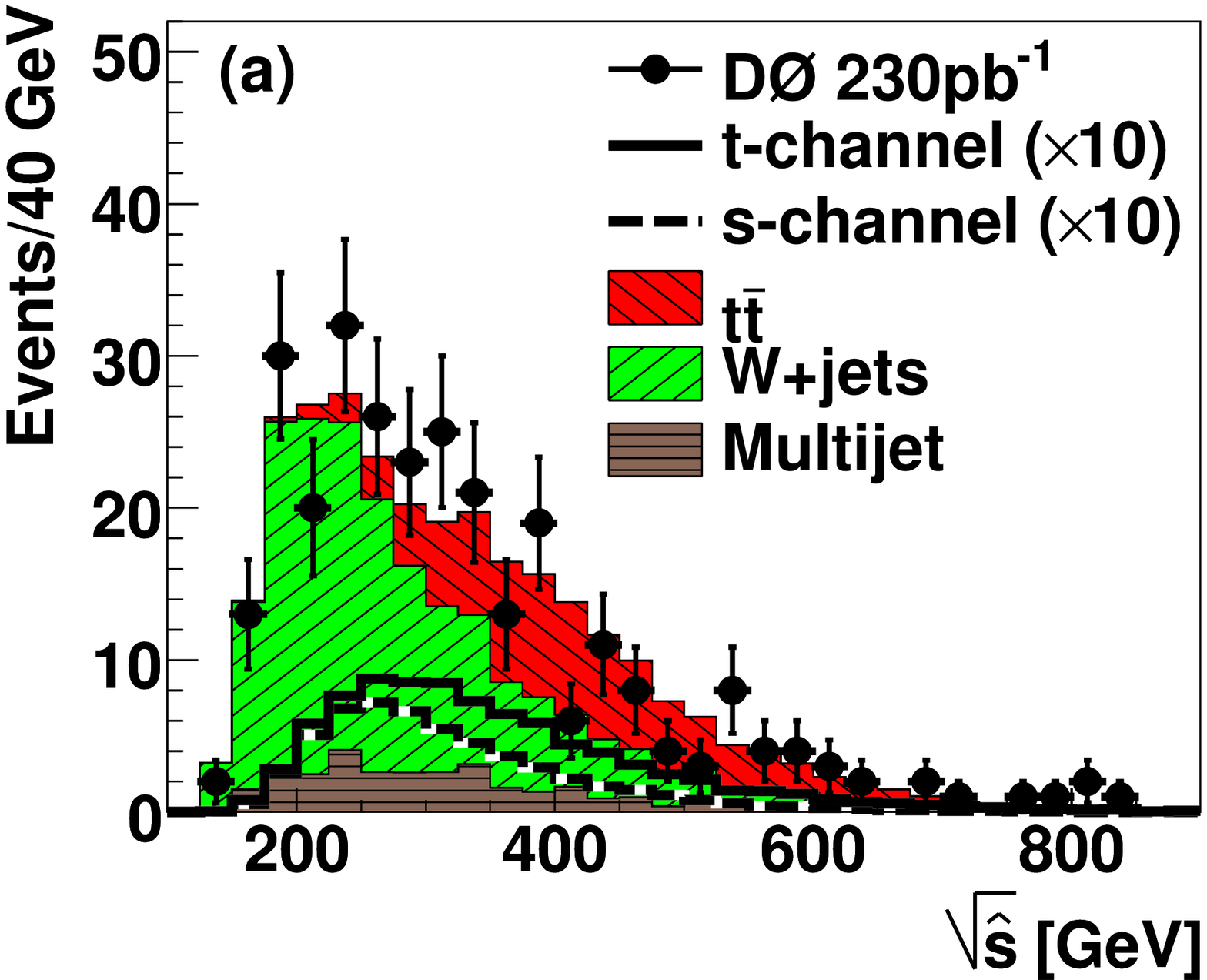}
\includegraphics[width=0.32\textwidth]
{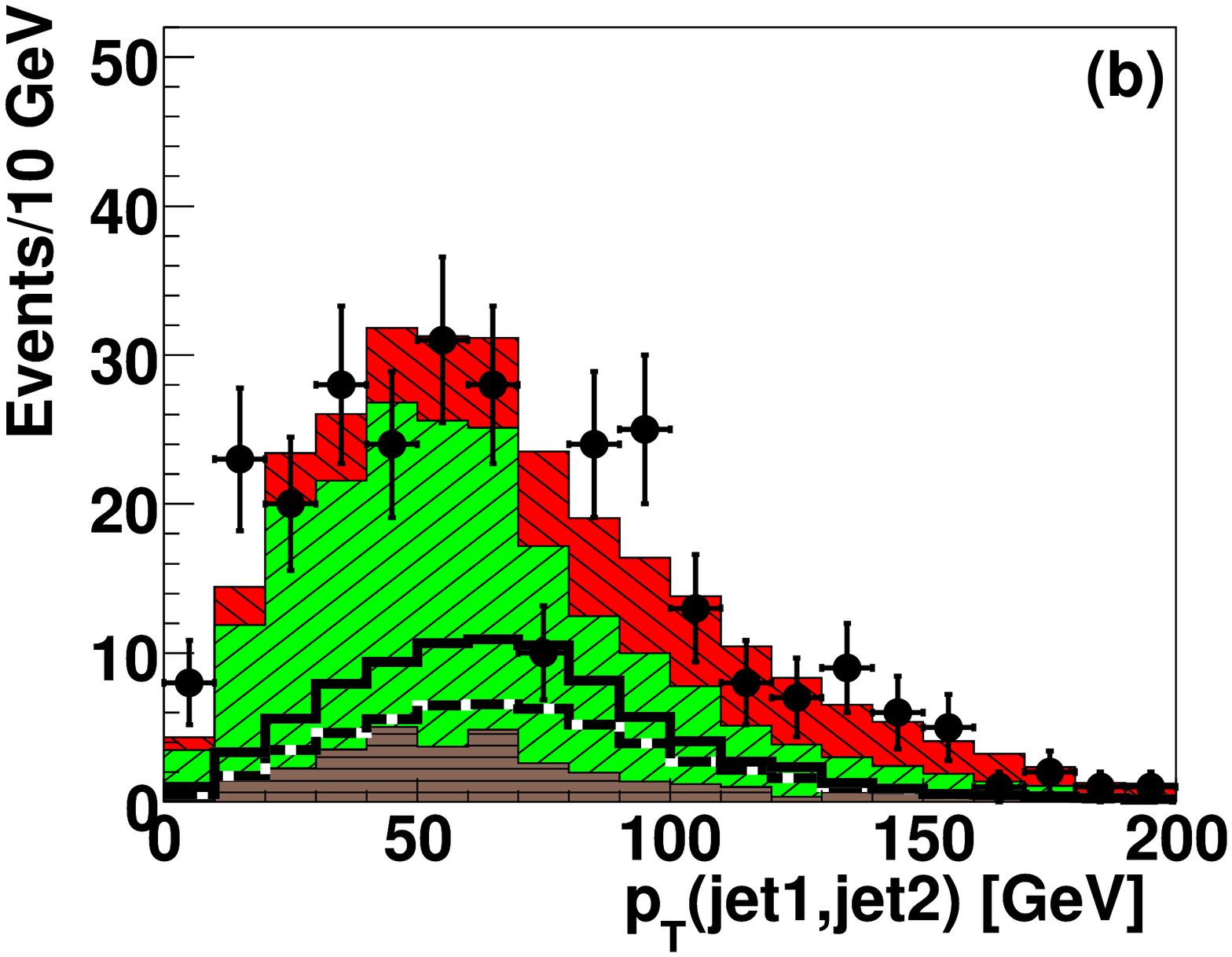}
\includegraphics[width=0.32\textwidth]
{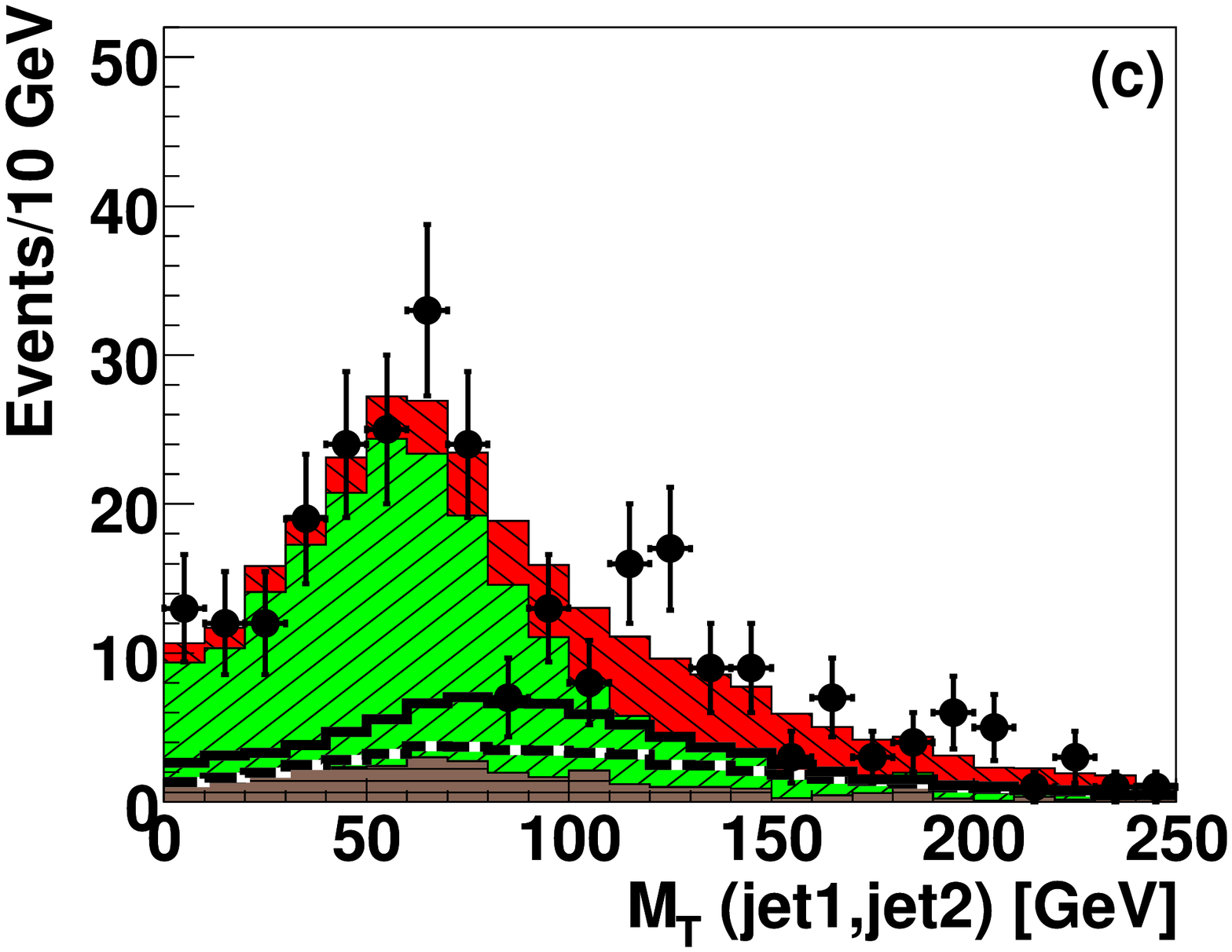}
\includegraphics[width=0.32\textwidth]
{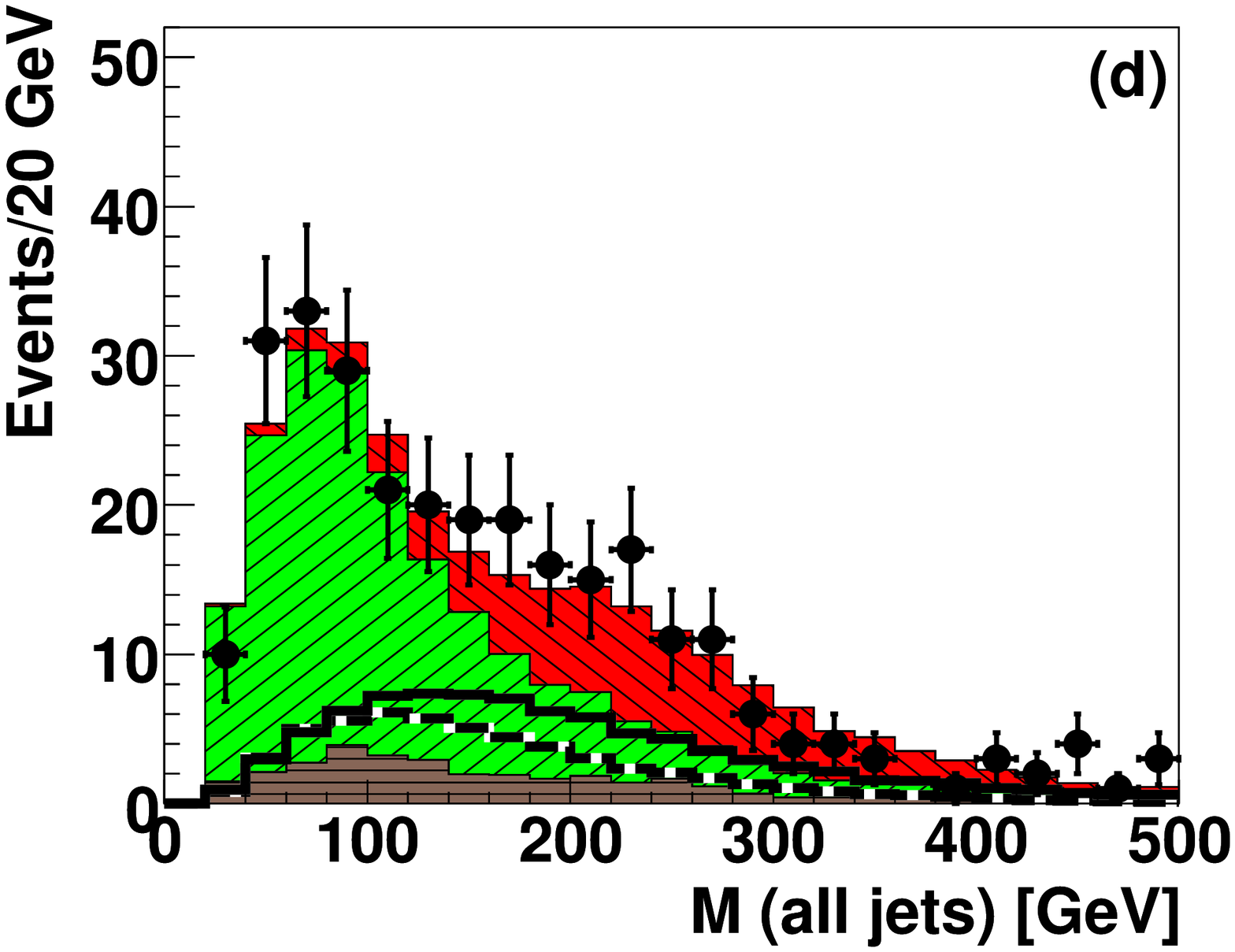}
\includegraphics[width=0.32\textwidth]
{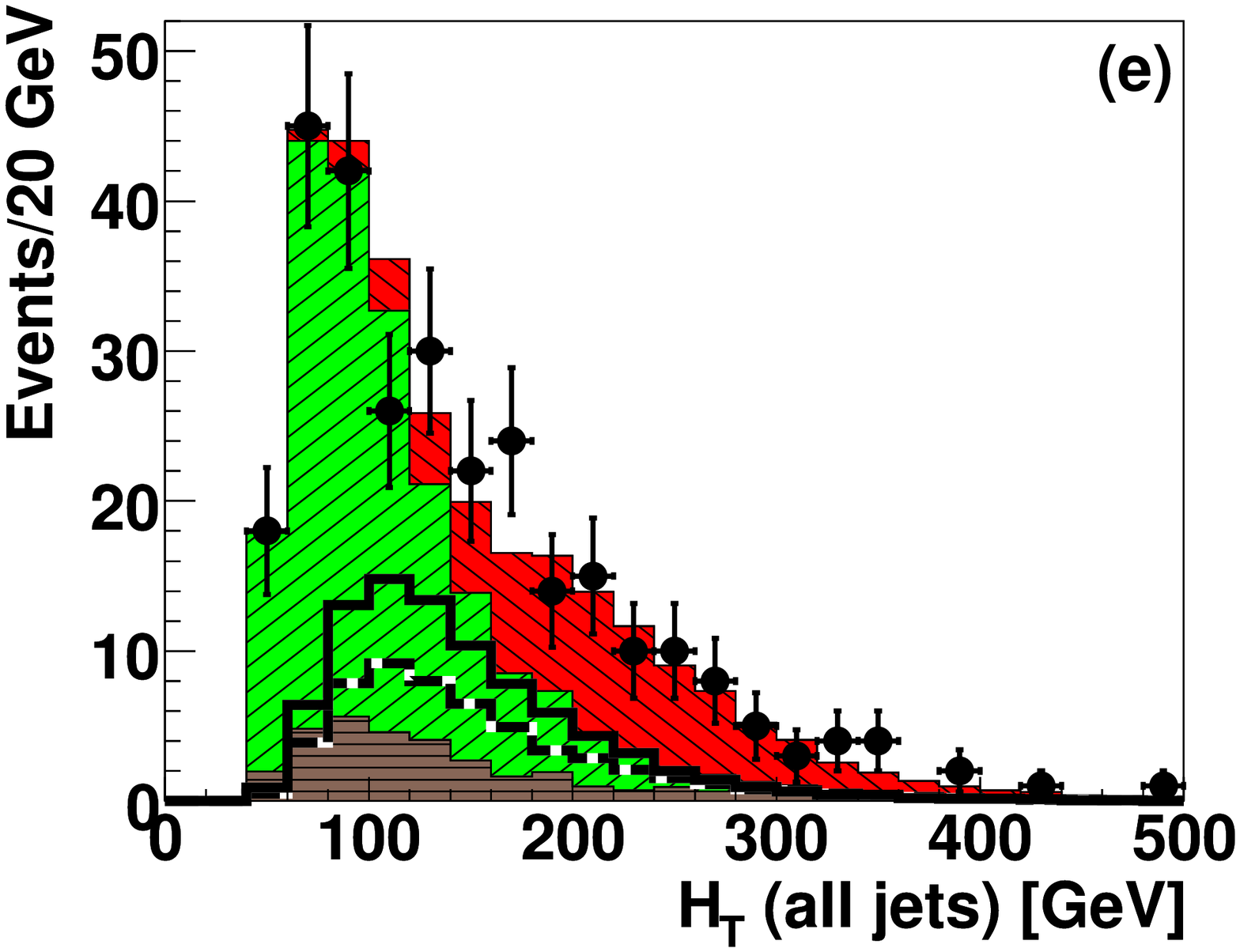}
\end{center}
\caption{Comparison of signal, backgrounds, and 
data after selection and requiring at least one $b$-tagged jet for five
discriminating event kinematic variables. Electron and muon channels are
combined.  Shown are (a) the invariant mass of all final state objects, (b) the
total transverse momentum of the leading two jets, (c) the transverse mass of the
leading two jets, (d) the invariant mass of all jets, and (e) the total
transverse energy of all jets.  Signals are multiplied by ten.}
\label{fig:variables-event1}
\end{figure*}

\begin{figure*}[!h!tbp]  
\begin{center}
\includegraphics[width=0.32\textwidth]
{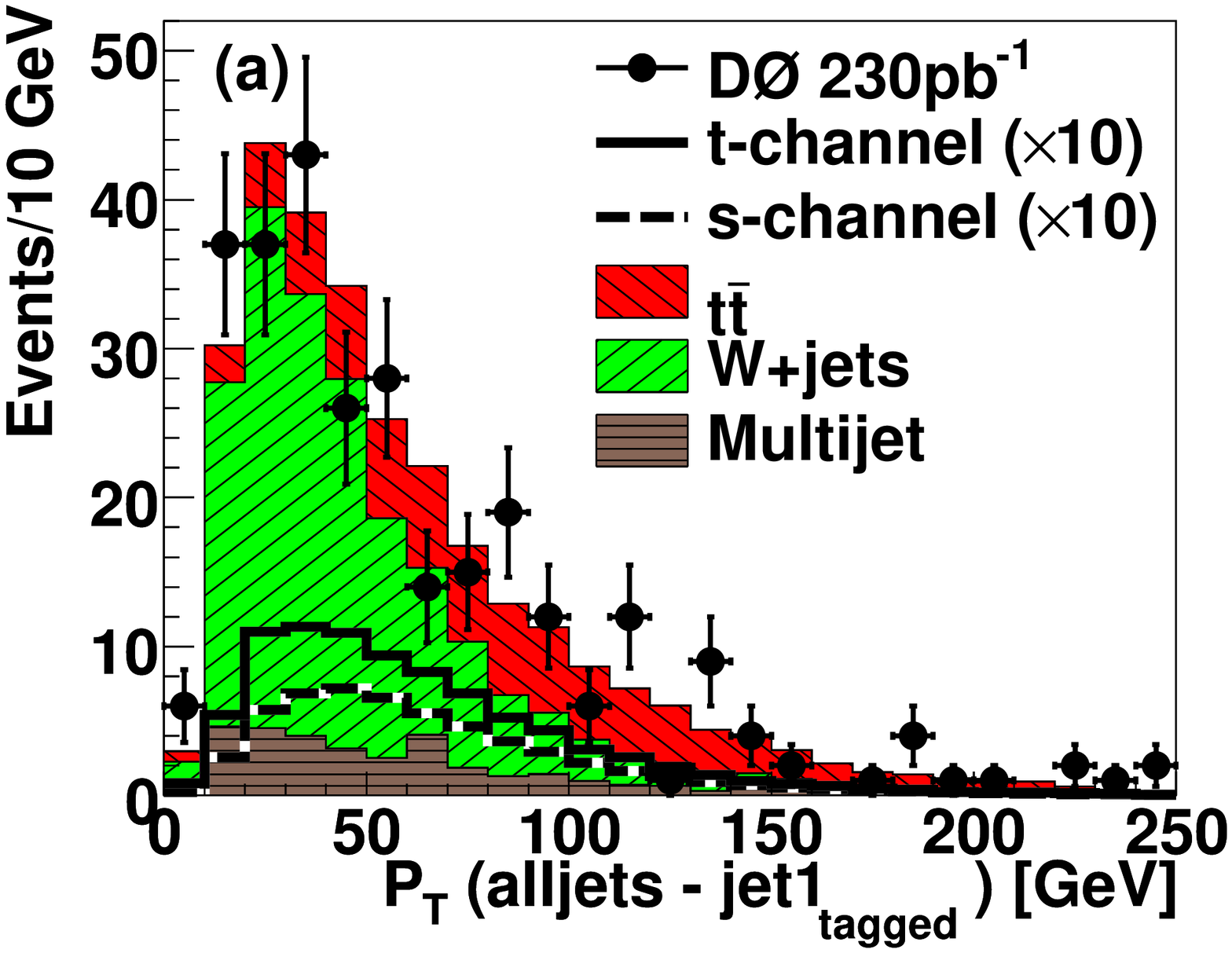}
\includegraphics[width=0.32\textwidth]
{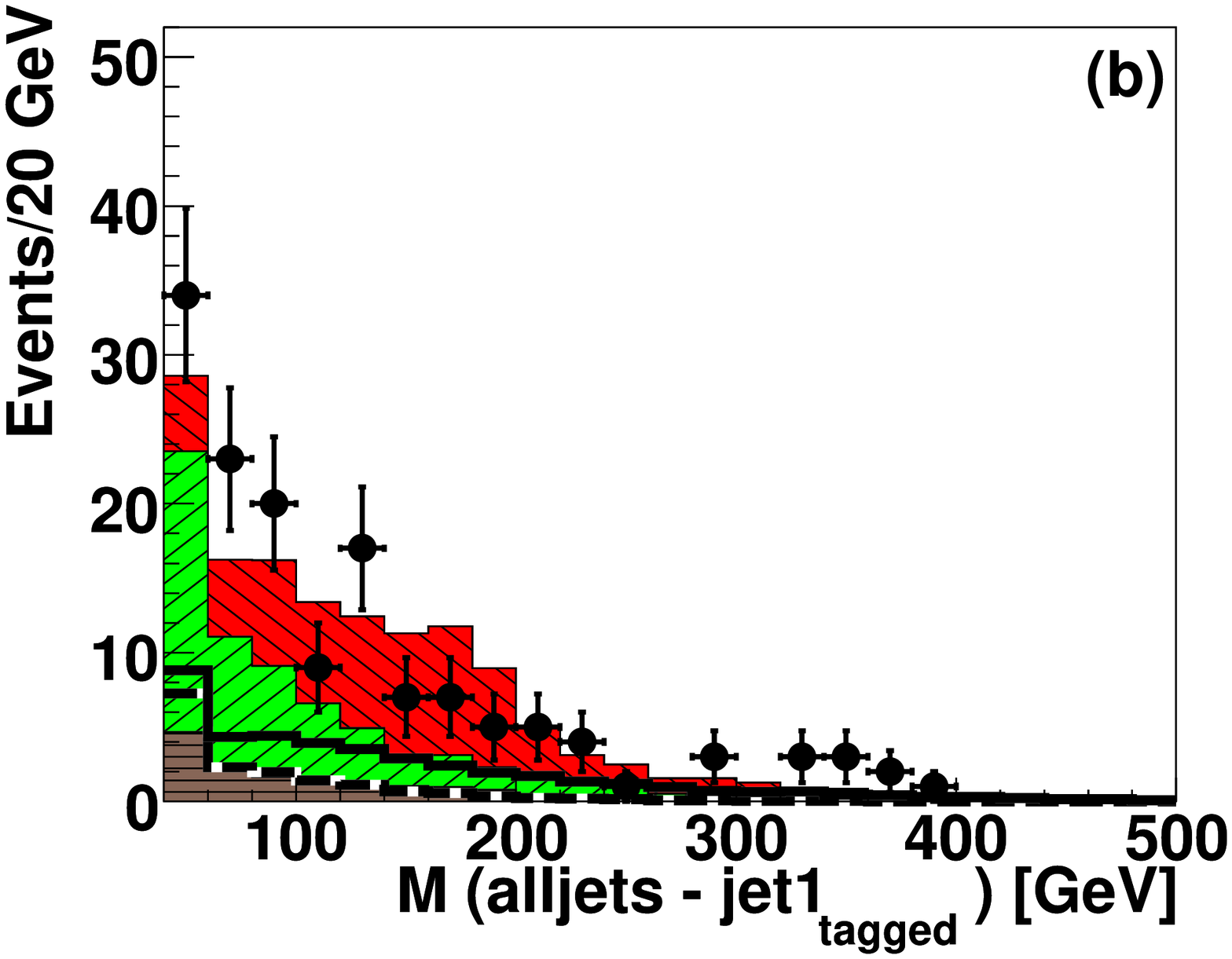}
\includegraphics[width=0.32\textwidth]
{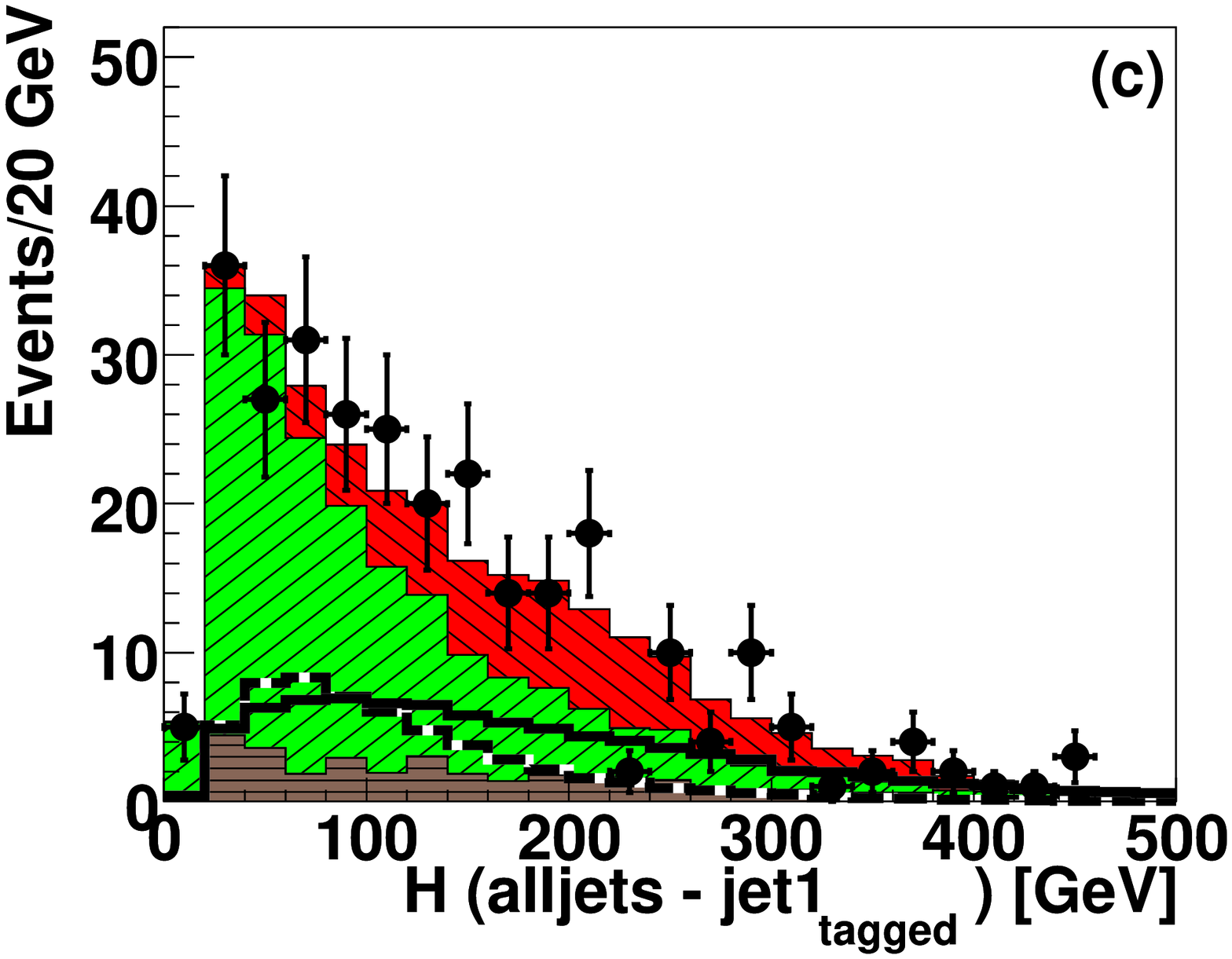}
\includegraphics[width=0.32\textwidth]
{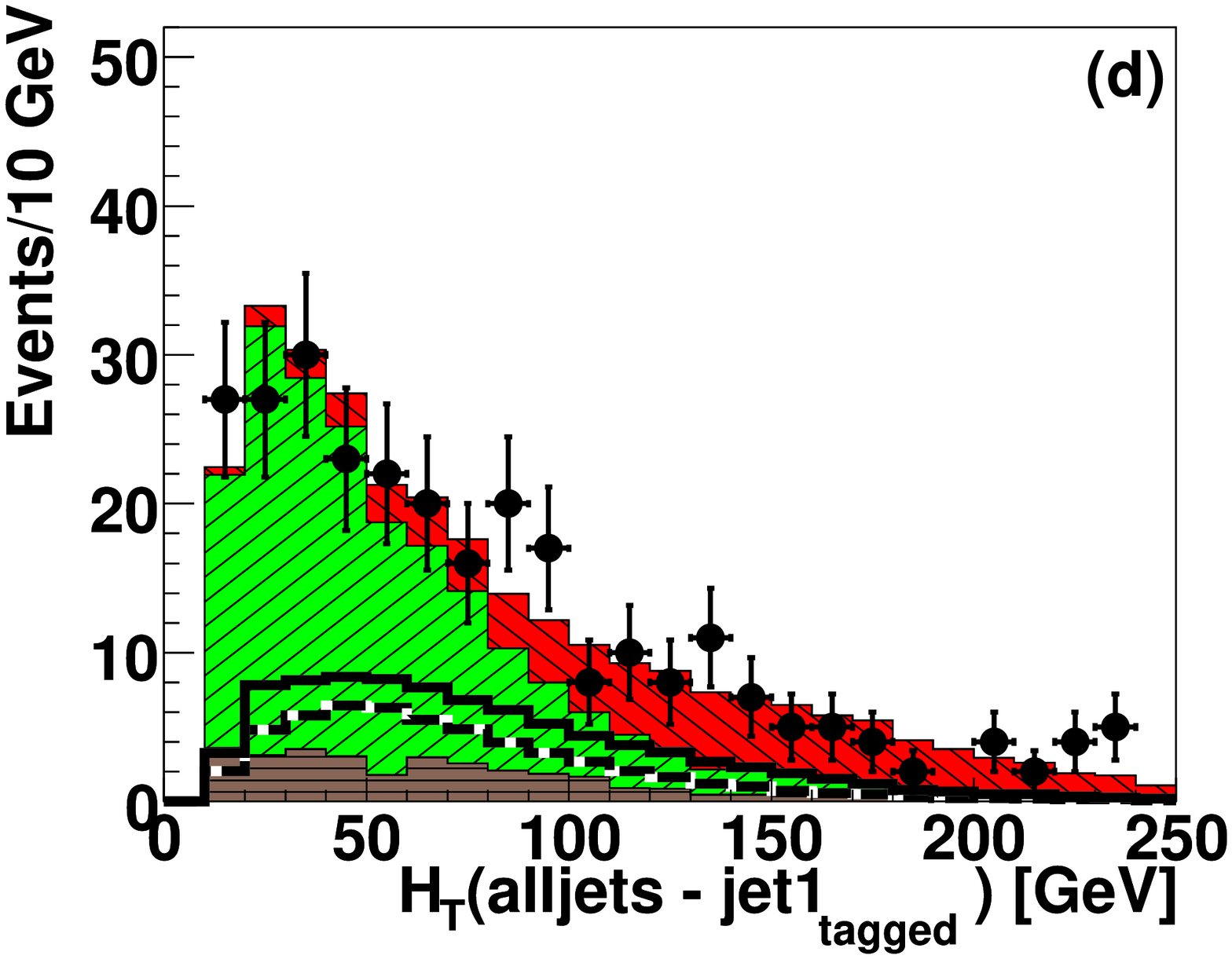}
\includegraphics[width=0.32\textwidth]
{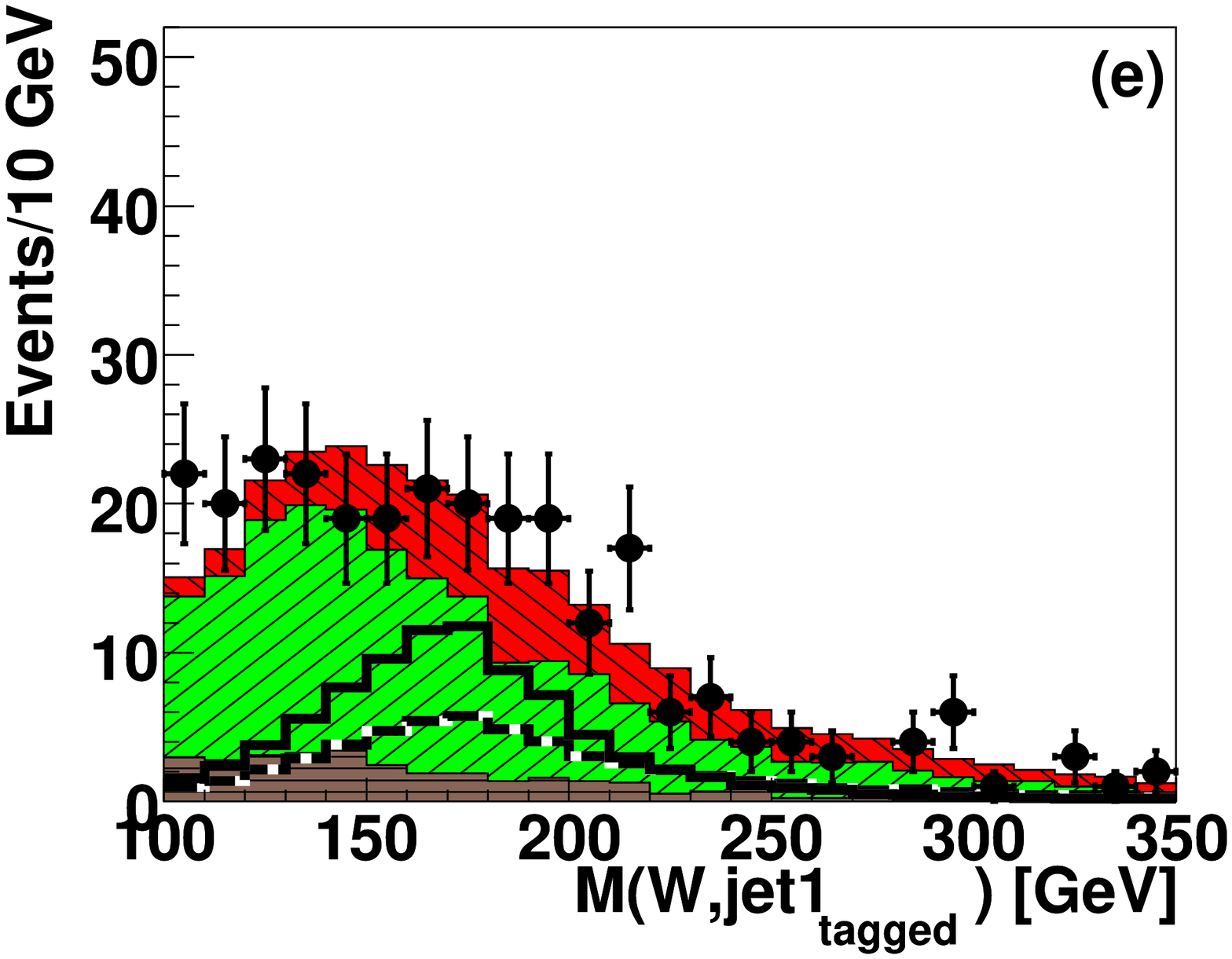}
\end{center}
\caption{Comparison of signal, backgrounds, and 
data after selection and requiring at least one $b$-tagged jet for five
discriminating event kinematic variables. Electron and muon channels are
combined.  Shown are (a) the transverse momentum of all jets except the leading
tagged jet, (b) the invariant mass of all jets except the leading tagged jet,
(c) the total energy of all jets except the leading tagged jet, (d) the
total transverse energy of all jets except the leading tagged jet, and (e) the
invariant mass of the top quark reconstructed from the reconstructed $W$~boson
and the leading tagged jet.  Signals are multiplied by ten.}
\label{fig:variables-event2}
\end{figure*}

\begin{figure*}[!h!tbp]  
\begin{center}
\includegraphics[width=0.32\textwidth]
{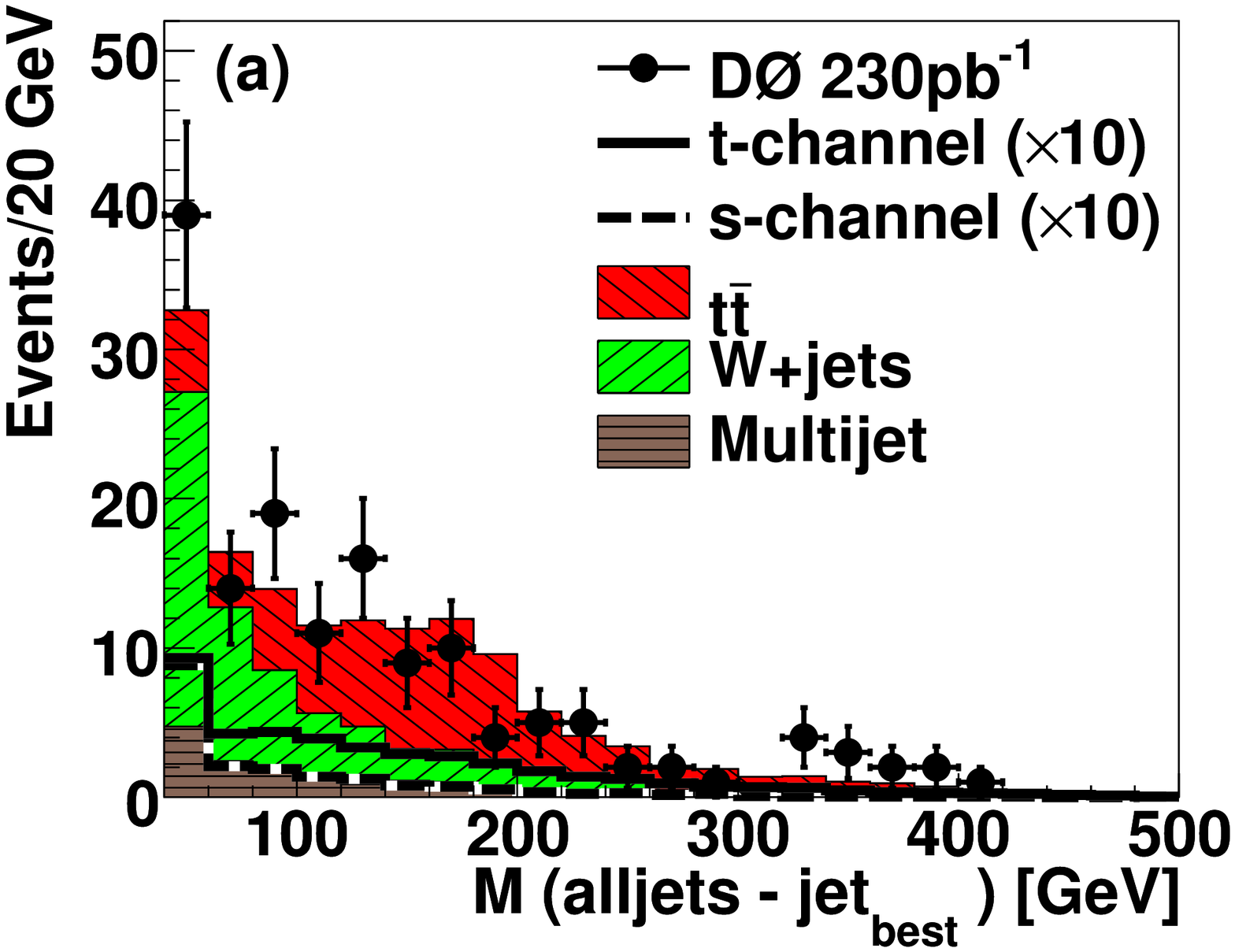}
\includegraphics[width=0.32\textwidth]
{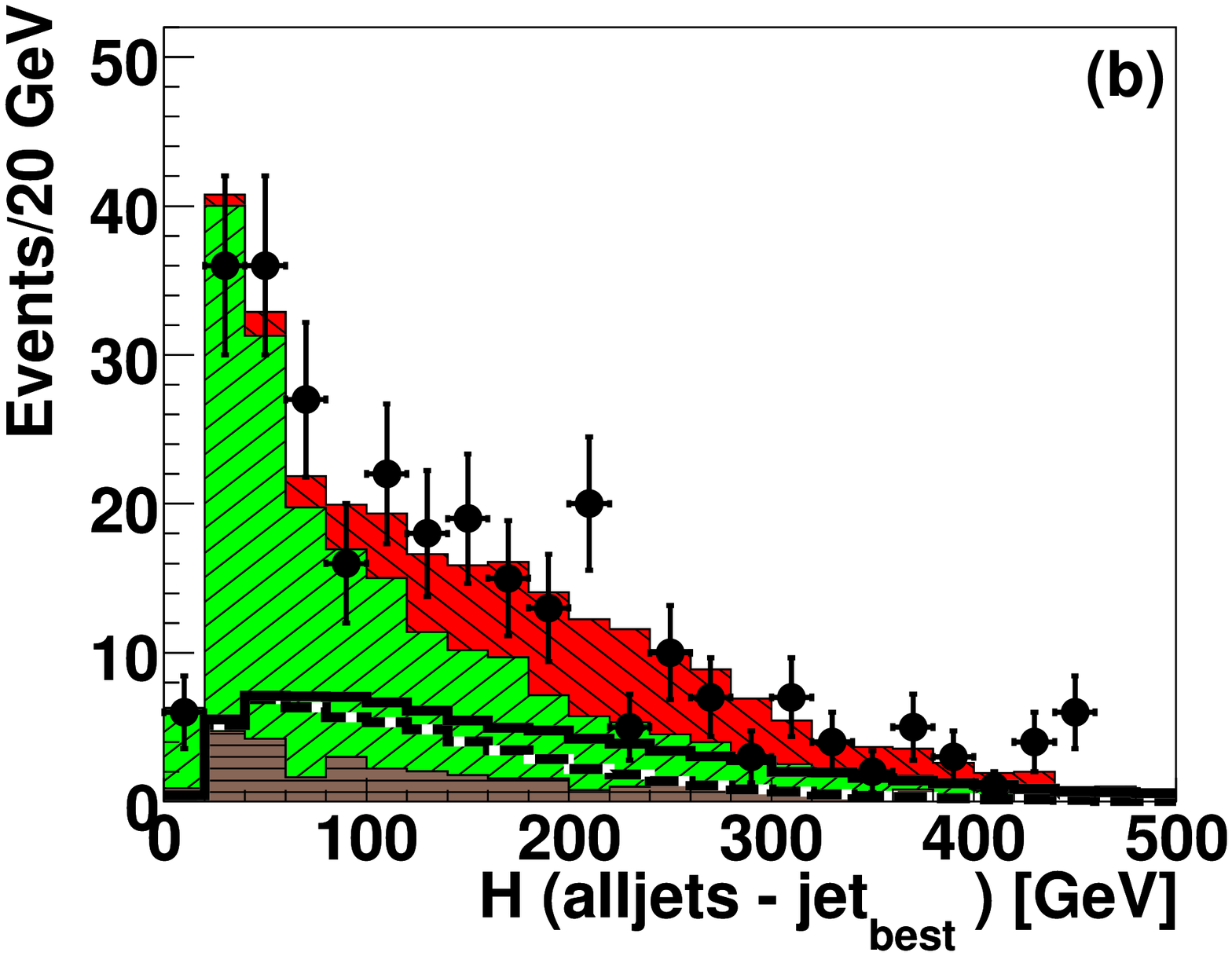}
\includegraphics[width=0.32\textwidth]
{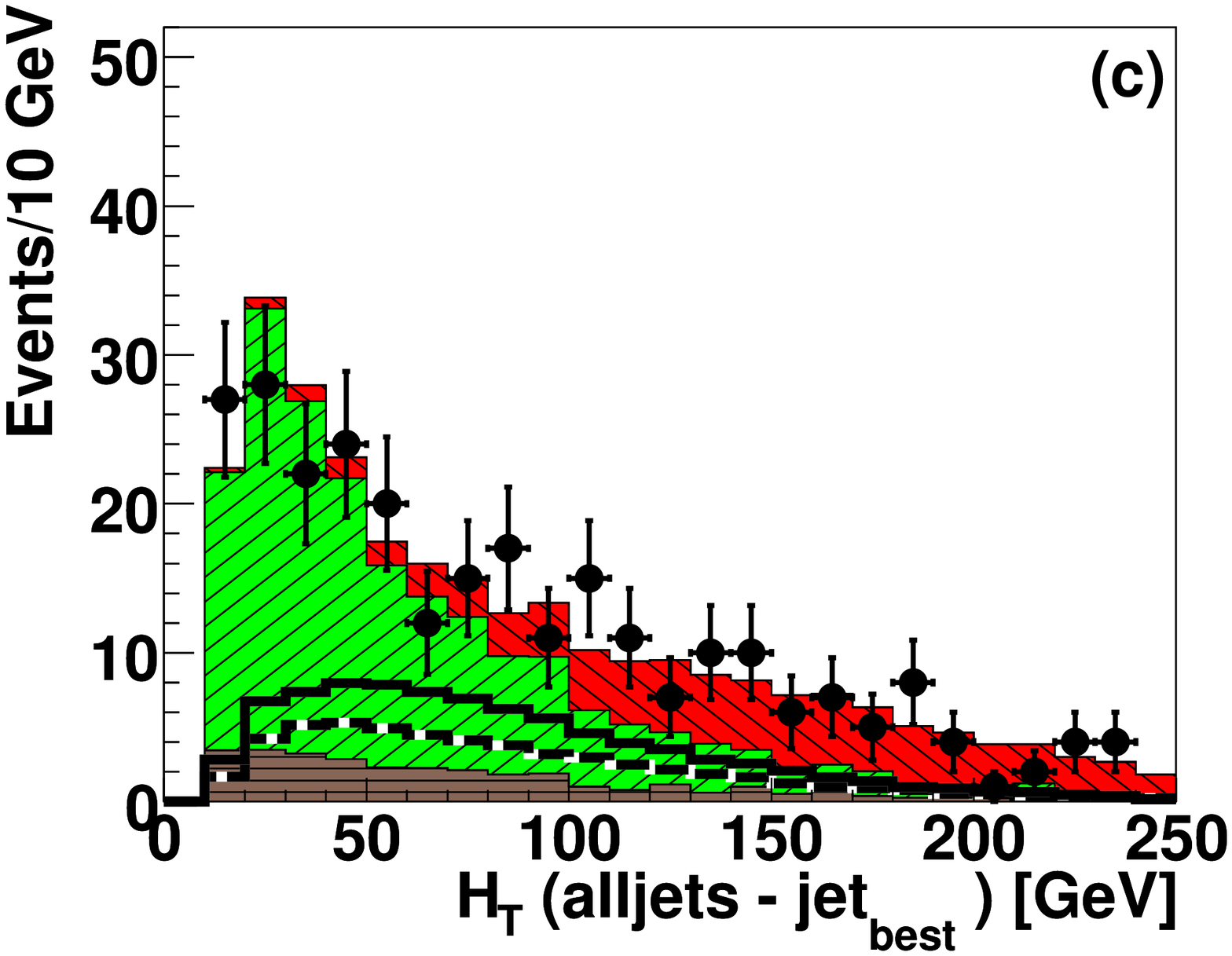}
\includegraphics[width=0.32\textwidth]
{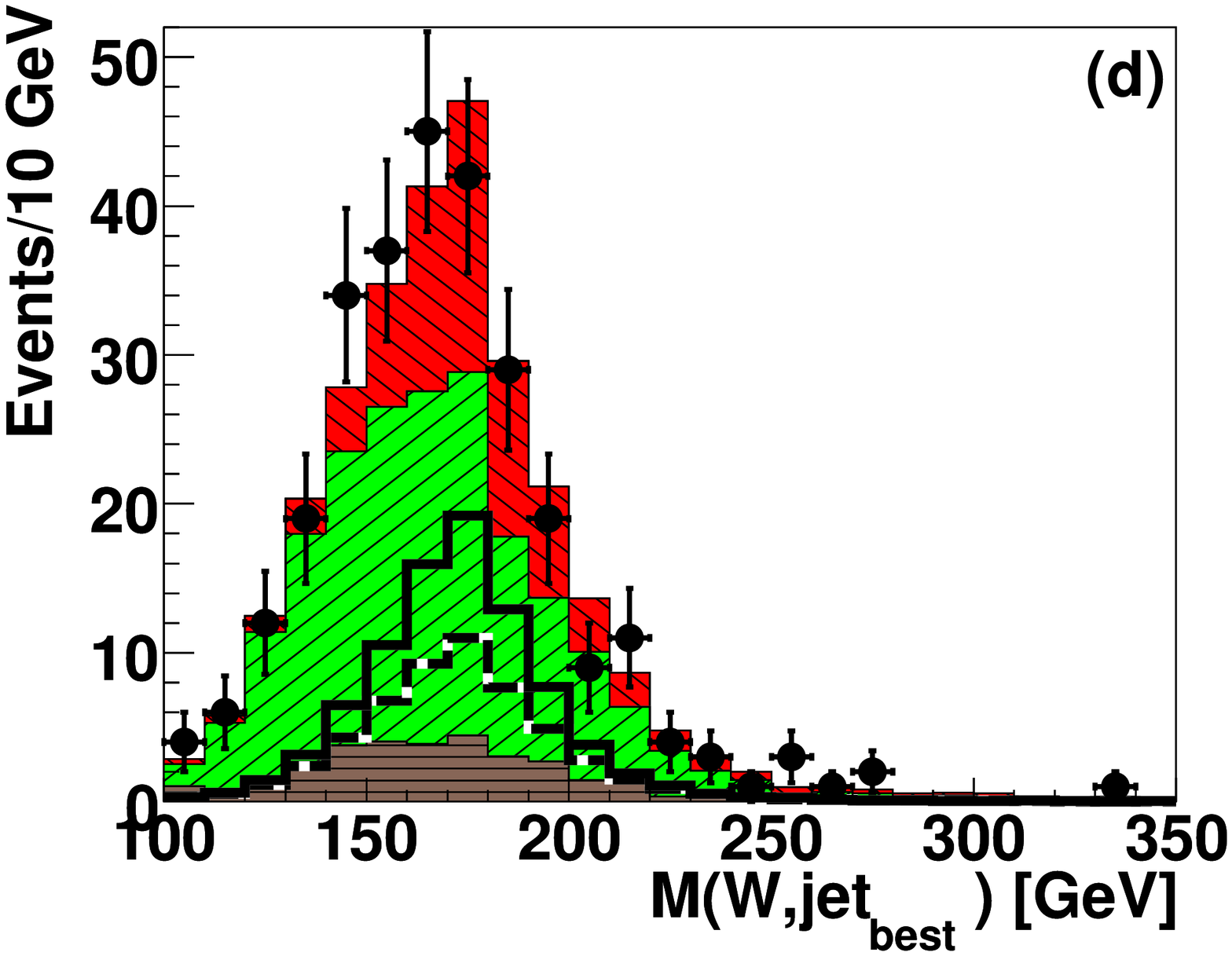}
\end{center}
\caption{Comparison of signal, backgrounds, and 
data after selection and requiring at least one $b$-tagged jet for four
discriminating event kinematic variables. Electron and muon channels are
combined.  Shown are (a) the invariant mass of all jets except the best jet, (b)
the total energy of all jets except the best jet, (c) the total transverse
energy of all jets except the best jet, and (d) the invariant mass of the top
quark reconstructed from the reconstructed $W$~boson and the best jet.  Signals
are multiplied by ten.}
\label{fig:variables-event3}
\end{figure*}

\begin{figure*}[!h!tbp]  
\begin{center}
\includegraphics[width=0.32\textwidth]
{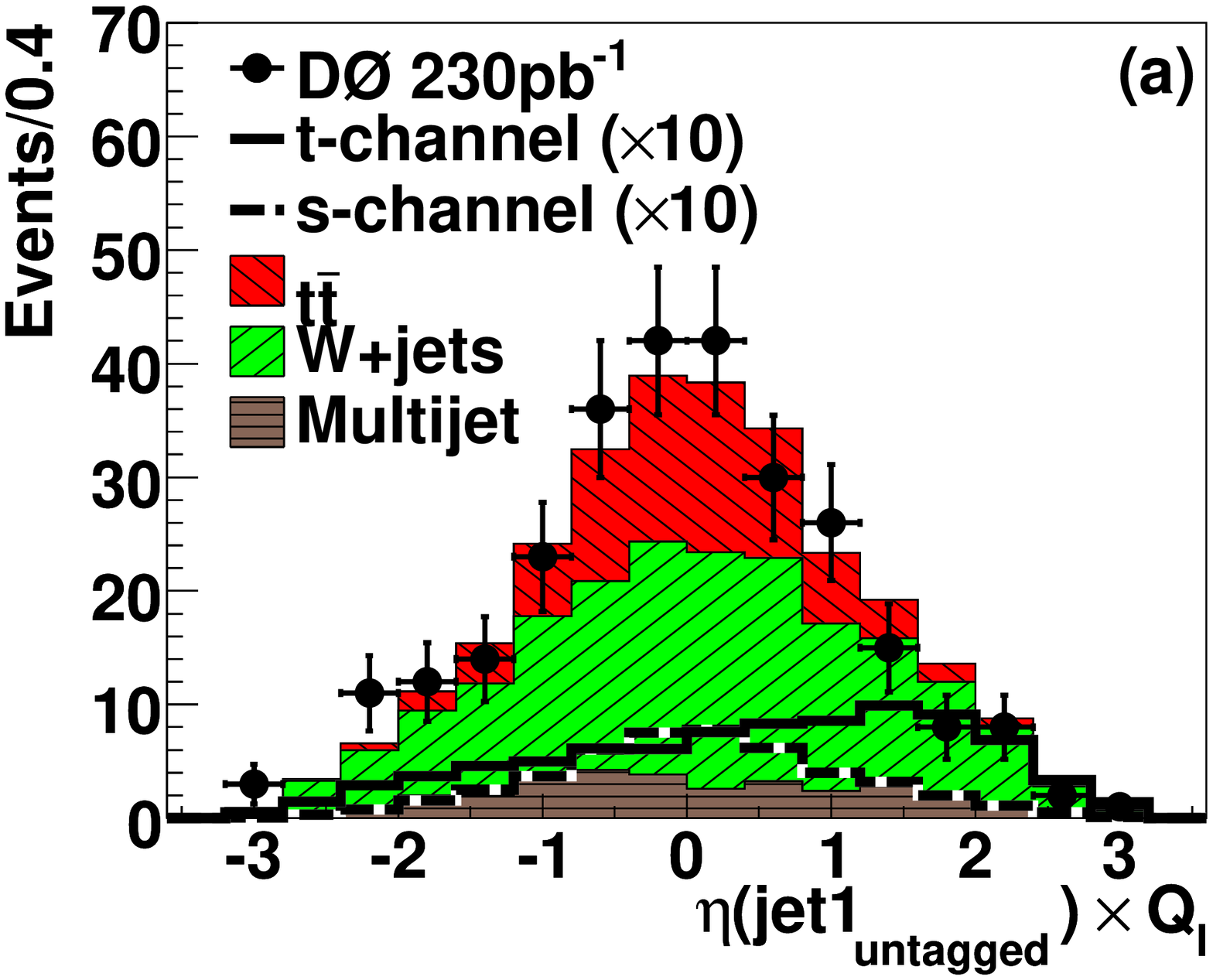}
\includegraphics[width=0.32\textwidth]
{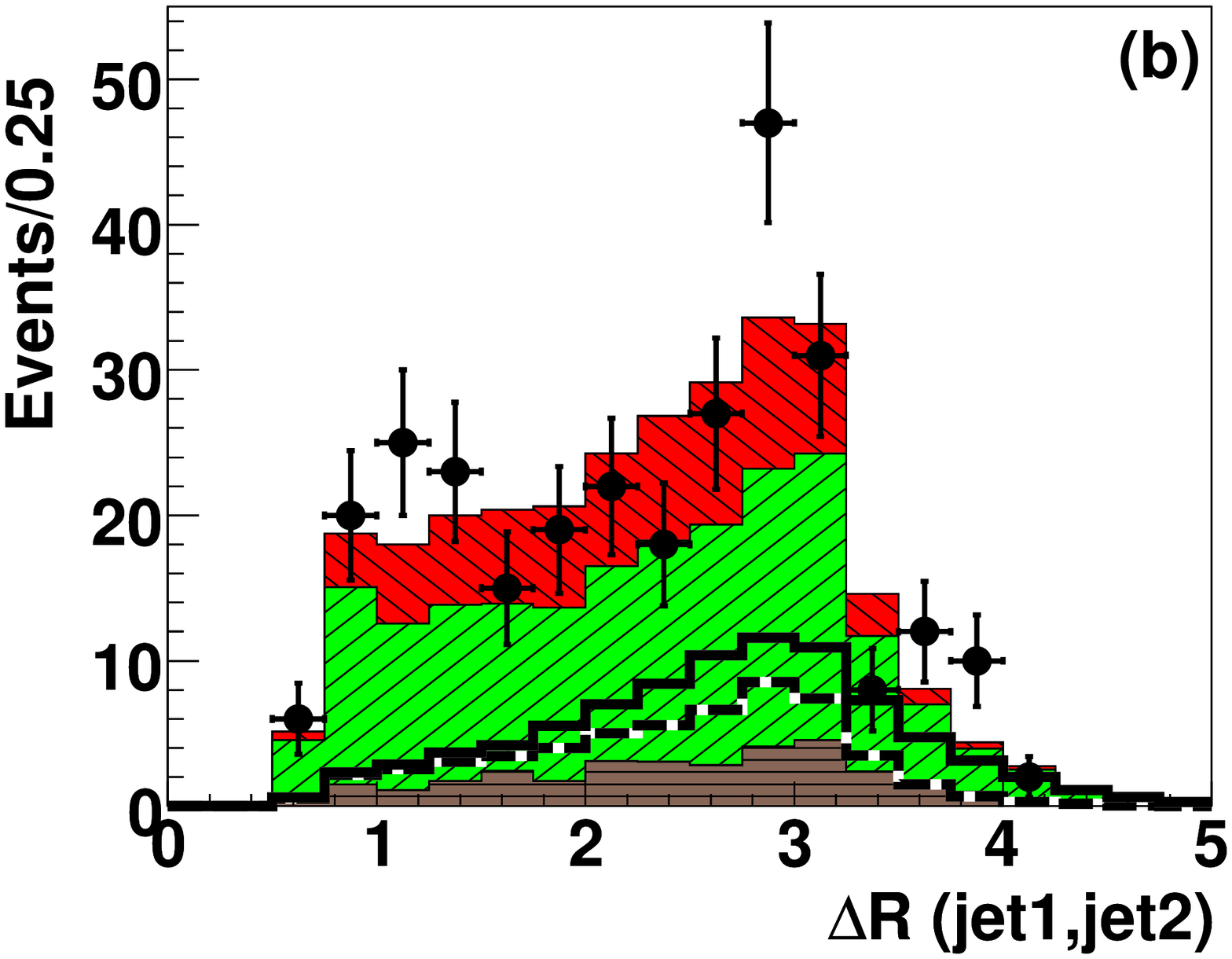}
\includegraphics[width=0.32\textwidth]
{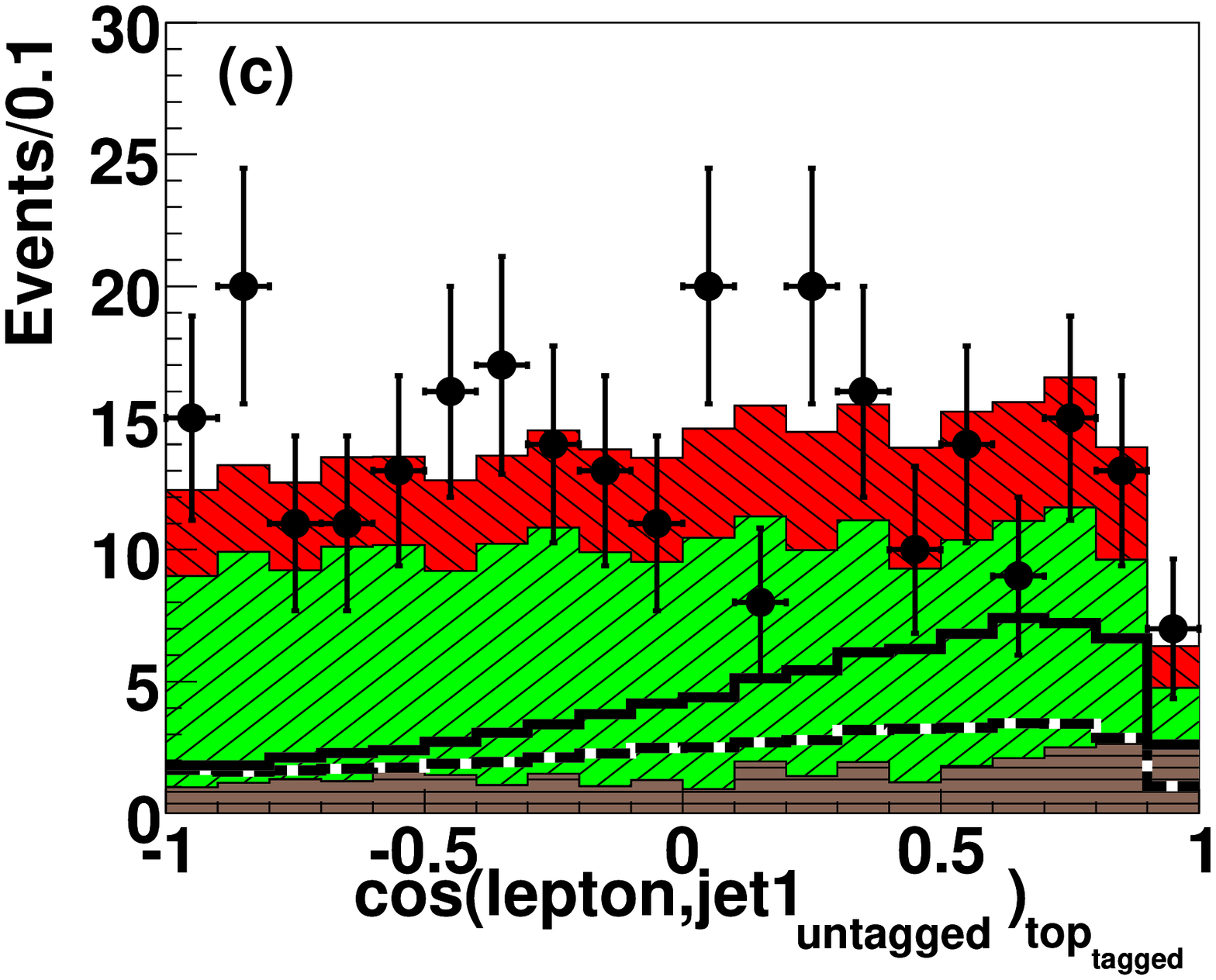}
\includegraphics[width=0.32\textwidth]
{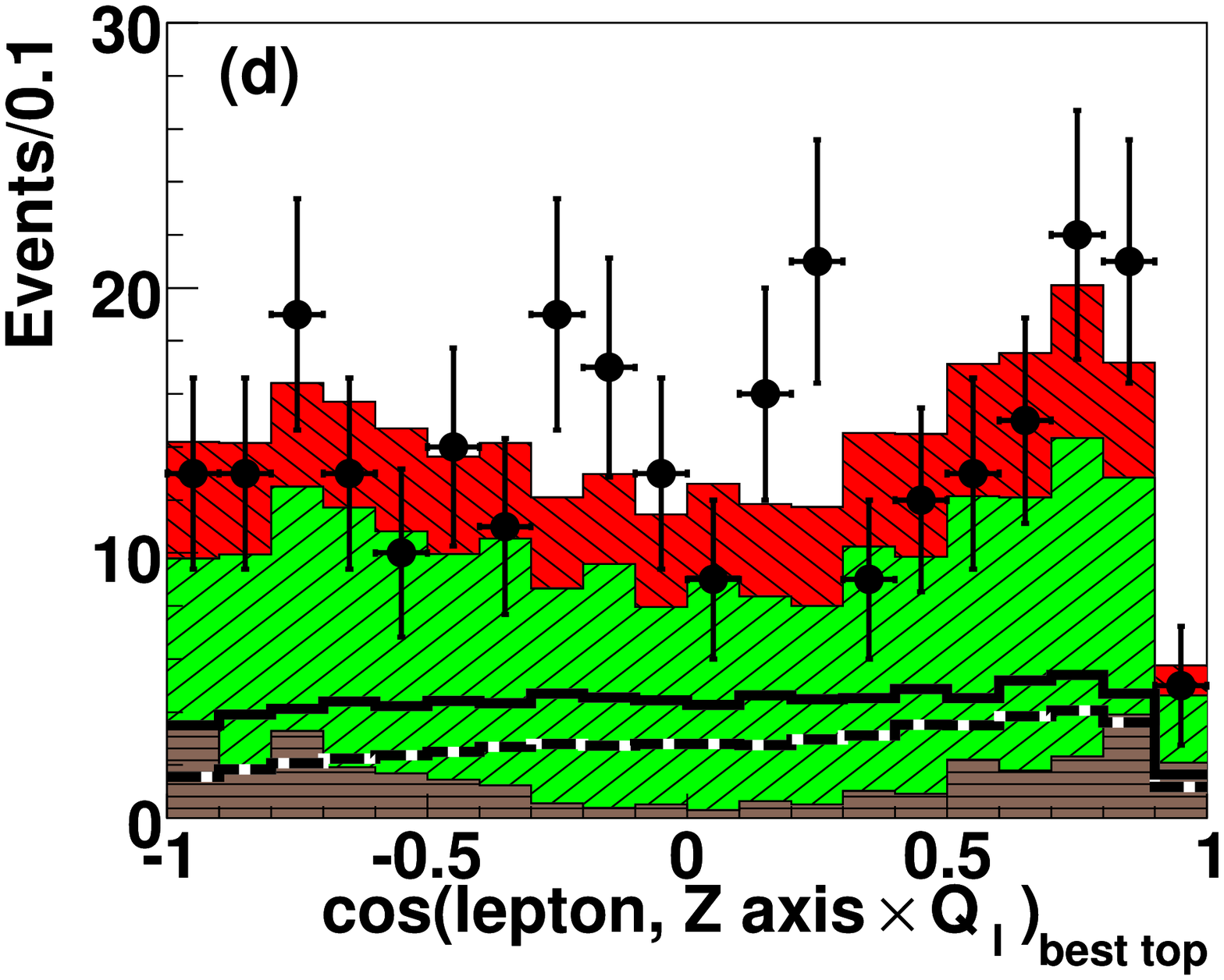}
\includegraphics[width=0.32\textwidth]
{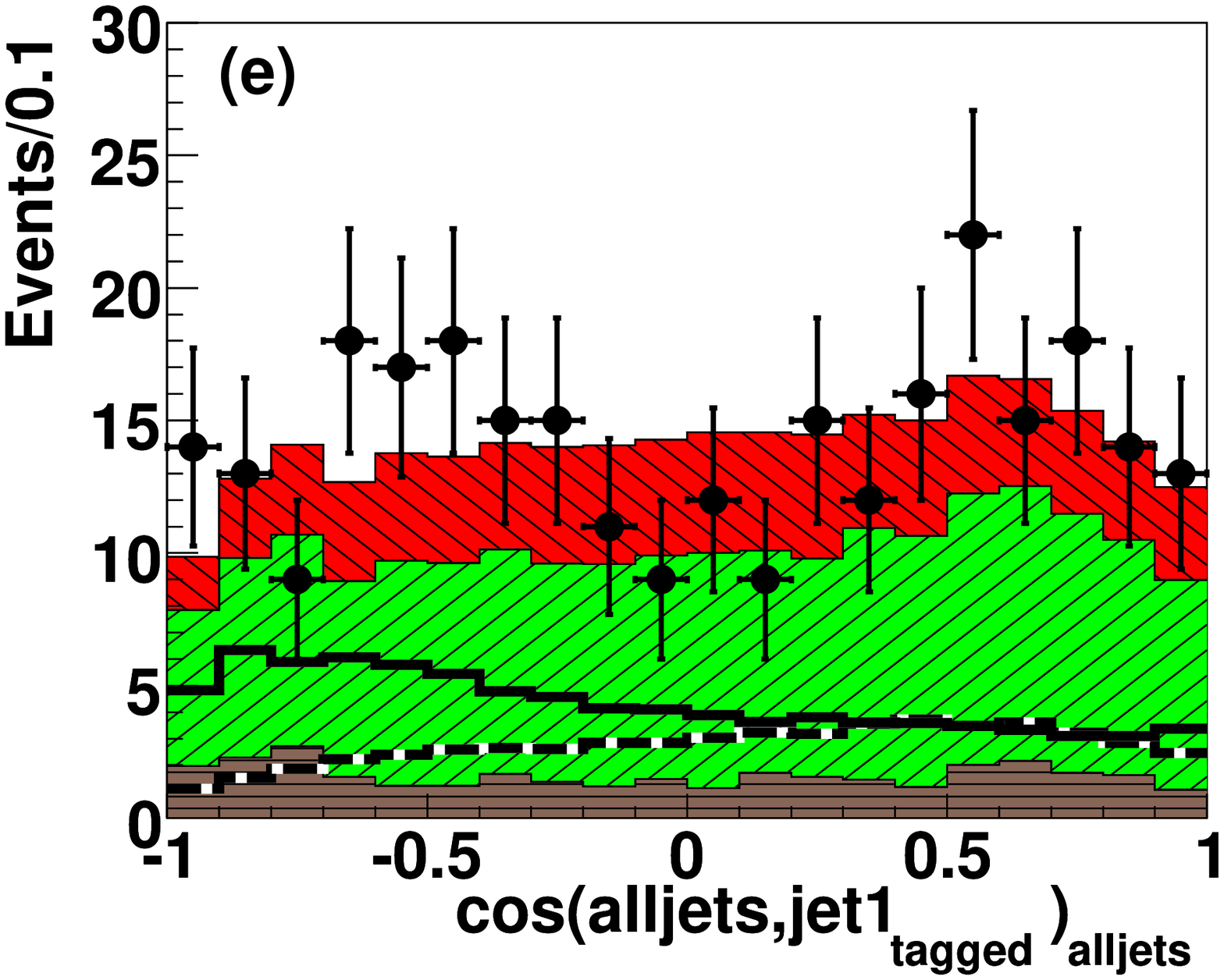}
\includegraphics[width=0.32\textwidth]
{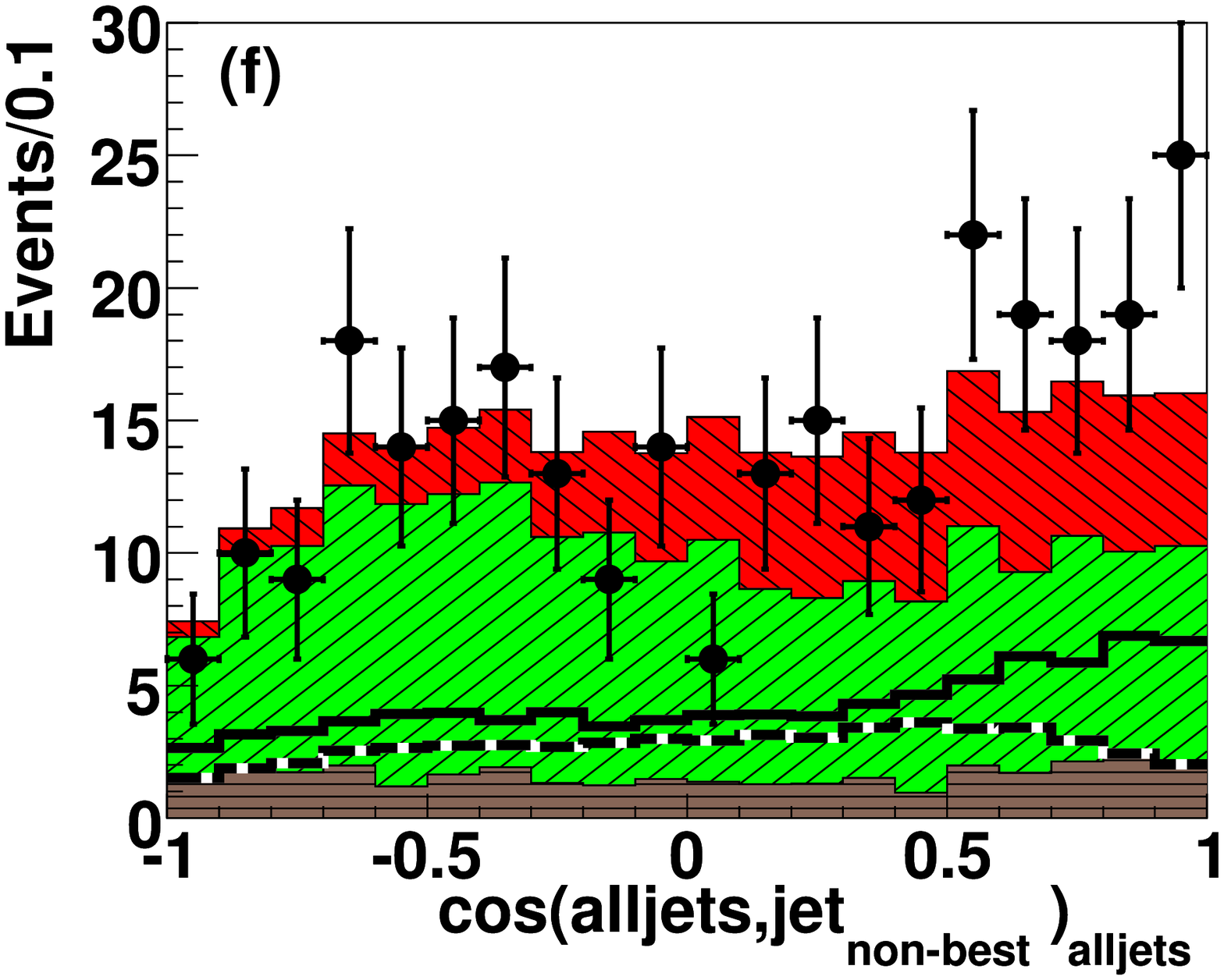}
\end{center}
\caption{Comparison of signal, backgrounds, and 
data after selection and requiring at least one $b$-tagged jet for six angular
correlation variables. Electron and muon channels are combined. Shown are (a)
the pseudorapidity of the leading untagged jet multiplied by the lepton charge;
(b) the angular separation between the leading two jets; (c) the top quark spin
correlation in the optimal basis for the $t$-channel; (d) the top quark spin
correlation in the optimal basis for the $s$-channel; (e) the cosine of the
angle between the leading tagged jet and alljets, in the alljets frame; and (f)
the cosine of the angle between the leading non-best jet and alljets, in the alljets
frame.  Signals are multiplied by ten.}
\label{fig:variables-angle}
\end{figure*}

Figures~\ref{fig:variables-object} to~\ref{fig:variables-angle} show all
discriminating variables used in this analysis, comparing the single top quark
signal distributions to those of the background sum and the data. Good agreement
between the data and the background model is seen in all cases.

\subsection{Cut-Based Analysis}
\label{cut_based_analysis}

This analysis takes the discriminating variables, chooses the best subsets, and
finds the optimal points to cut on them in order to improve the expected cross
section limits by increasing the signal to background ratio. 

Optimization of the cut positions is performed by using the signal Monte Carlo
events to seed the cut values scanned in the algorithm. The signal and
background pass rates are determined for each cut point, an expected limit on
the cross section is obtained from these, and the best result is used as the
operating point of the analysis.

The strategy is to look at the $s$- and $t$-channel processes separately to take
full advantage of the kinematical differences between the channels. For each
channel, there are four orthogonal analyses: two leptons ($e$, $\mu$) $\times$
number of tagged $b$~jets ($=1$, $\ge 2$).

The most critical part of this analysis is to find the combination of variables
and cuts that leads to the lowest expected cross section limit.
We first look at single-variable cuts
to determine which variables are most effective in each channel.
Once an ordered list of variables is found (ordered by their power to lower the
expected limit), sets of variables are formed starting with the best
variable and consecutively including one-by-one the rest of the
variables. For each set, the optimal cut position of each variable is
recalculated. Finally, the variable set that gives the lowest expected limit is
chosen. 
Table~\ref{rgs-vars-cuts} shows the optimal variable sets and
cuts found for each channel. Table~\ref{tab:rgs_yield} shows the numbers of
events and expected background and signal yields after these cuts
have been applied. 

\begin{table*}[!h!tbp]
\begin{center}
\begin{ruledtabular}
\caption[rgs-vars-cuts]{The best set of variables and cuts for
each analysis channel.}
\label{rgs-vars-cuts}
\begin{tabular}{lcccc}
 & \multicolumn{2}{c}{$s$-channel}
 & \multicolumn{2}{c}{$t$-channel}                          \\
Channel  &  Variables  &  Cuts  &  Variables  &  Cuts       \\
\hline
{\bf Electron}      &                           &                  &                                      &               \\
~~=1 tag      &  $p_T$(jet1$_{\rm tagged}$)        &  $> 27$~GeV      &  $H_{T}({\rm alljets})$              &  $> 71$~GeV   \\
              &  $M({\rm alljets-jet1_{tagged}})$  &  $< 70$~GeV      &  $M({\rm alljets})$                  &  $> 57$~GeV   \\
              &  $\sqrt{\hat{s}}$                  &  $> 196$~GeV     &  $\sqrt{\hat{s}}$                    &  $> 203$~GeV  \\
              &                                    &                  &  $\left|175-M(W,{\rm jet1_{tagged}})\right|$  &  $< 57$~GeV   \\
              &                                    &                  &  $p_T$(jet1$_{\rm tagged}$)                       &  $> 21$~GeV
\vspace{0.04in}                                                                                                     \\ 
~~$\ge$2 tags &  $p_T$(jet1$_{\rm tagged}$)        &  $> 42$~GeV      &  $p_T$(jet1$_{\rm tagged}$)                       &  $> 34$~GeV   \\
              &  $M({\rm alljets-jet1_{tagged}})$  &  $< 98$~GeV      &  $M({\rm alljets-jet1_{tagged}})$             &  $< 75$~GeV   \\
              &  $H({\rm alljets-jet_{best}})$     &  $< 304$~GeV     &  $H({\rm alljets-jet1_{tagged}})$             &  $< 504$~GeV  \\
              &  $H({\rm alljets-jet1_{tagged}})$  &  $< 304$~GeV     &  $H({\rm alljets-jet_{best}})$           &  $< 504$~GeV  \\
{\bf Muon}          &                                    &                  &                                      &               \\
~~=1 tag      &  $p_T$(jet1$_{\rm tagged}$)        &  $> 33$~GeV      &  $\left|175-M(W,{\rm jet1_{tagged}})\right|$  &  $< 60$~GeV   \\
              &  $M({\rm alljets-jet1_{tagged}})$  &  $< 74$~GeV      &  $\sqrt{\hat{s}}$                           &  $> 210$~GeV  \\
              &  $H({\rm alljets-jet_{best}})$     &  $< 504$~GeV     &  $M({\rm alljets})$                  &  $> 70$~GeV   \\
              &  $H({\rm alljets-jet1_{tagged}})$  &  $< 504$~GeV     &  $H_{T}({\rm alljets})$              &  $> 58$~GeV
\vspace{0.05in}                                                                                                     \\
~~$\ge$2 tags &  $p_T$(jet1$_{\rm tagged}$)            &  $> 33$~GeV      &  $\left|175-M(W,{\rm jet1_{tagged}})\right|$  &  $< 213$~GeV  \\
              &  $M({\rm alljets-jet1_{tagged}})$  &  $< 74$~GeV      &                                      &               \\  
              &  $H({\rm alljets-jet_{best}})$     &  $< 504$~GeV     &                                      &               \\
              &  $H({\rm alljets-jet1_{tagged}})$  &  $< 504$~GeV     &                                      &
\end{tabular}
\end{ruledtabular}
\end{center}
\end{table*}

\begin{table*}[!h!tbp]
\begin{center}
\begin{ruledtabular}
\caption[yields-rgs]{Event yields after the cut-based analysis selection.}
\label{tab:rgs_yield}
\begin{tabular}{lcccccccc}
 & \multicolumn{4}{c}{Electron Channel}
 & \multicolumn{4}{c}{Muon Channel} \\
 & \multicolumn{2}{c}{=1 Tag} & \multicolumn{2}{c}{$\ge$2 Tags}
 & \multicolumn{2}{c}{=1 Tag} & \multicolumn{2}{c}{$\ge$2 Tags} \\
 & $s$-channel & $t$-channel & $s$-channel  & $t$-channel
 & $s$-channel & $t$-channel & $s$-channel  & $t$-channel \\
\hline
{\bf Signals}    &        &        &        &        &        &        &        &       \\
~~$tb$      &   1.7  &  ---   &  0.45  &  0.12  &  1.9   &  ---   &  0.43  &  ---  \\
~~$tqb$     &   ---  &  3.4   &  ---   &  0.23  &  ---   &  3.1   &  ---   & 0.23 \\
{\bf Backgrounds} &        &        &        &        &        &        &        &       \\
~~$tb$      &   ---  &  1.6   &  ---   &  0.12  &  ---   &  1.4   &  ---   &  0.12 \\
~~$tqb$     &   2.5  &  ---   &  0.14  &  ---   &  2.8   &  ---   &  0.01 &  ---   \\
~~${\ttbar}{\rar}\ell$+jets
            &   3.8  & 18.5   &  1.14  &  4.61  &  9.7   & 17.8   &  0.61 &  5.20  \\
~~${\ttbar}{\rar}\ell\ell$
            &   4.3  &  4.1   &  1.15  &  0.62  &  5.8   &  4.3   &  1.12 &  0.73  \\
~~$Wb\bar{b}$     &   8.4  &  6.3   &  1.72  &  0.54  & 10.2   &  5.2   &  1.85 &  0.52  \\
~~$Wjj$     &  33.4  & 28.9   &  0.74  &  0.60  & 43.6   & 28.8   &  0.95 &  0.64  \\
~~$WW$      &   0.4  &  0.3   &  0.01  &  0.00  &  0.6   &  0.3   &  0.00 &  0.00  \\
~~$WZ$      &   0.4  &  0.3   &  0.09  &  0.01  &  0.4   &  0.2   &  0.09 &  0.01  \\
~~Multijet  &   6.8  &  6.9   &  0.20  &  0.14  &  10.1  &  9.9   &  0.11 &  0.01 \\
{\bf Summed signals}          &  4.3 &  4.9 & 0.59 & 0.35 &  4.7 &  4.5 & 0.53 & 0.35   \\
{\bf Summed backgrounds}       & 57.5 & 65.3 & 5.04 & 6.54 & 80.3 & 66.7 & 4.71 & 7.20   \\
{\bf Summed backgrounds+$tqb$} & 60.0 & 68.6 & 5.18 & 6.76 & 83.1 & 69.8 & 4.81 & 7.43   \\
{\bf Summed backgrounds+$tb$}  & 59.2 & 66.8 & 5.49 & 6.65 & 82.2 & 68.1 & 5.14 & 7.32 \\
{\bf Data}                    &  60  &  73  &  4   &  9   &  78  &  58  &  10  &  8
\end{tabular}
\end{ruledtabular}
\end{center}
\end{table*}

A summary of the yield estimates for the signal and backgrounds and the numbers
of observed events in data after the cut-based selection, including the systematic
uncertainties as described in Sec.~\ref{systematics}, is shown in
Table~\ref{tab:cut_yield_sys}.
\begin{table}[h!tbp] 
\begin{center}
\caption{Estimates of backgrounds and signal yields and the number
of observed events in data after the cut-based selection for the
electron and muon, =1 tag and $\geq$2 tags analyses combined.}
\label{tab:cut_yield_sys}
\begin{ruledtabular}
\begin{tabular}{lr@{$\,\pm\!\!\!\!$}lr@{$\,\pm\!\!\!\!$}l} 
Source           & \multicolumn{2}{c}{$s$-channel search} &
\multicolumn{2}{c}{$t$-channel search} \\ \hline
$tb$             &   4.5 & 1.0  &   3.2 & 0.8  \\ 
$tqb$            &   5.5 & 1.2  &   7.0 & 1.6  \\ 
$W$+jets         &  27.6 & 7.6  &  55.9 & 12.3 \\
$\ttbar$         & 102.9 & 13.7 &  72.6 & 9.7  \\
Multijet         &  17.2 & 2.0  &  17.0 & 2.0  \\ \hline
Total background & ~~~153.1 & 24.5	& ~~~148.7	& 24.8 \\
Observed events  & \multicolumn{2}{c}{152}  &  \multicolumn{2}{c}{148}     \\
\end{tabular}
\end{ruledtabular}
\end{center}
\end{table}

Ths $s$- and $t$-channel combined signal to background ratio improves from
around 1/20 after the basic selection (Table~\ref{tab:yield_sys}) to around 1/14
after these cuts have been applied. It is clear that more sophisticated
separation techniques are needed to isolate the signal better from the large
backgrounds.

\subsection{Neural Network Analysis}
\label{nn-analysis}
A neural network is a multivariate statistical technique for
separating signals from backgrounds.
We use the {\sc mlpfit}~\cite{mlpfit} package to construct and
implement the networks.
In order for a neural network to approach the
maximal signal-background separation, some optimization is
required. This occurs in three steps:
1) judicious choice of signal and background pairs,
2) selection of input variables, and
3) optimization of training parameters.

\subsubsection{Choice of Signal-Background Pairs}
\label{sec:nn-sig-back-pairs}
We have chosen to create networks trained on single
top quark signals against the two dominant backgrounds: $W$+jets
and {\ttbar}.  For $W$+jets, we train using
a $Wb\bar{b}$ Monte Carlo sample as this process best
represents all $W$+jets processes.  For {\ttbar},
we train on ${\ttbar}{\rar}\ell$+jets which is the
dominant background as opposed to the dilepton background
which is small.

\subsubsection{Choice of Input Variables}
\label{sec:nn-input-vars}
We start from a set of discriminating variables that each
show some signal-background separation as discussed
in Sec.~\ref{discrim_vars}.  Based on this, we
optimize the input variables for
each network by training with different combinations of variables 
and choosing the combination that produces the
minimum testing error, which corresponds to the
best signal-background separation.

We use the same variables
for the electron and muon channel. However, owing to different resolutions
and pseudorapidity ranges, we train the networks separately for the
two.

\subsubsection{Neural Network Training}
Each network is composed of three
layers of nodes: input, hidden, and output.  Testing and training event 
sets are created from simulated signal and background samples. 
We divide the input samples such that 60\% of the events are used for
training and the remaining 40\% for testing. Training is effected with
weighted events and the logarithm of all nonangular variables. 
We use a technique called early stopping~\cite{early_stopping} to 
determine the maximum number of epochs for training which prevents 
over-training.

Each network is further tuned by varying the number of hidden nodes
between 10 and 30 
and then selecting the number of hidden nodes that 
returns the smallest testing error.

\subsubsection{Neural Network Results}
The above procedure produces eight unique networks:
two signals ($s$-channel, $t$-channel) $\times$ two
backgrounds ($Wb\bar{b}$, ${\ttbar}{\rar}\ell$+jets) $\times$ two lepton
flavors ($e$, $\mu$). 

\begin{figure*}[!h!tbp]
\includegraphics[width=0.40\textwidth]
{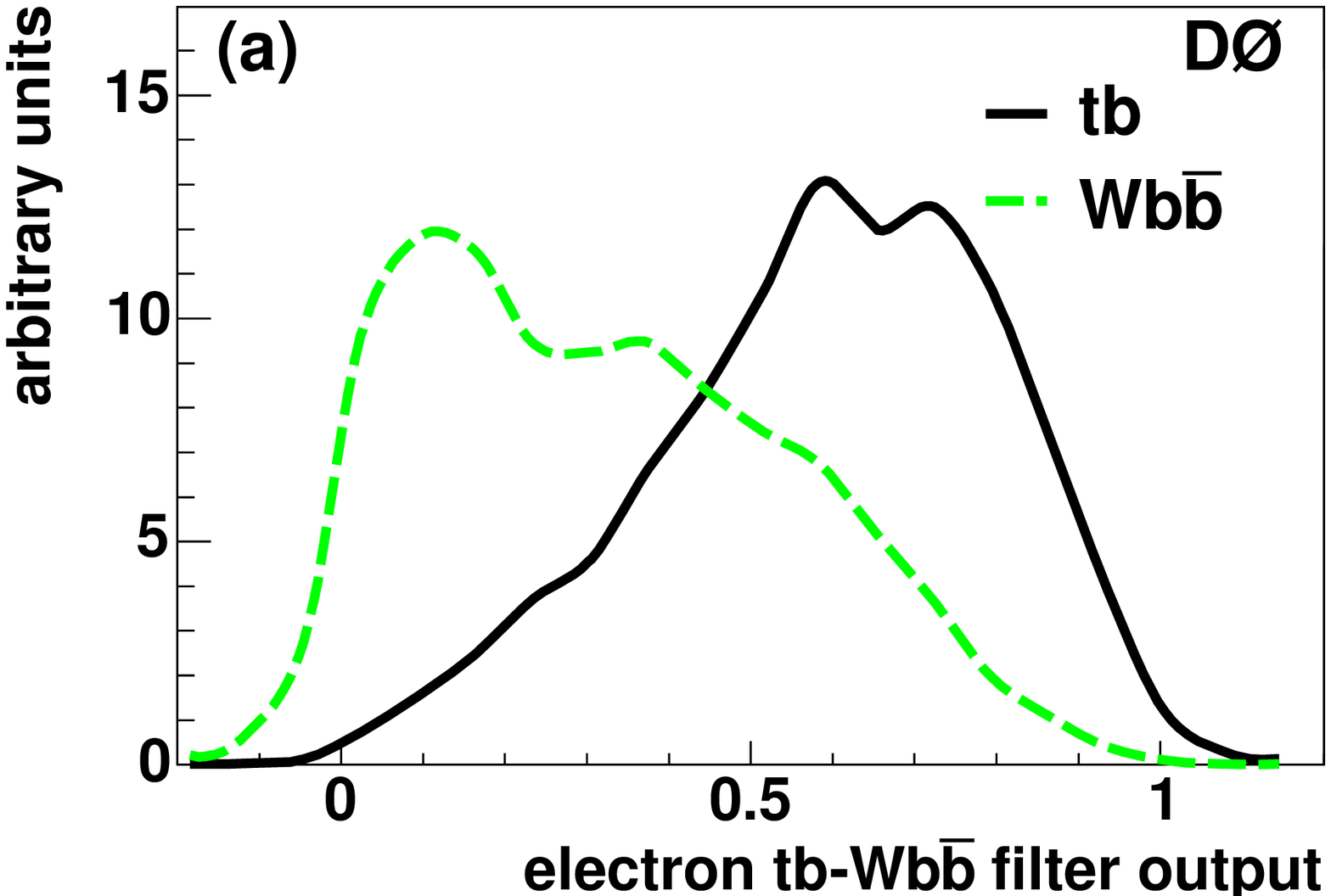}
\includegraphics[width=0.40\textwidth]
{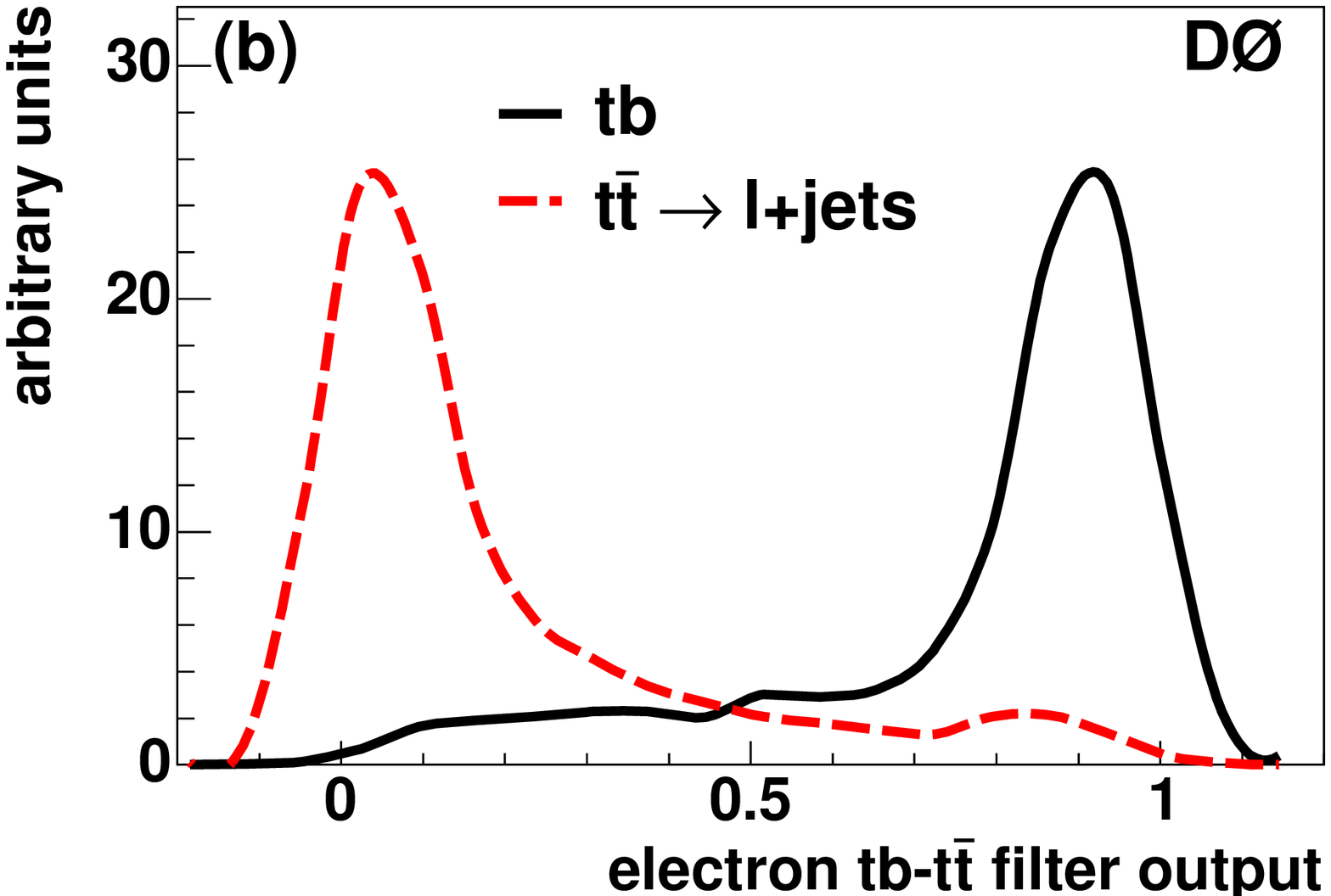}
\includegraphics[width=0.40\textwidth]
{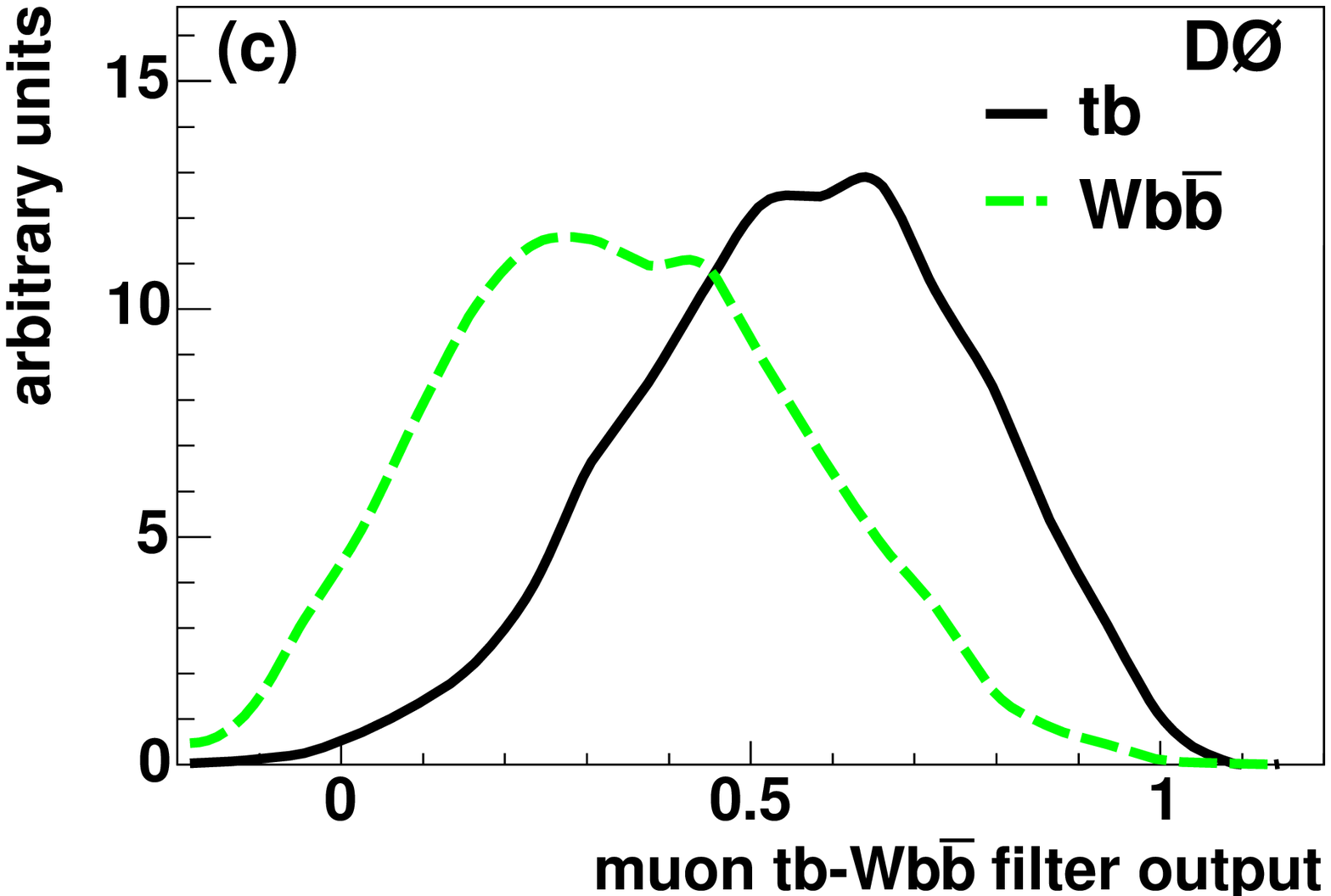}
\includegraphics[width=0.40\textwidth]
{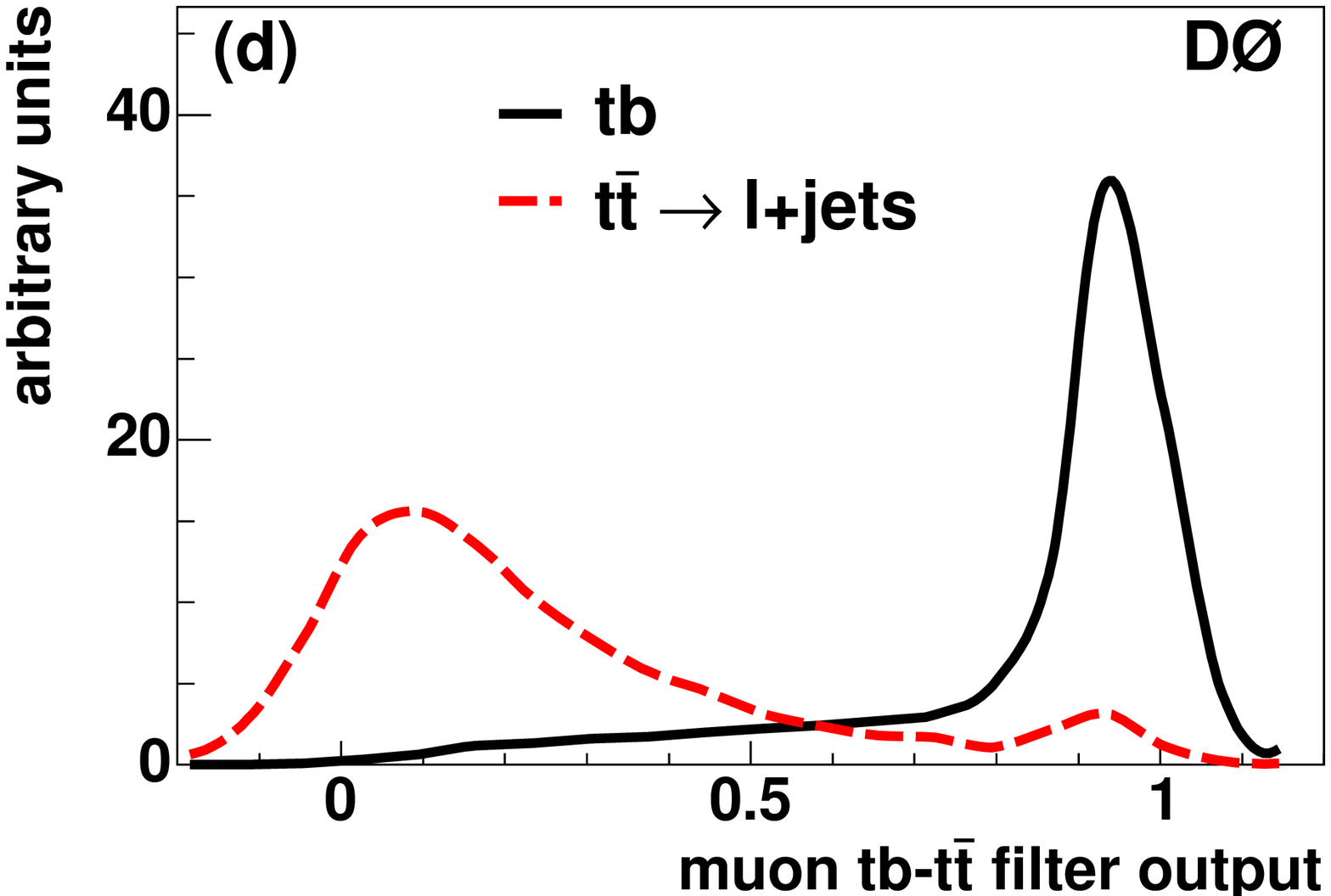}
\caption[nn-1d-output-tb]{Neural network outputs in the $s$-channel.
This figure shows the signal-background separation for
(a) the filter for $Wb\bar{b}$ in the electron channel,
(b) the filter for ${\ttbar}{\rar}\ell$+jets in the electron channel,
(c) the filter for $Wb\bar{b}$ in the muon channel, and
(d) the filter for ${\ttbar}{\rar}\ell$+jets in the muon channel
where the background is the dashed-lined and the top quark
signal is the solid line. All the curves are 
normalized to have equal area, so that the separation between signal 
and background can be best seen.}
\label{nn-1d-output-training_tb}
\end{figure*}

\begin{figure*}[!h!tbp]
\includegraphics[width=0.40\textwidth]
{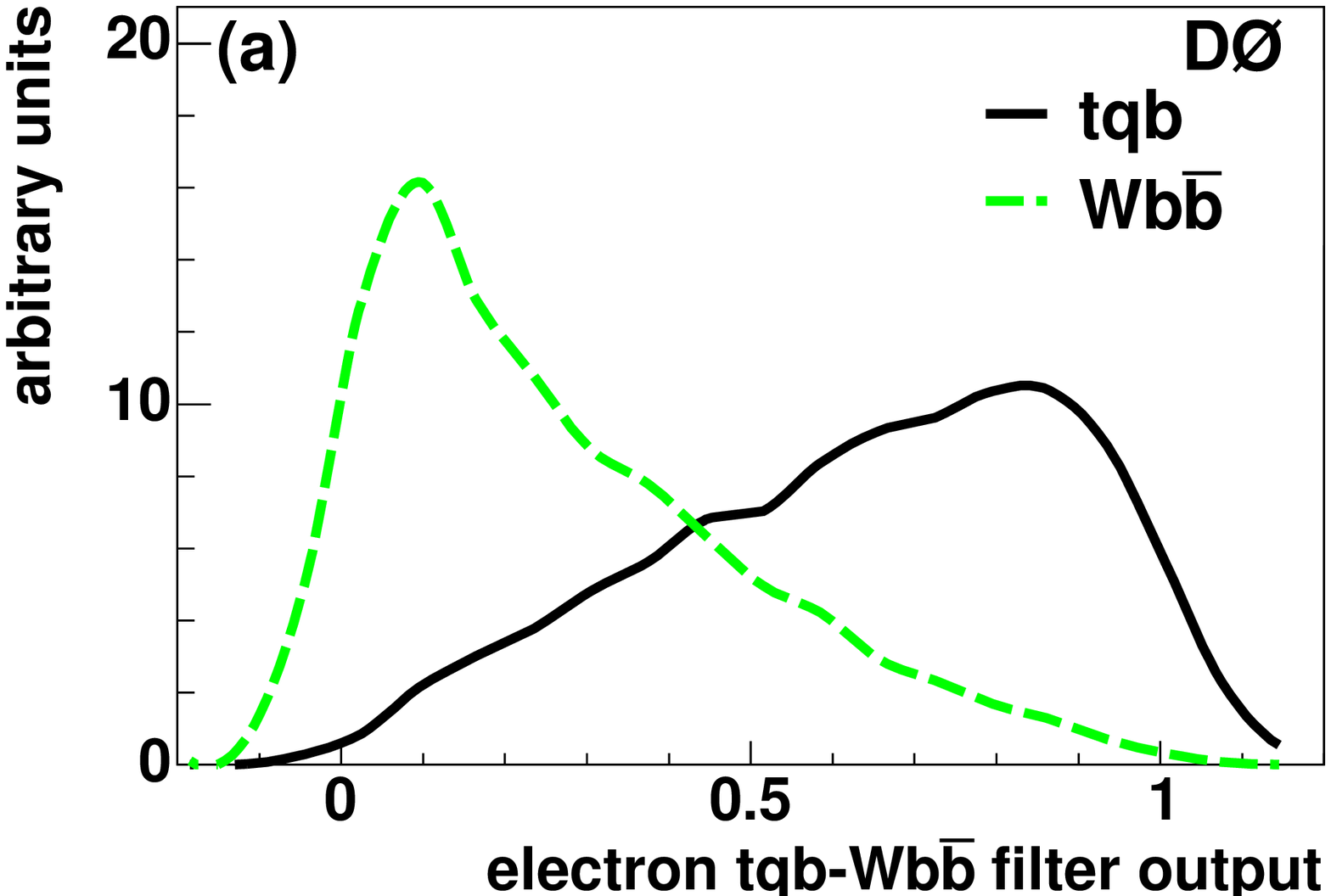}
\includegraphics[width=0.40\textwidth]
{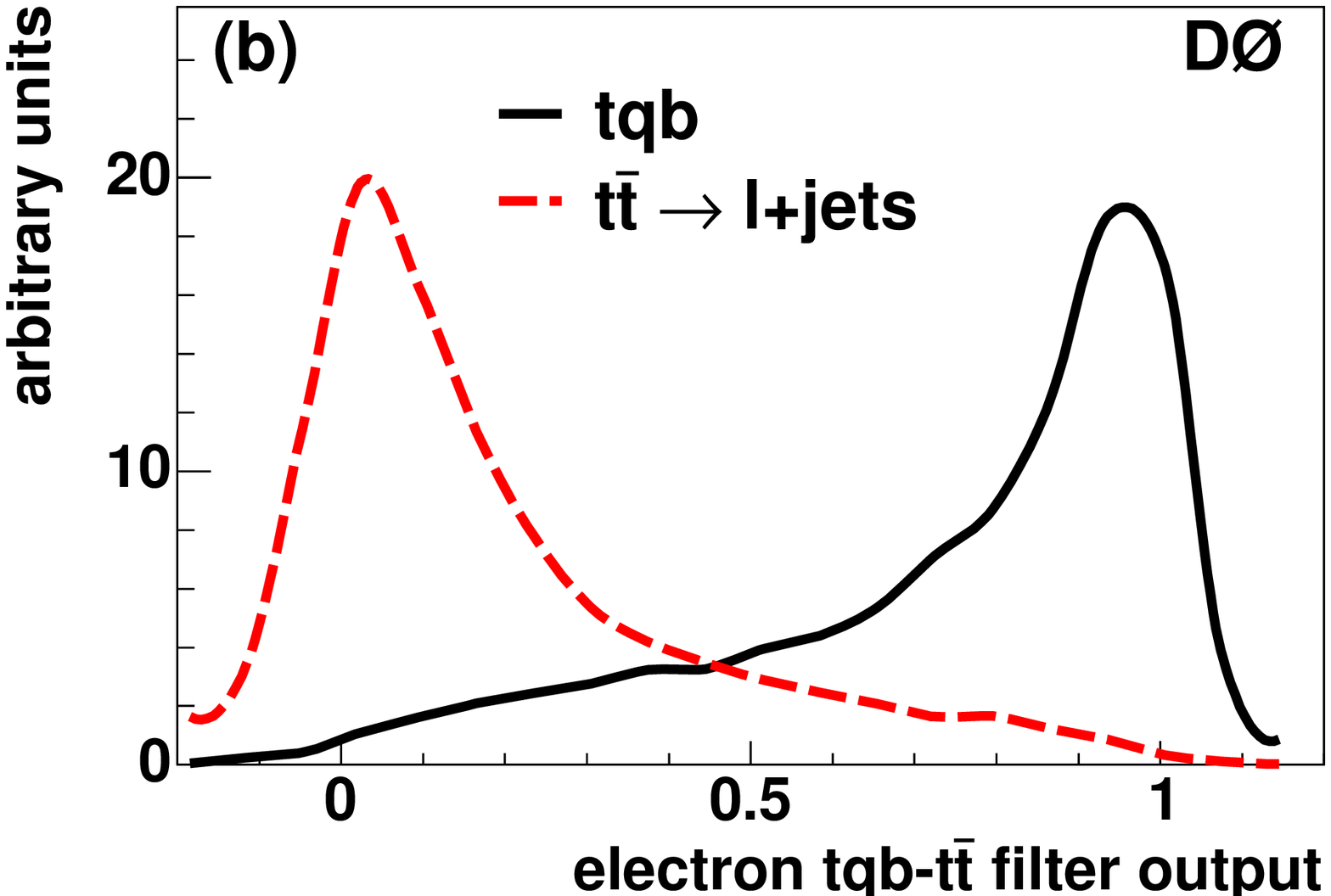}
\includegraphics[width=0.40\textwidth]
{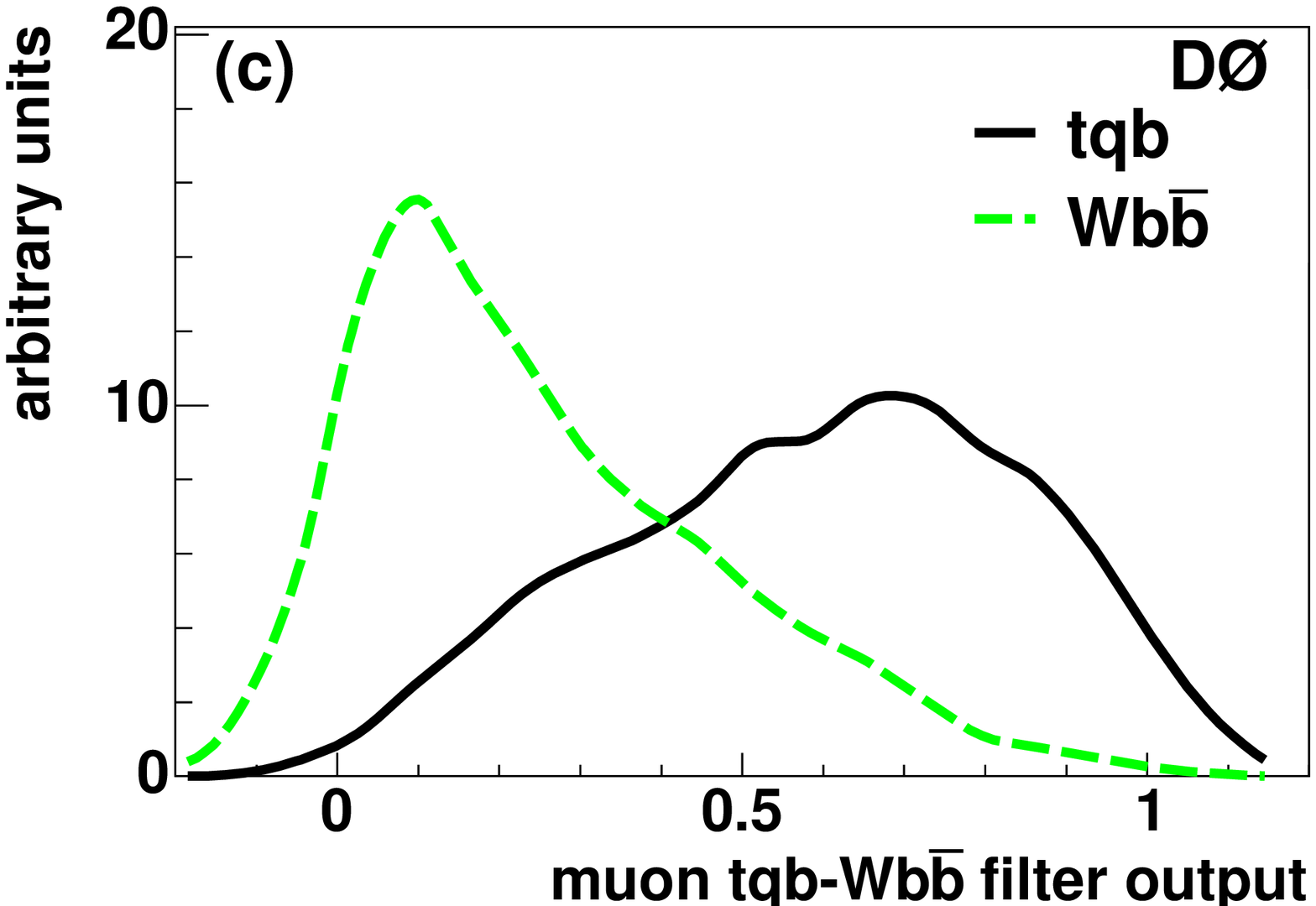}
\includegraphics[width=0.40\textwidth]
{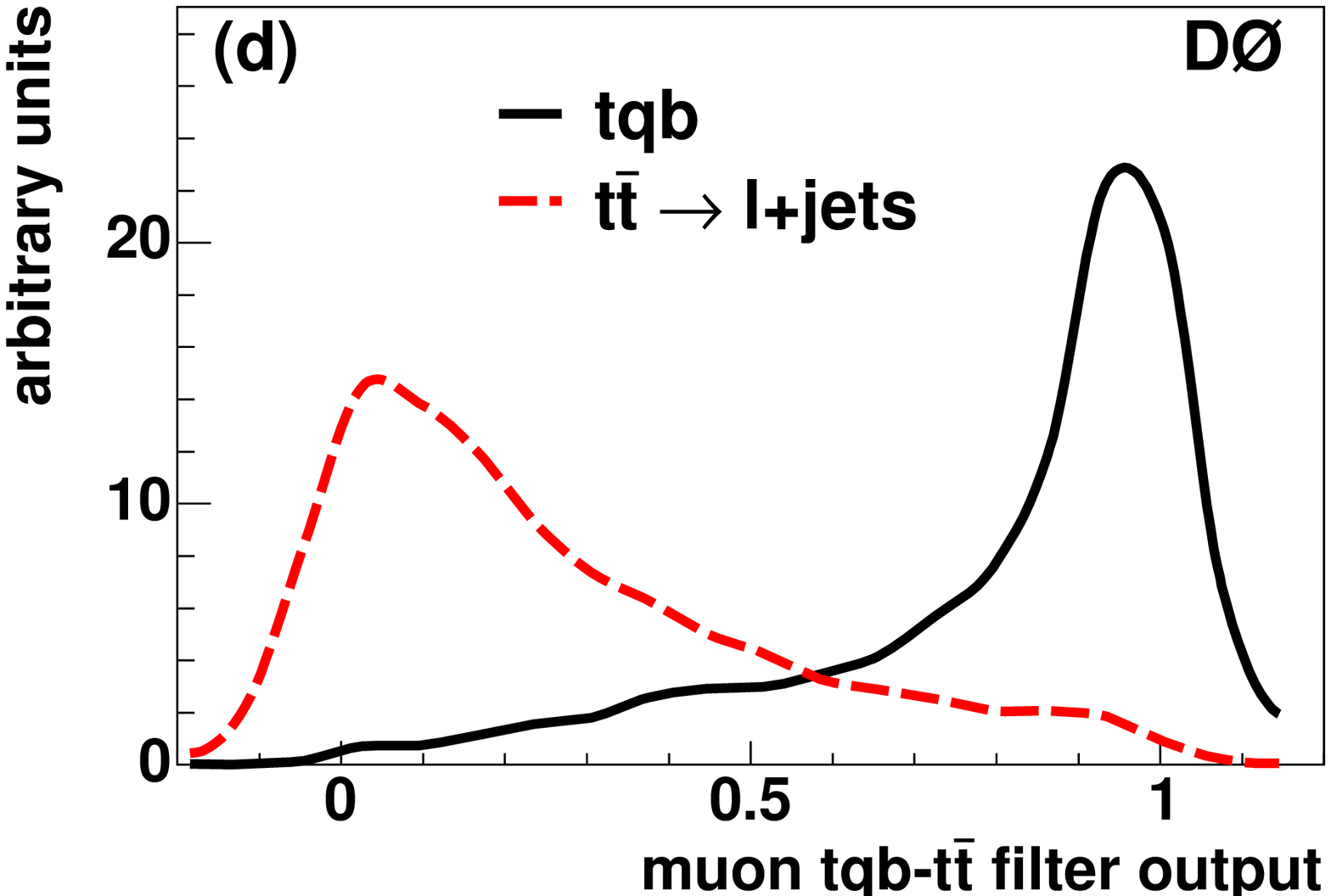}
\caption[nn-1d-output-tb]{Neural network outputs in the $t$-channel.
This figure shows the signal-background separation for
(a) the filter for $Wb\bar{b}$ in the electron channel,
(b) the filter for ${\ttbar}{\rar}\ell$+jets in the electron channel,
(c) the filter for $Wb\bar{b}$ in the muon channel, and
(d) the filter for ${\ttbar}{\rar}\ell$+jets in the muon channel
where the background is the dashed-lined and the top quark
signal is the solid line. All the curves are 
normalized to have equal area, so that the separation between signal 
and background can be best seen.}
\label{nn-1d-output-training_tqb}
\end{figure*}

Figures~\ref{nn-1d-output-training_tb} and~\ref{nn-1d-output-training_tqb}
show the output variable distributions
from the networks in the $s$-channel and $t$-channel searches for electrons 
and muons. From the figures, it can be seen that these
networks are highly efficient at separating the single top quark signal from the
${\ttbar}{\rar}\ell$+jets background. Studies have shown that these networks are
not as effective for the ${\ttbar}$ dilepton background, which is fortunately small.
The $s$-channel and $t$-channel networks are less efficient at separating
the single top quark signal from the $Wb\bar{b}$
background as compared to ${\ttbar}{\rar}\ell$+jets. In
addition, we find these networks are equally effective 
in separating the $Wjj$ and the misidentified lepton background
as compared to the $Wb\bar{b}$ background.
It should be noted that the output variable from {\sc mlpfit} networks is
not restricted to lie between zero and one.

\begin{figure*}[!h!tbp]
\includegraphics[width=0.45\textwidth]
{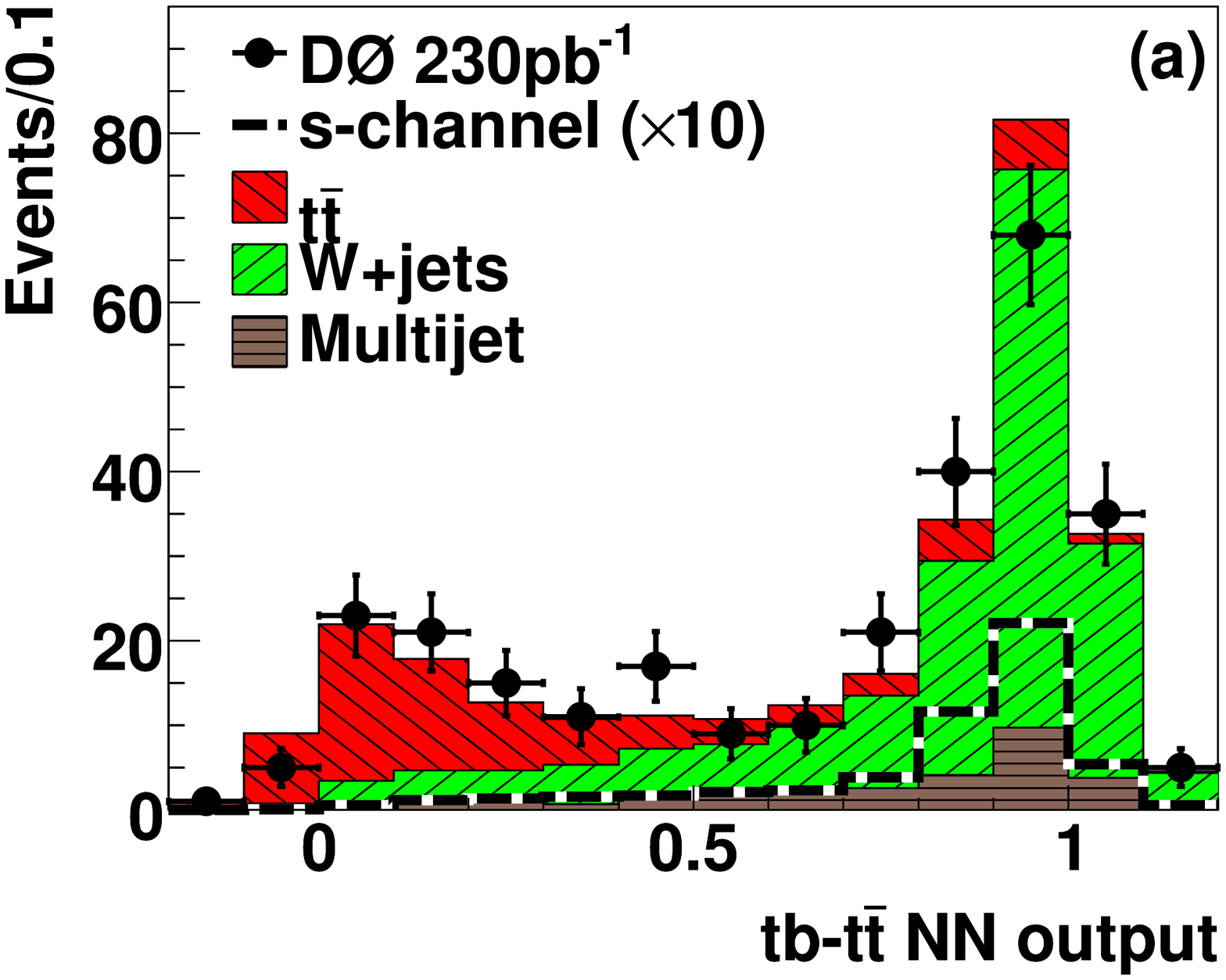}
\includegraphics[width=0.45\textwidth]
{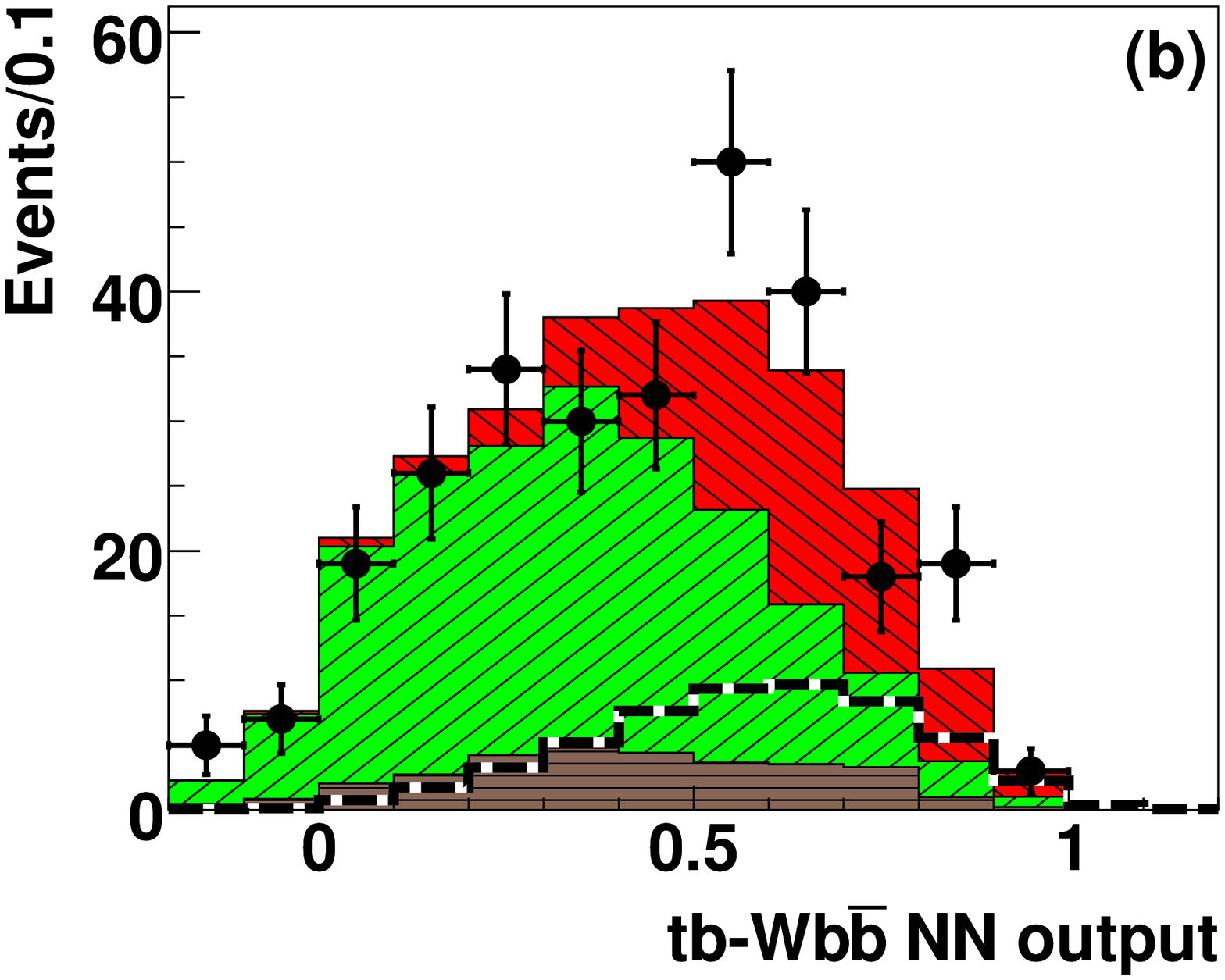}
\caption{Comparison of signal, background, and 
data for the neural network outputs in the $s$-channel, for the electron and muon
channels combined, requiring at least one $b$-tag.  This figure shows 
(a) the $\ttbar$ filter and 
(b) the $Wb\bar{b}$ filter.
Signals are multiplied by ten.}
\label{nn-yield-compare-schan}
\end{figure*}

\begin{figure*}[!h!tbp]
\includegraphics[width=0.45\textwidth]
{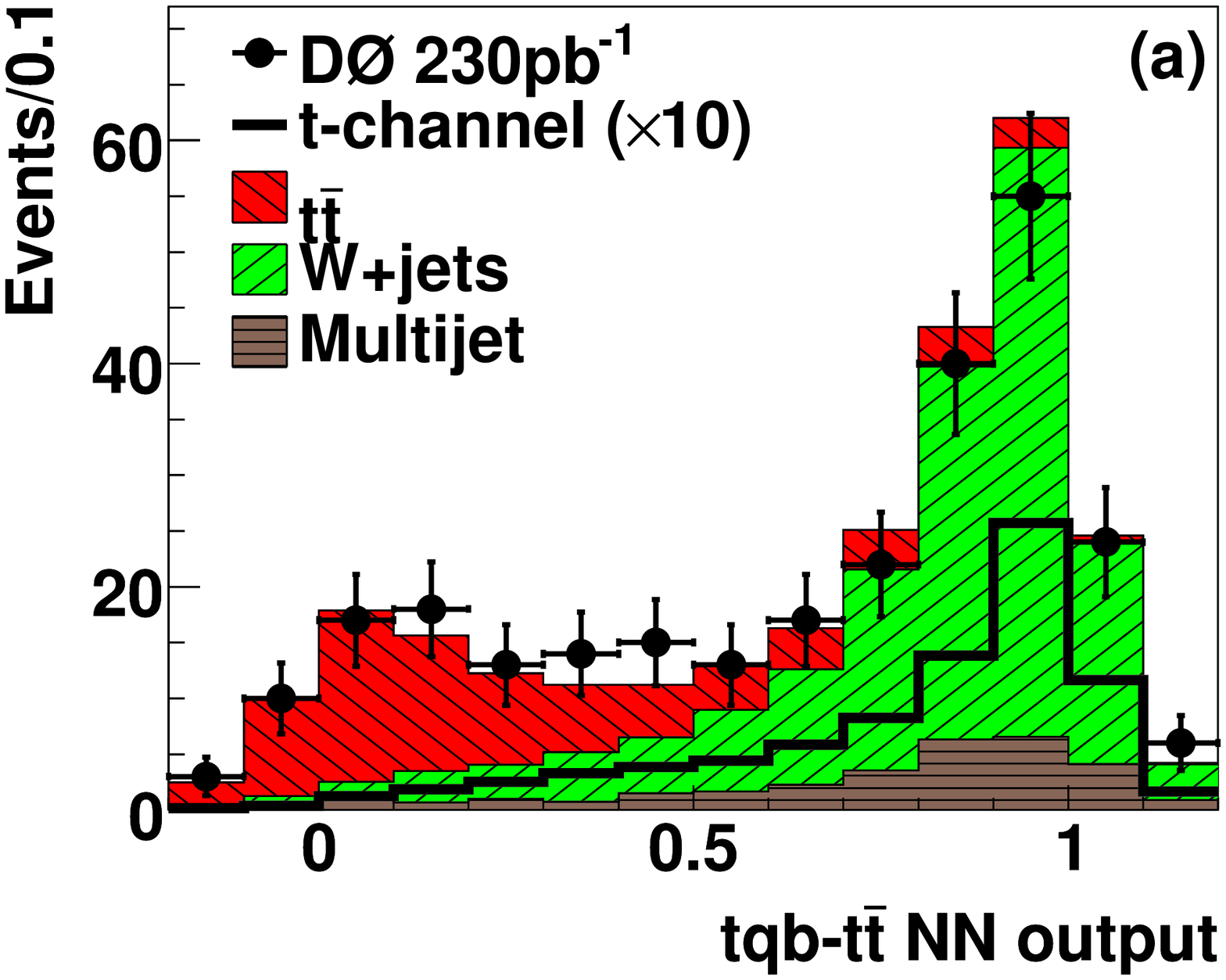}
\includegraphics[width=0.45\textwidth]
{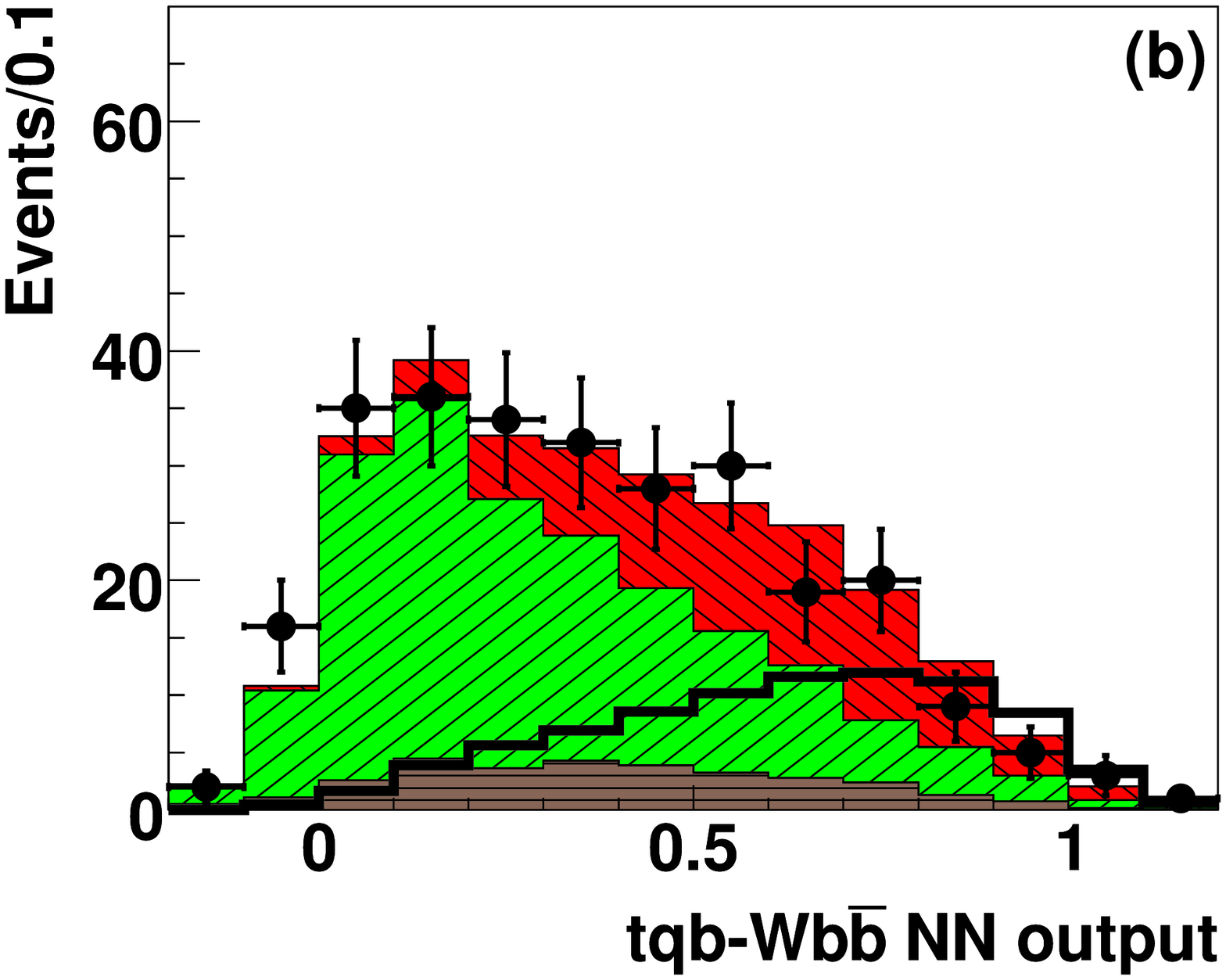}
\caption{Comparison of signal, background, and 
data for the neural network outputs in the $t$-channel, for the electron and muon
channels combined, requiring at least one $b$-tag.  This figure shows 
(a) the $\ttbar$ filter and 
(b) the $Wb\bar{b}$ filter.
Signals are multiplied by ten.}
\label{nn-yield-compare-tchan}
\end{figure*}

Figures~\ref{nn-yield-compare-schan} and \ref{nn-yield-compare-tchan}
show comparisons of the summed
backgrounds to data for the $s$-channel and $t$-channel searches, 
for electrons, muons, single-tagged, and double-tagged samples combined.
These distributions show that the background model reproduces the data very well. 
From the figures, it can be seen that the ${\ttbar}{\rar}\ell$+jets
filters do indeed separate the ${\ttbar}$ background which clusters
near zero, but does not affect the $W$+jets and multijet backgrounds,
which cluster near one.  Similarly, the $Wb\bar{b}$ filters
discriminate the $W$+jets and multijet backgrounds, which cluster
to the left of 0.5, but do not affect the ${\ttbar}$ 
background, which clusters to the right of 0.5.  
They also show that separation of the single top quark signal from 
background is not yet powerful enough since the background dominates
even in the regions where the signal peaks.

\begin{figure*}[!h!tbp]
\includegraphics[width=0.45\textwidth]
{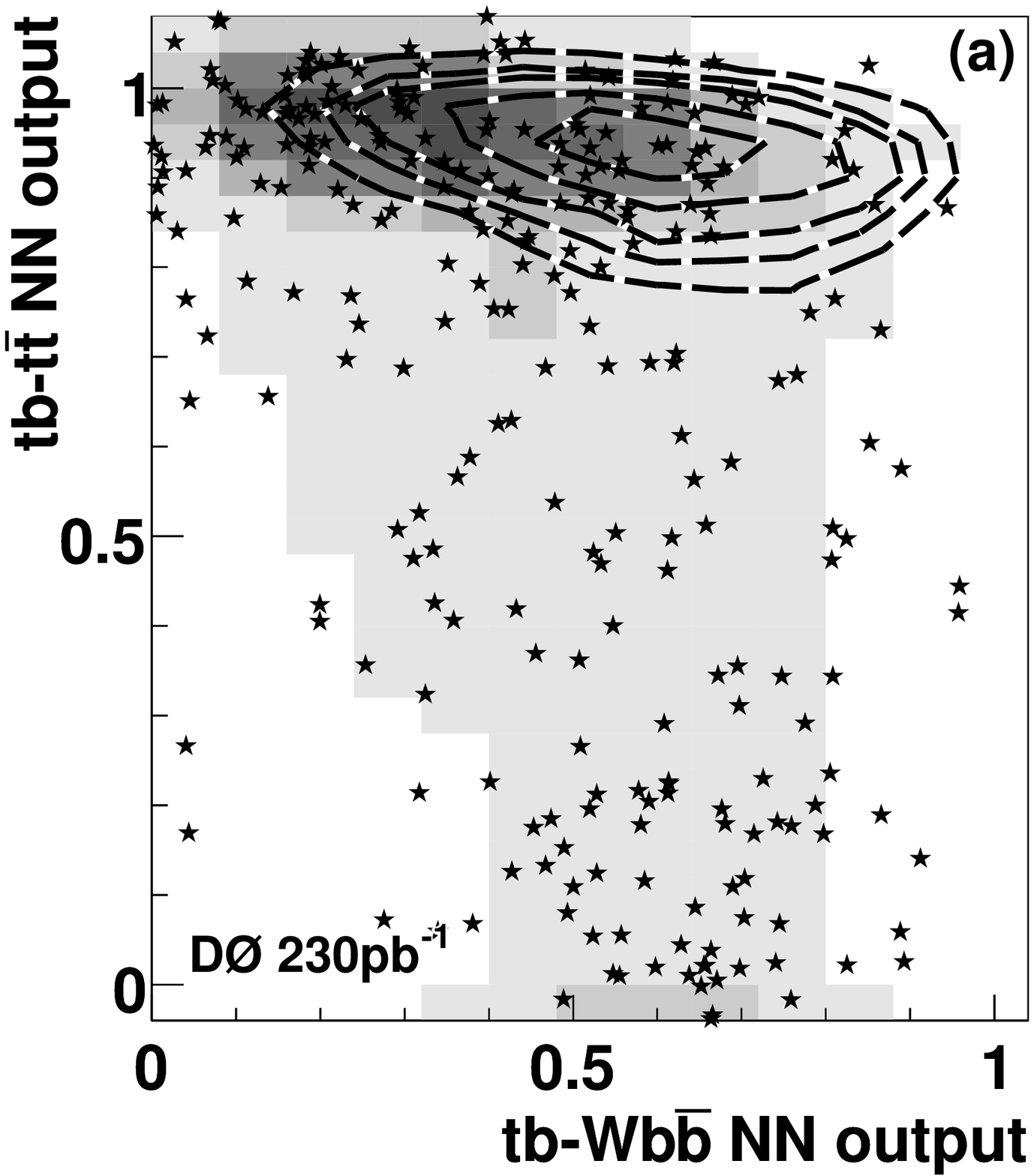}
\includegraphics[width=0.45\textwidth]
{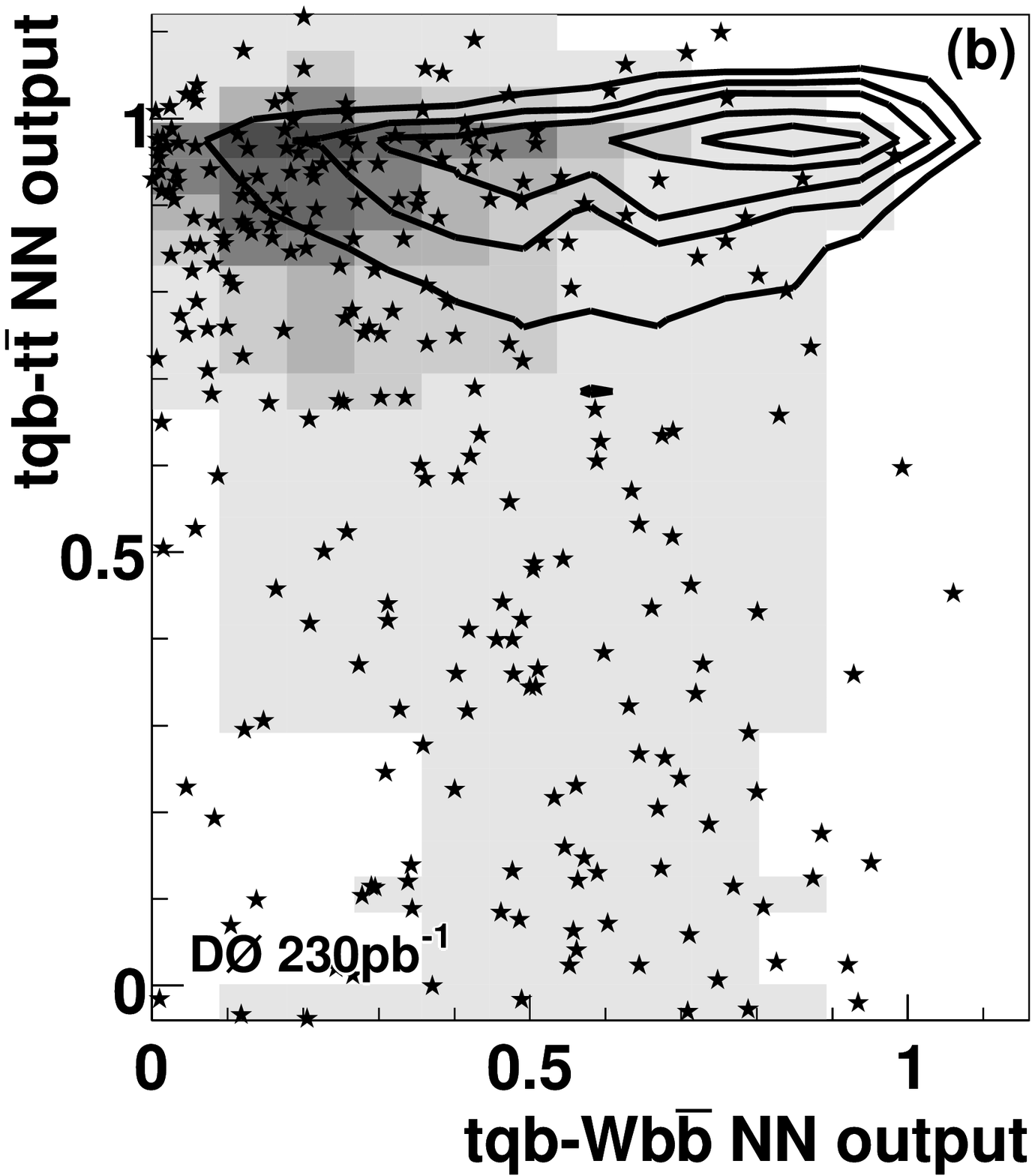}
\caption{Neural network outputs for both the $\ttbar$ versus $Wb\bar{b}$
filters in the (a) $s$-channel
(b) and $t$-channel analyses. The background sum is shown as the shaded
area, the signal as contour lines, and the data as stars.}
\label{nn-yield-2d-compare}
\end{figure*}

Figure~\ref{nn-yield-2d-compare} shows the output of the
$tb$-${\ttbar}$ network versus the $tb$-$Wb\bar{b}$ network, and similarly
for the $tqb$ networks, again for electrons, muons, single-tagged, and
double-tagged events combined.

\section{Systematic Uncertainties}
\label{systematics}

We consider several sources of systematic uncertainties in this
analysis, and study them separately for each signal and background
source. Some of the uncertainties affect acceptance for simulated
signals and backgrounds, others only affect background yield
estimates. This section lists the uncertainties for each signal and
background and their correlations.

We consider the following sources of systematic uncertainty:
\begin{itemize} 
\item The $b$-tag modeling uncertainty includes components for the
estimation of the $b$~tagging efficiency in data for the various quark
flavors, see Sec.~\ref{sec:bID}.
\item The jet energy calibration uncertainty reflects how well jet energies
measured in the simulation reflect jet energies measured in data, and includes
jet energy scale uncertainty as well as modeling of jet energy resolution
in the simulation, see Sec.~\ref{sec:objectreco}.
\item The trigger modeling uncertainty includes components for the 
estimation of the efficiency of the various trigger requirements in
data, see Sec.~\ref{sec:triggers}.
\item The jet fragmentation uncertainty covers the uncertainty in
modeling of initial- and final-state radiation as well as the
difference in the fragmentation model between {\pythia} and {\herwig}~\cite{d0runI}.
\item The uncertainty on the correction factor for simulated
samples to account for the jet identification efficiency as described in
Sec.~\ref{sec:objectreco}.
\item The uncertainty on the correction factor for lepton identification efficiency
in simulated samples as described in Sec.~\ref{sec:objectreco}.
\item The cross section and branching fraction
uncertainties from the yield normalization of simulated backgrounds.
\item The uncertainty on the normalization of the multijet and
$W$+jets background yields to the data.
\item The uncertainty on the integrated luminosity measurement.
\end{itemize}

The uncertainty on the multijet background normalization
includes two components: the estimate of the rate to
misidentify a jet as an isolated lepton in the data,
and the $b$-tagging probability in the multijet data sample.

The uncertainty for the $Wjj$ and $Wb\bar{b}$ backgrounds includes
several components: the normalization
of the $W$+jets background to data before $b$~tagging, the $b$-tagging probability
estimate, and the fraction of $Wb\bar{b}$ events in the $W$+jets sample.
Owing to the normalization to data, the
$Wb\bar{b}$ and $Wjj$ tagged yield estimates are not affected by any of the
systematic uncertainties that affect the other simulated samples. The exception
to this is $b$~tagging, which is applied after normalization. There
is still an effect on the shape of the $Wjj$ and $Wb\bar{b}$ 
distributions from uncertainty components that vary bin-by-bin. 

Table~\ref{tab:systematics} shows the systematic uncertainty
values for each signal and background component. The range is given
for the different analysis channels, electron and muon as well 
as single tags and double tags.

\begin{table*}[h!tbp]
\begin{center}
\caption{Range of relative systematic uncertainty values in percent for the various 
signal and background samples in the different
analysis channels. 
}
\begin{ruledtabular}
\begin{tabular}{lccccc}
 & $tb$  & $tqb$ & ${\ttbar}$ & $W$+jets & multijet \\
\hline
\multicolumn{4}{l}
{\bf Signal and background acceptance}                  &         &  \\
~~$b$-tag modeling      &  5 -- 20& 8 -- 20 & 6 -- 20& 7 -- 20 & ---  \\ 
~~Jet energy calibration&  6 -- 20& 6 -- 15 & 3 -- 11& ---     & ---  \\ 
~~Trigger modeling      &  2 -- 6 &  2 -- 6 &  ---   &  ---    & ---  \\
~~Jet fragmentation     &  5      &  5      &  7     &   ---   & ---  \\
~~Jet identification    &  1 -- 13&  5 -- 11& 1 -- 4 &    ---  & ---  \\
~~Lepton identification &  4      &   4     &    4   &     --- & ---  \\ 

\multicolumn{4}{l}
{\bf Background normalization}                           &         &      \\
~~Theory cross sections &  16     &  15     & ---    &  ---    & ---  \\
~~Normalization to data &   ---   & ---     & ---    & 5 -- 16 &  5 -- 16  \\
{\bf Luminosity}              &   6.5   &  6.5    &  6.5   &  ---    & ---  \\ 

\end{tabular}
\end{ruledtabular}
\label{tab:systematics}
\end{center}
\end{table*}

Note that the $W$+jets background includes small contributions from
$WW$ and $WZ$, whose uncertainties are also included in the limit setting calculation.
Furthermore, the normalization for $Wb\bar{b}$ and $Wjj$ accounts for the
other simulated backgrounds and thus their uncertainties in principle also
affect $Wb\bar{b}$ and $Wjj$.  However, the other simulated backgrounds only
contribute about 3\% to the pretagged yield, which means their
uncertainties are negligible compared to the overall normalization
uncertainties.

\section{Cross Section Limits}
\label{section_limits}
We use a Bayesian approach \cite{IainTM2000} to calculate limits on the cross
section for single top quark production in the $s$-channel and $t$-channel
modes. The limits are derived from a likelihood function that is proportional to
the probability to obtain the number of observed events. In the cut-based
analysis, we count the total number of observed events, and in the neural
network analysis, we use the two-dimensional distributions of the $t\bar{t}$
versus $Wb\bar{b}$ network outputs.

\vspace{0.1in}
\subsection{Bayesian Approach}

We assume that the probability to observe a count $D$, if the mean 
count is $d$, is given by the Poisson distribution:
\begin{equation}
\label{eq:poisson}
p(D|d) = \frac{e^{-d} \, d^D}{\Gamma(D+1)} \, ,  
\end{equation}
where $\Gamma$ is the gamma function. The mean count $d$ is a sum of the 
predicted contributions from the signal and background sources:
\begin{equation}
\label{eq:MeanCount}
d = \alpha \, \mathcal{L} \, \sigma + \sum_{i=1}^N b_i \equiv a \sigma +
\sum_{i=1}^N b_i \, ,
\end{equation}
where $\alpha$ is the signal acceptance, $\mathcal{L}$ the integrated
luminosity, $\sigma$ the signal cross section (the quantity of
interest), $b_i$ the mean count for background source $i$, and $a
\equiv \alpha \, \mathcal{L}$ is the effective luminosity for the signal. For the 
$s$-channel ($t$-channel) search, the background $b_i$ includes the 
$t$-channel ($s$-channel) process. The likelihood function 
$L(D|d)$ is proportional to $p(D|d)$. 

For two or more independent channels, we simply replace the 
single channel likelihood by a product of likelihoods:
\begin{equation}
\label{eq:channellLikelihood}
L({\bf D}|{\bf d}) \equiv L({\bf D}|\sigma, {\bf a}, {\bf b}) = \prod_{i=1}
^{M} \, L(D_{i}|d_{i}) \, ,
\end{equation}
where ${\bf D}$ and ${\bf d}$, respectively, represent vectors of the observed counts and the 
mean counts for the sources of signal and background, in the $M$ different channels. 
In addition, given $K$ bins of any distribution, 
we calculate the likelihood for each channel as the product of the individual likelihoods in each bin:
\begin{equation}
\label{eq:overallLikelihood}
L(D_{i}|d_{i}) = \prod_{j=1}^K \, L(D_{i, j}|d_{i, j}) \, ,
\end{equation}
which is true if the probability to observe a count in a given bin is
independent of the count in other bins. 

We use Bayes' theorem to compute the posterior probability density of the
parameters, $p(\sigma, {\bf a}, {\bf b}| {\bf D})$, which is then integrated
with respect to the parameters ${\bf a}$ and ${\bf b}$ to obtain the posterior
density for the signal cross section, given the observed distribution of counts
${\bf D}$:
\begin{equation}
\label{eq:posterior}
p(\sigma | {\bf D}) = \frac{1}{\mathcal{N}} \int \int L({\bf D} | \sigma, 
{\bf a}, {\bf b}) \pi(\sigma, {\bf a}, {\bf b}) \, d{\bf a} \, d{\bf b}\, .
\end{equation}
Here $\mathcal{N}$ is an overall normalization obtained 
from the requirement $\int { p(\sigma|{\bf \rm D})}d\sigma = 1$, and 
$\pi(\sigma, {\bf a}, {\bf b})$ is the prior probability that encodes 
what we know about the parameters $\sigma$, ${\bf a}$ and
${\bf b}$. We assume that any prior knowledge of ${\bf a}$ and 
${\bf b}$ is independent of the cross section $\sigma$, in which case we may 
write the prior density as
\begin{eqnarray}
\label{eq:prior}
\nonumber \pi(\sigma, {\bf a}, {\bf b}) &=& \pi({\bf a}, {\bf b} | \sigma) \,\pi(\sigma) \, \\
&=& \pi({\bf a}, {\bf b}) \,\pi(\sigma) \, .
\end{eqnarray}
We use a flat prior for $\sigma$: $\pi(\sigma) = 1/\sigma_{\rm max}$, 
where $\sigma_{\rm max}$ is any sufficiently high upper bound on the 
cross section. The posterior probability density for 
the signal cross section is therefore
\begin{equation}
\label{eq:finalPosterior}
p(\sigma | {\bf D}) = \frac{1}{\mathcal{N}} \int \int L({\bf D} | \sigma, 
{\bf a}, {\bf b}) \pi({\bf a}, {\bf b}) \, d{\bf a} \, d{\bf b} \, . 
\end{equation}
The Bayesian upper limit $\sigma_{\rm UL}$ at confidence level $\beta$ 
is the solution of
\begin{equation}
\label{eq:BayesUpperLimit}
\int_0^{\sigma_{\rm UL}} p(\sigma|{\bf D}) \, d\sigma \,= \beta  \label{CLlimit} \, .
\end{equation}
The integral in Eq.~\ref{eq:finalPosterior} is done numerically using Monte
Carlo importance sampling: we generate a large number $K$ of randomly sampled
points $({\bf a}_k, {\bf b}_k)$ that represents the prior density $\pi({\bf a},
{\bf b})$, and estimate the posterior using
\begin{equation}
\label{eq:MCIntegration}
\int \int L({\bf D} | \sigma, 
{\bf a}, {\bf b}) \pi({\bf a}, {\bf b}) \, d{\bf a} \, d{\bf b} = 
\frac{1}{K} \sum_{k=1}^K \, L({\bf D}|\sigma, {\bf a}_k,
{\bf b}_k) \, .
\end{equation}

\subsection{Definition of the Prior Probability}
The prior $\pi({\bf a}, {\bf b})$ encodes our knowledge of the effective signal
luminosities and the background yields: we have estimates of the parameters and
the associated uncertainties from the different systematic effects discussed in
Sec.~\ref{systematics}. In the case of the cut-based analysis, since we consider
the total yield for any source of signal or background, the different
uncertainties affect the overall normalization only. In the neural network
analysis, since we consider distributions, we separate the uncertainties into
two classes: those that alter only the overall normalization, such as the
luminosity measurement and theory cross sections; and those that also alter the
shapes of distributions, such as the trigger modeling, jet energy calibration,
jet energy resolution, jet identification, and $b$-tag modeling.

The normalization effects are modeled by sampling the effective
signal luminosities {\bf a} and the background yields {\bf b} from 
a multivariate Gaussian, with a vector of means given by the estimates of 
the yields, and covariance matrix computed from the associated 
uncertainties. The covariance matrix takes into account the correlations 
of the systematic uncertainties across the different sources of signal and background.  
Each entry in the covariance matrix is calculated as 
follows: 
\begin{equation}
\label{eq:CovarianceMatrix}
c_{ij} = y_i y_j \sum_{k} f_{ik}f_{jk} \, ,
\end{equation}
where $y_i$ ($y_j$) is the yield for the $i^{\rm th}$ ($j^{\rm th}$) source of
background or signal from Table~\ref{tab:yields}, and $f_{ik}$ is the
corresponding fractional uncertainty from the $k^{\rm th}$ component of
systematic uncertainty, for the $i^{\rm th}$ source.

The shape effects are modeled by shifting, one by one, the trigger 
modeling, jet energy calibration, $b$-tag modeling, and so on, by 
plus or minus one standard deviation with respect to their nominal 
values. For each systematic effect, we have three distributions: 
the nominal, and those from the plus and minus shifts. The  
systematic uncertainty in each bin is then sampled from a 
Gaussian distribution with mean defined 
by the nominal yield in that bin, and width defined by the plus and minus shifts.
The sampled shifts are added linearly to the yields 
generated from the sampling of the normalization-only systematic uncertainties. 
We assume that any shape-changing systematic is 100$\%$ 
correlated across all bins and sources.

\section{Results}
\label{results}

For both the $s$-channel and $t$-channel searches, we compute an observed limit as
well as an expected limit. We define the latter as the limit obtained if the
observed counts were equal to the background prediction. The different tag
multiplicities ($=1$~tag and $\geq2$~tags) and lepton flavor (electron and muon)
are combined as shown in Eq.~\ref{eq:channellLikelihood}.

\begin{figure}[!h!tbp]
\begin{center}
\includegraphics[width=0.45\textwidth]
{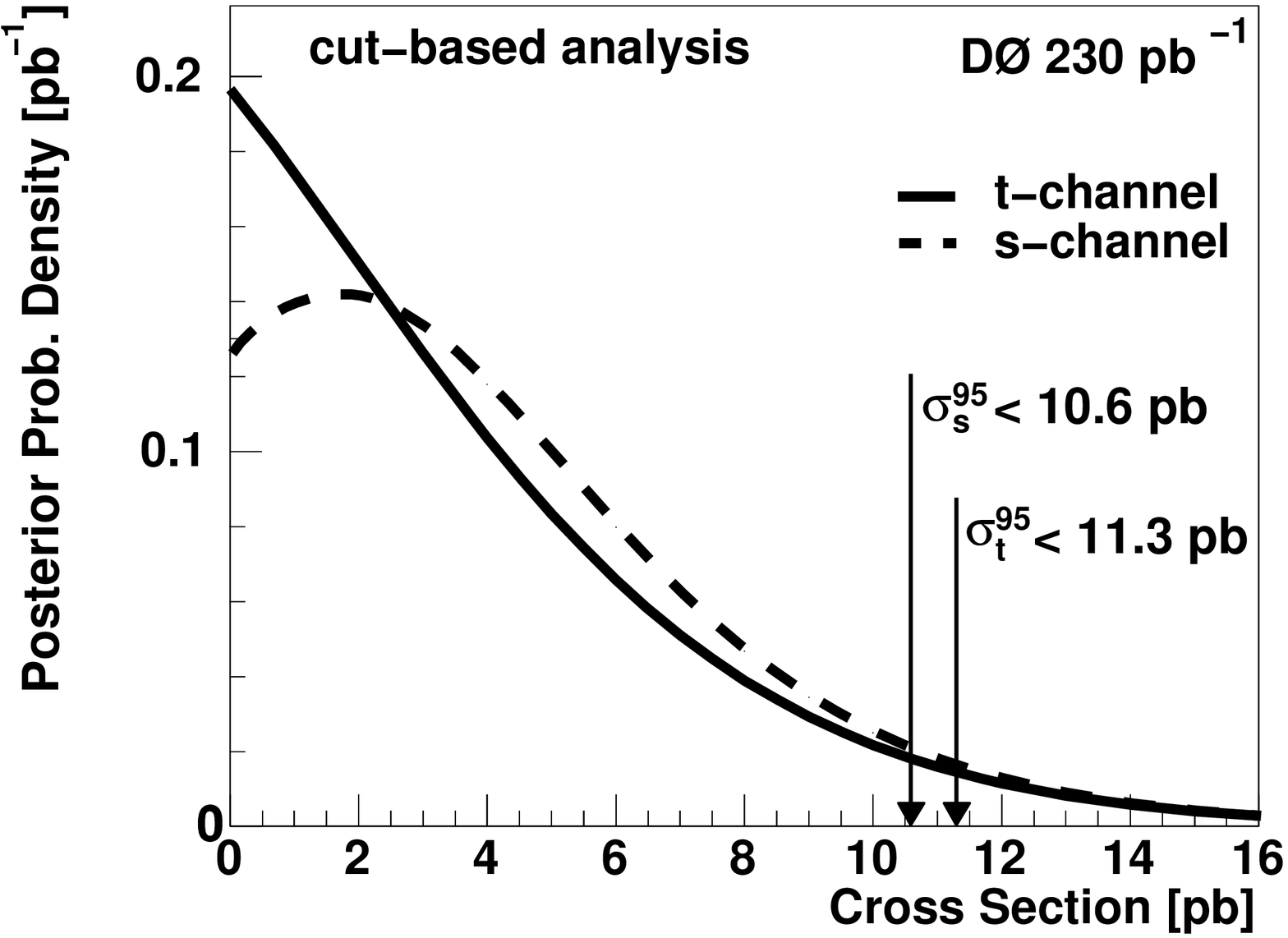}
\end{center}
\vspace{-0.5cm}
\caption{The observed posterior probability density 
as a function of the single top quark cross section for the 
cut-based analysis, for the electron and muon channels 
combined in the $s$-channel and the $t$-channel searches.}
\label{cb-posterior-1d}
\end{figure}

\begin{figure}[!h!tbp]
\begin{center}
\includegraphics[width=0.45\textwidth]
{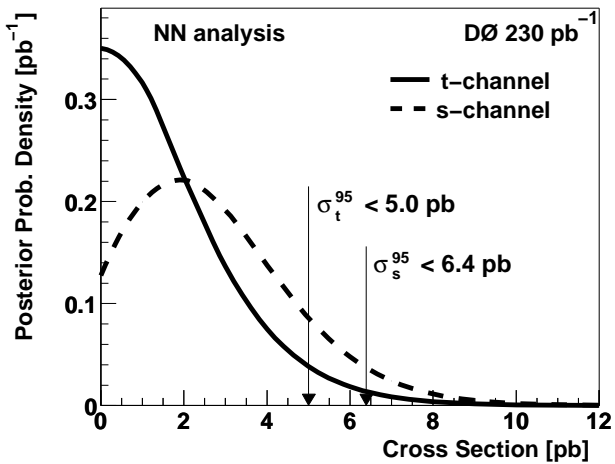}
\end{center}
\vspace{-0.5cm}
\caption{The observed posterior probability density 
as a function of the single top quark cross section for the 
neural network analysis, for the electron and muon channels 
combined in the $s$-channel and the $t$-channel searches.}
\label{nn-posterior-1d}
\end{figure}

The expected and observed upper limits at the 95\% confidence level, after the
initial event selection, and from the cut-based and neural network analyses, are
shown in Table~\ref{limits} for the electron and muon channels combined, and
with all systematic effects included. We see that the limits improve upon
applying cuts on the discriminating variables, but that tighter limits are
obtained when the variables are combined using our neural networks method. The
observed posterior probability densities as a function of the $s$-channel and
$t$-channel cross sections are shown in Fig.~\ref{cb-posterior-1d} for the
cut-based analysis and in Fig.~\ref{nn-posterior-1d} for the neural network
analysis.

\begin{table}[!h!tbp]
\begin{center}
\caption[counting-limits-cuts]{Expected and observed upper limits (in picobarns) at 
the 95\% confidence level, on the production cross sections of single top quarks
in the $s$-channel ($tb$) and $t$-channel ($tqb$) searches, for the electron and
muon channels combined, with all systematic effects included.}
\label{limits}
\begin{ruledtabular}
\begin{tabular}{lcccc}
& \multicolumn{2}{l}{Expected Limits} & \multicolumn{2}{l}{Observed Limits} \\
& $tb$ & $tqb$ & $tb$ & $tqb$ \\ \hline
Initial selection & 14.5 & 16.5 & 13.0 & 13.6\\
Cut-based         & 9.8  & 12.4 & 10.6 & 11.3 \\
Neural networks   & 4.5  &  5.8 &  6.4 & 5.0 \\   
\end{tabular}
\end{ruledtabular}
\end{center}
\end{table}

The method described so far yields limits on
the $s$-channel or $t$-channel cross sections separately. This
requires some assumptions about whichever of the two signal 
processes is not being considered. In this particular analysis,
we have assumed that in the
$s$-channel ($t$-channel) search, the $t$-channel ($s$-channel)
contributes as a SM background.
This assumption is, however, not necessary. Instead, we can
set limits on both the $s$-channel and $t$-channel cross
sections simultaneously. We accomplish this by generalizing the
likelihood so that it depends explicitly on the two cross sections 
$\sigma_s$ and $\sigma_t$. 
Equation~\ref{eq:MeanCount} for the mean count $d$ then becomes:
\begin{equation}
\label{eq:tb_tqb_MeanCount}
d =  \alpha_{s} \, \mathcal{L} \, \sigma_{s} +  \alpha_{t} \, \mathcal{L} \, \sigma_{t} +
\sum_{i} b_i \, .
\end{equation}
The backgrounds $b_i$ now include only the non-single top quark sources. 

In order to exploit the sensitivity to both the $s$-channel and
$t$-channel signals, we combine the output of the neural networks in
both searches. We calculate a signal probability $P$ in each bin of
the histograms in Fig.~\ref{nn-yield-2d-compare}:

\begin{equation}
  P_{s(t)} = \frac{n_{s(t)}}{n_s + n_t + \sum_{i} b_i} \, ,
\end{equation}
for the $s$-channel ($t$-channel) search, where $n_s$ and $n_t$ are the yields
for the $tb$ and $tqb$ samples, respectively, and the sum in the denominator is
over all the non-single top quark backgrounds in that bin. We then evaluate
$P_s$ and $P_t$ simultaneously for each event and fill histograms of $P_s$
versus $P_t$. As before, we consider a Poisson probability for the likelihood in
each bin. We assume a flat prior in the plane of $\sigma_{s}$ versus
$\sigma_{t}$, which is equivalent to flat priors for either cross
section. Equations~\ref{eq:finalPosterior} and \ref{eq:MCIntegration} can then
be used to define the posterior probability density for different values of the
$s$- and $t$-channel cross sections. The limit at a fixed confidence level is
then given by a contour of constant probability enclosing a fraction of volume
corresponding to this confidence level using an equation analogous to
Eq.~\ref{CLlimit}, but in two dimensions.

\begin{figure}[!h!tbp]
\includegraphics[width=0.50\textwidth,height=0.30\textheight]
{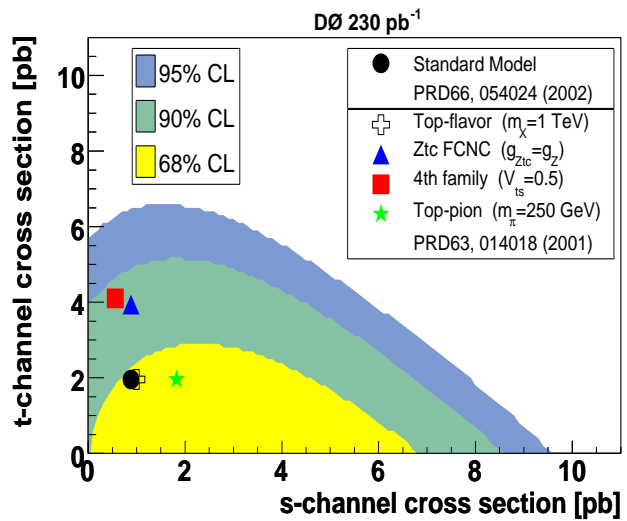}
\vspace{-0.3in}
\caption[nn-posterior-2d-nonsm]{ Exclusion contours at the 68\%, 90\%, and 95\%
confidence levels on the observed posterior density distribution as a function of
both the $s$-channel and $t$-channel cross sections in the neural networks 
analysis. Several representative non-standard model
contributions from Ref.~\cite{Tait:2000sh} are also shown.}
\label{nn-posterior-2d-nonsm}
\end{figure}

Figure~\ref{nn-posterior-2d-nonsm} shows contours of observed posterior 
density in the $\sigma_s$ versus $\sigma_t$ plane for the neural network analysis. 
To illustrate the sensitivity of this analysis to different
contributions, the expected SM cross section as well as
several representative non-SM contributions are also shown~\cite{Tait:2000sh}.

\vspace{-0.1in}

\section{Summary}
\label{sec:conclusions} 
We have analyzed electron+jet and muon+jet events containing exactly one or more
than one $b$ jet, identified with a secondary-vertex algorithm, and find no
evidence for the electroweak production of single top quarks in 230~pb$^{-1}$ of
data collected by the \dzero detector at $\sqrt{s}=1.96$~TeV. The upper limits
at the 95\% confidence level on the cross section for $s$-channel and $t$-channel
processes are 10.6 pb and 11.3 pb, respectively, using event counts in a
cut-based analysis, and 6.4 pb and 5.0 pb, respectively, using binned
likelihoods in a neural network analysis. The neural network-base
limits presented here and in Ref.~\cite{RunII:d0_result} are
significantly more stringent than those previously 
published~\cite{d0runI,Acosta:2001un,RunII:cdf_result}. They are also
close to the sensitivity required to probe models of physics beyond
the standard model.

\section{Acknowledgements}
%
We thank the staffs at Fermilab and collaborating institutions, 
and acknowledge support from the 
DOE and NSF (USA);
CEA and CNRS/IN2P3 (France);
FASI, Rosatom and RFBR (Russia);
CAPES, CNPq, FAPERJ, FAPESP and FUNDUNESP (Brazil);
DAE and DST (India);
Colciencias (Colombia);
CONACyT (Mexico);
KRF and KOSEF (Korea);
CONICET and UBACyT (Argentina);
FOM (The Netherlands);
PPARC (United Kingdom);
MSMT (Czech Republic);
CRC Program, CFI, NSERC and WestGrid Project (Canada);
BMBF and DFG (Germany);
SFI (Ireland);
The Swedish Research Council (Sweden);
Research Corporation;
Alexander von Humboldt Foundation;
and the Marie Curie Program.
%


\begin{thebibliography}{99}
\bibitem{fiducial_endnote}
Pseudorapidity is defined as $\eta = - \ln(\tan \frac{\theta}{2})$, 
where $\theta$ is the polar angle with the origin at the primary vertex.
\end{thebibliography}
\end{document}